\documentclass[a4paper,11pt]{article}
\pdfoutput=1
\usepackage{jheppub}
\usepackage[T1]{fontenc}
\usepackage{caption}
\usepackage{subcaption}
\usepackage{bm}
\title{\boldmath A new method for obtaining a Born cross section using visible
  cross section data from $e^+e^-$ colliders} \author[a,b]{S.S. Gribanov,}
\author[a,b]{A.S. Popov} \affiliation[a]{Budker Institute of Nuclear Physics, SB
  RAS,\\Prospekt Akademika Lavrent’yeva, 11, Novosibirsk, 630090 Russia}
\affiliation[b]{Novosibirsk State University, Physics Department,\\Pirogova, 2,
  Novosibirsk, 630090 Russia} \emailAdd{S.S.Gribanov@inp.nsk.su} \abstract{In
  this paper, we propose a new method for obtaining a Born cross section using
  visible cross section data. It is assumed that the initial state
  radiation is taken into account in a visible cross section, while in a Born
  cross section this effect is ommited. Since the equation that connects Born
  and visible cross sections is an integral equation of the first kind, the
  problem of finding its numerical solution is ill-posed. Various
  regularization-based approaches are often used to solve ill-posed problems,
  since direct methods usually do not lead to an acceptable result. However, in
  this paper it is shown that a direct method can be successfully used to
  numerically solve the considered equation under the condition of a small beam
  energy spread and uncertainty. This naive method is based on finding a
  numerical solution to the integral equation by reducing it to a system of
  linear equations. The naive method works well because the kernel of the
  integral operator is a rapidly decreasing function of the variable $x$. This
  property of the kernel leads to the fact that the condition number of the
  matrix of the system of linear equations is of the order of unity, which makes
  it possible to neglect the ill-posedness of the problem when the above
  condition is satisfied. The advantages of the naive method are its model
  independence and the possibility of obtaining the covariance matrix of a Born
  cross section in a simple way.

  It should be noted that there are already a number of methods for obtaining a
  Born cross section using visible cross section data, which are commonly used
  in $e^+e^-$ experiments. However, at least some of these methods have various
  disadvantages, such as model dependence and relative complexity of
  obtaining a Born cross section covariance matrix. It should be noted that this
  paper focuses on the naive method, while conventional methods are
  hardly covered. The paper also discusses solving the problem using the
  Tikhonov regularization, so that the reader can better understand the
  difference between regularized and non-regularized solutions. However, it
  should be noted that, in contrast to the naive method, regularization methods
  can hardly be used for precise obtaining of a Born cross section. The reason
  is that the regularized solution is biased and the covariance matrix of this
  solution do not represent the correct covariance matrix of a Born cross
  section.}

\begin{document}
\maketitle
\flushbottom

\section{Introduction}
\subsection{Relationship between Born and visible cross sections}
\begin{figure}[tbp]
  \centering
  \includegraphics[width=\textwidth]{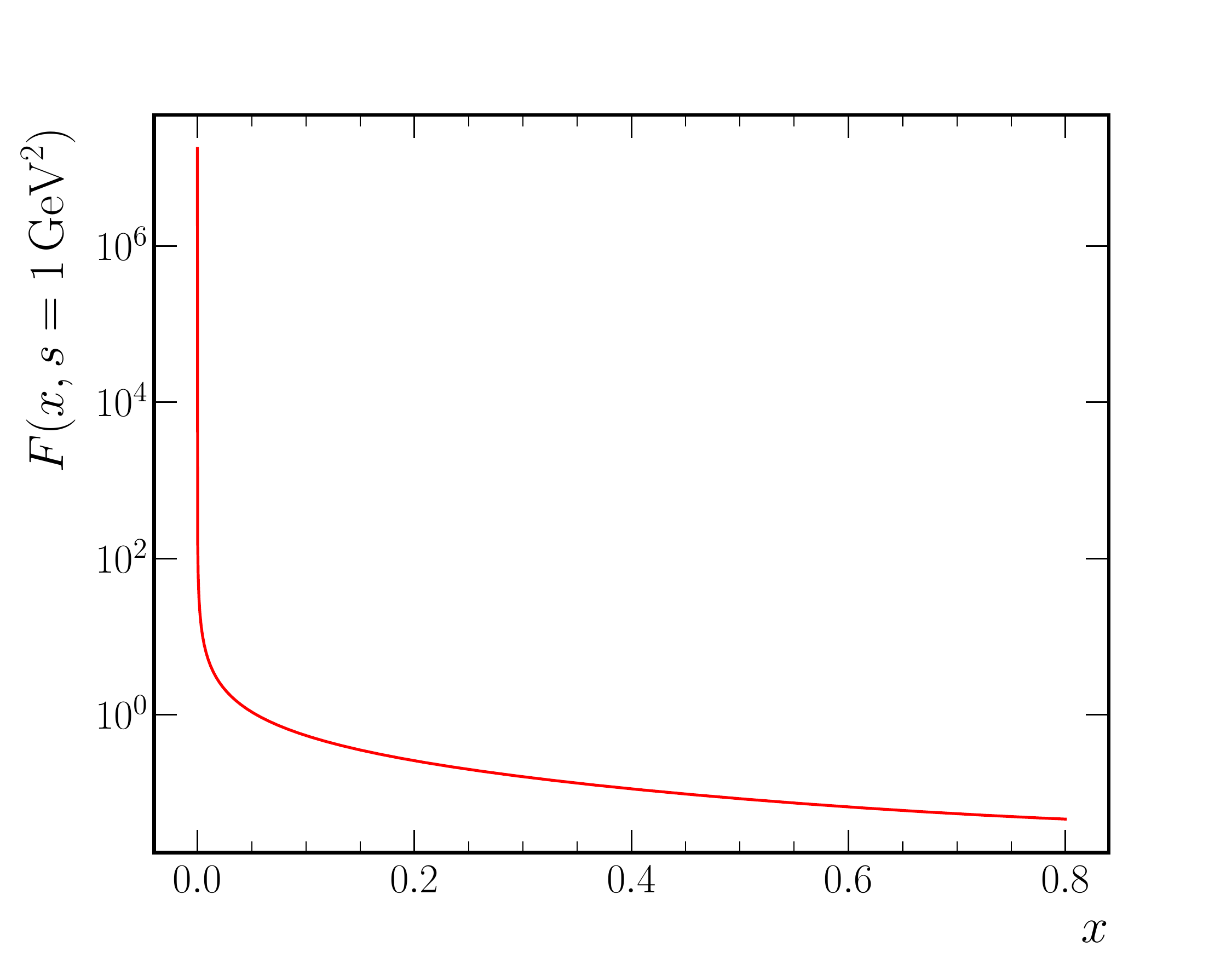}
  \caption{\label{fig:kernel-kuraev-fadin} The dependence of kernel function
    $F(x, s)$ on $x$ at $\sqrt{s}=1\text{ GeV}$.}
\end{figure}

The precise measurement of the inclusive cross section of an $e^+e^-$
annihilation to hadrons is one of the goals of experiments carried out with
electron-positron colliders. This cross section is of interest in connection
with the measurement of the fine structure constant and the problem of the muon
anomalous magnetic moment. The inclusive cross section of an $e^+e^-$
annihilation to hadrons is considered as the sum of the cross sections for
exclusive processes of an $e^+e^-$ annihilation into different hadronic states.
In such experiments, a visible cross section $\sigma_{\rm vis}$ is measured,
while a Born cross section $\sigma_{\rm Born}$ is of interest. In this paper, we
assume that the initial state radiation is taken into account in a visible
cross section, while in a Born cross section this effect is omitted. Visible and
Born cross sections are related by the following integral equation:
\begin{equation}
  \label{eq:visible-Born-relationship-kuraev-fadin}
  \sigma_{\rm vis}(s) = \int\limits^\epsilon_{0}
  F(x, s)\varepsilon(x, s)\sigma_{\rm Born}(s(1-x))dx,
\end{equation}
where $\varepsilon(x, s)$ is the detection efficiency, that depends on
center-of-mass energy (c.m.\ energy) $\sqrt{s}$ and the energy fraction carried away by initial
state radiation. The upper limit $\epsilon$ of the integral in
eq.~\eqref{eq:visible-Born-relationship-kuraev-fadin} is determined by the
experimental conditions. The visible cross section of a certain process can be
found as the ratio of the event yield $N(s)$ of this process to the integral
luminosity $L_{\rm int}(s)$:
\begin{equation}
  \label{eq:visible-cross-section}
  \sigma_{\rm vis}(s) = \frac{N(s)}{L_{\rm int}(s)}.
\end{equation}
The kernel function $F(x, s)$ from
eq.~\eqref{eq:visible-Born-relationship-kuraev-fadin} has the following form:
\begin{equation}
  \label{eq:kernel-kuraev-fadin}
  \begin{split}
  F(x, s) &= \beta x^{\beta - 1}\left(1 + \frac{\alpha}{\pi}
  \left(\frac{\pi^2}{3} - \frac{1}{2}\right) + \frac{3}{4}\beta - \frac{1}{24} \beta^2
  \left(\frac{1}{3}L  + 2 \pi^2 - \frac{37}{4}\right)\right)\\
& - \beta\left(1 - \frac{1}{2}x\right) + 
  \frac{1}{8} \beta^2 \left(4 \left(2 - x \right)
    \ln\frac{1}{x} + \frac{1}{x} \left(1 + 3
      \left(1 - x\right)^2\right)
    \ln\frac{1}{1 - x} - 6 + x\right)\\
  & + \frac{\alpha^2}{\pi^2}
  \left[\frac{1}{6x}\left(x - \frac{4m}{\sqrt{s}}\right)^{\beta}
  \left(\ln\frac{s x^2}{m^2} - \frac{5}{3}\right)^2
  \left.\bigg(2 - 2x + x^2 + \frac{1}{3} \beta \left(\ln\frac{s x^2}{m^2} -
          \frac{5}{3}\right)\right)\right.\\
      &\left. + \frac{1}{2} L^2
  \left(\frac{2}{3} \frac{1 - {\left(1 - x\right)}^3}
  {1 - x} - \left(2 - x\right) \ln\frac{1}{1 - x} +
  \frac{1}{2} x\right)\right]\theta\left(x - \frac{4m}{\sqrt{s}}\right),
\end{split}
\end{equation}
where $\alpha$ is the fine structure constant, $m$ is the electron mass, $L =
\ln\frac{s}{m^2}$ and $\beta = \frac{2\alpha}{\pi}(L - 1)$. The function $F(x,
s)$ has an integrable singularity at $x=0$. The dependence of this function on
$x$ is shown in figure~\ref{fig:kernel-kuraev-fadin} at $\sqrt{s} = 1\text{
  GeV}$. The values of the $F(x, s)$ at $\sqrt{s} = 1\text{ GeV}$ are also
listed in table~\ref{tab:kernel-kuraev-fadin}. The relationship between visible
and Born cross sections, as well as the form of the function $F(x, s)$, were
first obtained by E.A.\ Kuraev and V.S.\ Fadin in the work~\cite{Kuraev:1985hb}.
\begin{table}[tbp]
  \centering
  \begin{tabular}[t]{|cc|}
    \hline
    $x$ & $F(x, s)$ \\
    \hline
    $1\times10^{-4}$ & $377.886$\\
    $1\times10^{-3}$ & $43.9340$\\
    $1\times10^{-2}$ & $5.07188$\\
    $0.1$ & $0.54227$\\
    $0.2$ & $0.25807$\\
    $0.3$ & $0.16101$\\
    $0.4$ & $0.11261$\\
    $0.5$ & $0.08422$\\
    $0.6$ & $0.06611$\\
    $0.7$ & $0.05408$\\
    $0.8$ & $0.04619$\\
    \hline
  \end{tabular}
  \caption{\label{tab:kernel-kuraev-fadin} The dependence of kernel function
    $F(x, s)$ on $x$ at $\sqrt{s} = 1\text{ GeV}$.}
\end{table}

Taking into account that a Born cross section is equal to zero at the energies
below the threshold, eq.~\eqref{eq:visible-Born-relationship-kuraev-fadin} can be
rewritten as follows:
\begin{equation}
  \label{eq:visible-Born-relationship-kuraev-fadin-2}
  \sigma_{\rm vis}(s) = \int\limits^{1 - s_{\rm T}/s}_{0}
  F(x, s)\varepsilon(x, s)\sigma_{\rm Born}(s(1-x))dx,
\end{equation}
where $s_{\rm T}$ is the square of the threshold energy. The fact that the upper
integral limit $\epsilon$ form
eq.~\eqref{eq:visible-Born-relationship-kuraev-fadin} can be less than $1 -
s_{\rm T} / s$ is taken into account in 
eq.~\eqref{eq:visible-Born-relationship-kuraev-fadin-2} due to detection
efficiency $\varepsilon(x, s)$. In this work, we assume that the efficiency is
always equal to unity, unless otherwise stated. It should also be noted that
eq.~\eqref{eq:visible-Born-relationship-kuraev-fadin-2} is the main equation
considered in this paper. Further, it will be shown that due to the properties
of the kernel, this equation can be solved numerically with good accuracy
without using regularization.

It should be noted that the beam energy in colliders has a spread. For instance,
this spread is of the order of $1\text{ MeV}$ in the case of the
VEPP-$2000$~\cite{Danilov:860802,vepp-berkaev-2012}. It should also be noted
that the beam energy is measured with some uncertainty. In order to take these
effects into account, eq.~\eqref{eq:visible-Born-relationship-kuraev-fadin-2}
needs to be modified as follows:
\begin{equation}
  \label{eq:visible-Born-relationship-kuraev-fadin-2-blured}
  \begin{split}
    \sigma_{\rm vis}(s) = \frac{1}{\sqrt{2\pi\sigma^2_E(s)}}\int\limits^{\infty}_{-\infty}
    \exp\left(-\frac{(E - \sqrt{s})^2}{2\sigma^2_E(s)}\right)\times\\
    \int\limits^{1 - s_{\rm T}/E^2}_{0}
    F(x, E^2)\varepsilon(x, E^2)\sigma_{\rm Born}(E^2(1-x))dx dE,
  \end{split}
\end{equation}
where $\sigma^2_E(s)$ is the sum of the squares of the c.m.\ energy spread and
the uncertainty of this energy. The c.m.\ energy spread is usually much greater
than the uncertainty of its measurement. Therefore, below we refer to the
parameter $\sigma^2_E(s)$ simply as the c.m.\ energy spread. It should be noted
that the value of this parameter depends on the experimental conditions, i.e.
this parameter can be different for different c.m.\ energy points.

\subsection{\label{sec:overview-other-methods}Brief overview of conventional methods for obtaining a Born cross section}
Let us take a quick look at some of the techniques that are commonly used to
obtain a Born cross section using visible cross section data. One of the most
frequently used methods is the fit to a visible cross section using
eq.~\eqref{eq:visible-Born-relationship-kuraev-fadin-2} or
eq.~\eqref{eq:visible-Born-relationship-kuraev-fadin-2-blured}, where a Born
cross section is specified in a certain model. The main disadvantage of this
approach is its model dependence. In addition, the result of this method is a
set of parameters with their uncertainties, on which a Born cross section
depends in the considered model. However, it is customary to present a Born
cross section in the form of points with error bars. In this form, a Born cross
section is more convenient to compare with results obtained in other
experiments. In order to obtain a Born cross section in the form of points with
error bars, it is first necessary to calculate the radiative correction. This
can be done using the following equation:
\begin{equation}
  \label{eq:radiative-correction}
  \varepsilon(s)(1 + \delta(s)) = \frac{\sigma^{\rm model}_{\rm vis}(s)}{\sigma^{\rm model}_{\rm Born}(s)},
\end{equation}
where $\sigma^{\rm model}_{\rm Born}$ and $\sigma^{\rm model}_{\rm vis}$ are
model Born and visible cross sections, respectively. In the case of using the
relationship between visible and Born cross sections given in the form of
eq.~\eqref{eq:visible-Born-relationship-kuraev-fadin-2} or
eq.~\eqref{eq:visible-Born-relationship-kuraev-fadin-2-blured}, the product
$\varepsilon(s)(1 + \delta(s))$ in eq.~\eqref{eq:radiative-correction} is
calculated as a whole. In eq.~\eqref{eq:radiative-correction}, $\varepsilon(s)$
means the average detection efficiency for each c.m.\ energy point, and
$\delta(s)$ means the radiative correction. A model visible cross
section is determined using the relationship between Born and visible
cross sections given by eq.~\eqref{eq:visible-Born-relationship-kuraev-fadin-2}
or eq.~\eqref{eq:visible-Born-relationship-kuraev-fadin-2-blured}, while the
dependence of a model Born cross section on c.m.\ energy and model parameters is
given. Further, a Born cross section $\sigma_{\rm Born}$ in the form of points
with error bars can be calculated using the following equation:
\begin{equation}
  \label{eq:bcs-using-vcs-fit}
\sigma_{\rm Born}(s_k)  = \frac{\sigma_{\rm vis}(s_k)}{\varepsilon(s_k)(1 + \delta(s_k))},
\end{equation}
where $k$ is the index of the c.m.\ energy point and $\sigma_{\rm vis}(s_k)$ is
the experimental data of a visible cross section. Next, the error propagation
formula can be used in order to calculate the statistical uncertainties of a
Born cross section at each c.m.\ energy point. The simplest way to do this is as
follows:
\begin{equation}
  \label{eq:error-propogation}
  \Delta\sigma_{\rm Born}(s_k)  = \frac{\Delta\sigma_{\rm vis}(s_k)}{\varepsilon(s_k)(1 + \delta(s_k))},
\end{equation}
where $\Delta\sigma_{\rm Born}(s_k)$ is the statistical uncertainty of a Born
cross section and $\Delta\sigma_{\rm vis}(s_k)$ is the statistical uncertainty
of a visible cross section. However, the statistical uncertainty of a Born cross
section found using eq.~\eqref{eq:error-propogation} is not correct. The reason
is that the values of a Born cross section at different c.m.\ energy points are
related by means of an integral equation. Consequently, the statistical
uncertainties of a Born cross section at different c.m.\ energy points are not
independent. To take into account the fact that the uncertainties are not
independent, an additional concept of the radiative correction uncertainty is
sometimes introduced. It should be noted, however, that the most reliable way to
obtain the statistical uncertainties of a Born cross section is to find the
corresponding covariance matrix. In the case of the considered method, the
covariance matrix can be estimated using the Monte Carlo technique. In this
approach, a visible cross section is randomly generated several times at each
c.m.\ energy point according to its statistical uncertainties. Each time a
visible cross section is generated, a Born cross section is calculated using the
algorithm described above. The values of a Born cross section obtained in this
way can be used to estimate the sample covariance matrix of this cross section.

Another commonly used method for obtaining a Born cross section using visible
cross section data is based on an iterative procedure for finding the radiative
correction. At the first step of this procedure, the fit to the Born cross
section $\sigma^{\rm external}_{\rm Born}(s)$ obtained in some of the previous
experiments is substituted into the integral from
eq.~\eqref{eq:visible-Born-relationship-kuraev-fadin-2} or
eq.~\eqref{eq:visible-Born-relationship-kuraev-fadin-2-blured} as the zero
approximation of a Born cross section. Further, the zero approximation of the
radiative correction is calculated through the ratio of the integral from
eq.~\eqref{eq:visible-Born-relationship-kuraev-fadin-2} or
eq.~\eqref{eq:visible-Born-relationship-kuraev-fadin-2-blured} to the zero
approximation of a Born cross section. At the $i$-th step of the iterative
procedure, the radiative correction and the next approximation of a Born cross
section are calculated using the following equations:
\begin{equation}
  \begin{split}
    \sigma^{(0)}_{\rm Born}(s) &= \sigma^{\rm external}_{\rm Born}(s),\\
    \varepsilon^{(i)}(s)(1 + \delta^{(i)}(s)) &= \frac{\sigma^{(i)}_{\rm vis}(s)}{\sigma^{(i)}_{\rm Born}(s)},\\
    \sigma^{(i+1)}_{\rm Born}(s) &= \frac{\sigma_{\rm vis}(s)}{\varepsilon^{(i)}(s)(1 + \delta^{(i)}(s))},
  \end{split}
\end{equation}
where the visible cross section $\sigma^{(i)}_{\rm vis}$ at the $i$-th step is
calculated using eq.~\eqref{eq:visible-Born-relationship-kuraev-fadin-2} or
eq.~\eqref{eq:visible-Born-relationship-kuraev-fadin-2-blured} and the $i$-th
approximation of a Born cross section $\sigma^{(i)}_{\rm Born}$. As in the case
of a first method, the term $\varepsilon^{(i)}(s)(1 + \delta^{(i)}(s))$ is calculated
as a whole. The main disadvantage of this method is that, as in the case of the
first method, the covariance matrix of a Born cross section can be estimated
using the Monte Carlo technique, which has a fairly high computational
complexity.

\subsection{\label{sec:first-kind-integral-eq}Integral equation of the first kind as an example of ill-posed problem}
Note that eq.~\eqref{eq:visible-Born-relationship-kuraev-fadin} is the
Fredholm integral equation of the first kind, and
equations~\eqref{eq:visible-Born-relationship-kuraev-fadin-2} and
\eqref{eq:visible-Born-relationship-kuraev-fadin-2-blured} can be written in the form of the
Volterra integral equation of the first kind. There is a well-known fact that
finding a numerical solution to Fredholm or Volterra integral equations of
the first kind is an ill-posed problem. A problem is said to be well-posed in
the sense of Hadamard~\cite{Tikhonov:1977,Hadamard:1923} if:
\begin{enumerate}
\item a solution exists;
\item that solution is unique;
\item the solution is changes continuously with changes in the data.
\end{enumerate}
Otherwise, the problem is ill-posed.

Before discussing integral equations
~\eqref{eq:visible-Born-relationship-kuraev-fadin-2} and
\eqref{eq:visible-Born-relationship-kuraev-fadin-2-blured} in detail, let us
consider an arbitrary Fredholm integral equation of the first kind:
\begin{equation}
  \label{eq:fredholm-integral-equation}
  \int^{b}_{a}K(t,x)q(x)dx = f(t),
\end{equation}
where $f(t)$ is the given function, $q(x)$ is the unknown function, $K(t, x)$ is
the given kernel function, $a$ and $b$ are the integration limits. The fact that
the problem given by eq.~\eqref{eq:fredholm-integral-equation} is ill-posed
means mainly that any arbitrarily small perturbations of the function $f(t)$ can
lead to the significant changes in the solution of the integral equation. This
property of the integral equation~\eqref{eq:fredholm-integral-equation} can be
explained using the Riemann-Lebesgue lemma~\cite{Serov:Riemann-Lebesgue}. A
corresponding explanation can be found in the book~\cite{Hansen2010DiscreteIP}.
It is assumed that the dependence of the function $f(t)$ on the variable $t$ is
measured experimentally, so that the function $f(t)$ is always perturbed due to
the presence of uncertainties in its measurement. This means that due to the
ill-posedness of the problem given by eq.~\eqref{eq:fredholm-integral-equation},
the numerical solution of this equation may differ significantly from the exact
function $q(x)$.

Linear integral equation can be approximately reduced to a system of linear
equations. Let us consider a discrete problem that corresponds to the integral
equation~\eqref{eq:fredholm-integral-equation}:
\begin{equation}
  \label{eq:fredholm-integral-equation-discrete}
  \hat{\mathcal{K}}\bm{q} = \bm{f}.
\end{equation}
Despite the fact that other options are possible, for the purposes of this work,
it is sufficient to consider the case when the $\hat{\mathcal{K}}$ is a full
rank square matrix. In this case, the system of linear
equations~\eqref{eq:fredholm-integral-equation-discrete} can be naively solved
as follows $\bm{q} = \hat{\mathcal{K}}^{-1}\bm{f}$. Suppose that the right side
of eq.~\eqref{eq:fredholm-integral-equation-discrete} is perturbed, which
leads to a perturbation of the left side. If we neglect the perturbations of the
matrix $\hat{\mathcal{K}}$, it is easy to show that the relative perturbations
of $\bm{q}$ and $\bm{f}$ are related by means of the following
inequality~\cite{Neumaier:INA}:
\begin{equation}
  \label{eq:linear-system-cond-ineq}
  \frac{\|\Delta\bm{q}\|_2}{\|\bm{q}\|_2}\leq
  cond(\hat{\mathcal{K}})\frac{\|\Delta\bm{f}\|_2}{\|\bm{f}\|_2},
\end{equation}
where $\Delta\bm{f}$ is the perturbation of the right side, $\Delta\bm{q}$ is
the perturbation of the numerical solution of
eq.~\eqref{eq:fredholm-integral-equation-discrete} and
$cond(\hat{\mathcal{K}})=\|\hat{\mathcal{K}}\|_2\|\hat{\mathcal{K}}^{-1}\|_2$ is the
condition number of the matrix $\hat{\mathcal{K}}$. In a more general case, when the
matrix $\hat{\mathcal{K}}$ is also perturbed, the following
inequality~\cite{Angst:CSLE} holds:
\begin{equation}
  \label{eq:linear-system-cond-ineq-general}
  \frac{\|\Delta\bm{q}\|_2}{\|\bm{q}\|_2}\leq
  \frac{cond(\hat{\mathcal{K}})}
  {1 - cond(\hat{\mathcal{K}})\frac{\|\Delta\hat{\mathcal{K}}\|_2}{\|\hat{\mathcal{K}}\|_2}}
  \left(
  \frac{\|\Delta\bm{f}\|_2}{\|\bm{f}\|_2} +
  \frac{\|\Delta\hat{\mathcal{K}}\|_2}{\|\hat{\mathcal{K}}\|_2}
  \right),
\end{equation}
where $\Delta\hat{\mathcal{K}}$ is the perturbation of the matrix
$\hat{\mathcal{K}}$. The inequality~\eqref{eq:linear-system-cond-ineq-general}
is obtained under the assumption
$\|\hat{\mathcal{K}}^{-1}\|_2\|\Delta\hat{\mathcal{K}}\|_2<1$. Note that in this
paper only the spectral norm of the matrices~\footnote{The spectral norm
$\|\hat{K}\|_2=\sup\limits_{\|\bm{x}\|_2=1}{\|\hat{K}\bm{x}\|_2}=\sigma_{\rm max}$
of an arbitrary matrix $\hat{K}\in\mathbb{F}^{m\times n}$ is induced by the $l^2$
vector norm $\|\bm{x}\|_2=\sqrt{\sum\limits^n_{i = 1}|x_i|^2}$. Value
$\sigma_{\rm max}$ is the maximum singular value of the matrix $\hat{K}$ and
field $\mathbb{F}$ is the field of real ($\mathbb{R}$) or complex ($\mathbb{C}$)
numbers.} is used. Almost always, in the case of ill-posed problems, the matrix
$\hat{\mathcal{K}}$ is ill-conditioned, i.e.\ $cond(\hat{\mathcal{K}})\gg1$.
Equations~\eqref{eq:linear-system-cond-ineq} and
\eqref{eq:linear-system-cond-ineq-general} show that even for small
perturbations of the right-hand side of
eq.~\eqref{eq:fredholm-integral-equation-discrete}, the perturbations in the
solution of this equation can be large if the matrix $\hat{\mathcal{K}}$ is
ill-conditioned.

Another way to understand the nature of perturbations in the solution of
eq.~\eqref{eq:fredholm-integral-equation-discrete} is to consider the
singular value decomposition (SVD) of the matrix $\hat{\mathcal{K}}$. The SVD
decomposition of the matrix $\hat{\mathcal{K}}$ can be written in the following
form:
\begin{equation}
  \hat{\mathcal{K}} = \hat{\mathcal{U}}\hat{\Sigma}\hat{\mathcal{V}}^{\intercal},
\end{equation}
where $\hat{\mathcal{U}}$ and $\hat{\mathcal{V}}$ are orthogonal matrices, the
matrix $\hat{\Sigma}$ is a diagonal matrix whose diagonal elements
$\sigma_1,\sigma_2,\ldots,\sigma_{n}$ are always non-negative numbers and called
singular values. Let us choose the numbering of singular values in such a way
that these values decrease as the index increases. Using the SVD decomposition
of the matrix $\hat{\mathcal{K}}$, the direct solution to
eq.~\eqref{eq:fredholm-integral-equation-discrete} can be written in the
following form:
\begin{equation}
  \label{eq:svd-solution}
  \bm{q} = \sum\limits_{i=1}^{n}\frac{(\bm{u}_i,\bm{f})}{\sigma_i}\bm{v}_i,
\end{equation}
where vectors $\bm{u}_i$ and $\bm{v}_i$ are the columns of the matrices
$\hat{\mathcal{U}}$ and $\hat{\mathcal{V}}$, respectively.
Eq.~\eqref{eq:svd-solution} shows that terms with small singular values in the
denominator can make a larger contribution to the solution than terms with large
singular values. It is also important to note the well-known fact that the
frequency of oscillations of the components of the vector $\bm{v}_i$ increases
as the index $i$ increases, i.e.\ as the corresponding singular values decrease.
Thus, with a large condition number $cond(\hat{\mathcal{K}})=\sigma_n/\sigma_1$,
high-frequency perturbations make a large contribution to the solution.

The simplest way to avoid high-frequency perturbations in the solution is to
exclude the terms with small singular numbers from the sum in
eq.~\eqref{eq:svd-solution}. This approach to the solution regularization is
called the truncated SVD method (TSVD). The solution $\bm{q}_{\rm TSVD}$ in this case is given by
the following formula:
\begin{equation}
  \label{eq:tsvd-solution}
  \bm{q}_{\rm TSVD} = \sum\limits_{i=1}^{n^{\prime}}\frac{(\bm{u}_i,\bm{f})}{\sigma_i}\bm{v}_i,
\end{equation}
where $n^{\prime} < n$.

However, in practice, the Tikhonov regularization method is most often used.
This method consists in finding the minimization of a functional of the
following form:
\begin{equation}
  \label{eq:tikhonov-functional}
  \mathcal{L}[\bm{q}] = \|\hat{\mathcal{K}}\bm{q} - \bm{f}\|^2_2 + \lambda\|\bm{q}\|^2_2,
\end{equation}
where $\lambda$ is a fixed parameter that has the meaning of the strength of the
regularization. The functional $\mathcal{L}[\bm{q}]$ is minimized by the
components of the vector $\bm{q}$. In a more general case, Tikhonov
regularization method consists in minimizing a slightly more complex functional
given by the following formula:
\begin{equation}
  \label{eq:tikhonov-functional-general}
  \mathcal{L}_{\rm G}[\bm{q}] = \|\hat{\mathcal{K}}\bm{q} - \bm{f}\|^2_2 + \lambda\|\hat{\mathcal{B}}\bm{q}\|^2_2,
\end{equation}
where $\hat{\mathcal{B}}$ is some matrix. The simplest examples of such a matrix
are the identity matrix or a matrix that links the vector $\bm{q}$ with the
vector of numerical derivatives of the function $q(x)$ with respect to its
argument. There are a number of semi-heuristic methods for choosing the optimal
value of the regularization parameter $\lambda$. One of these methods is the
L-curve criterion. This criterion is considered in
section~\eqref{sec:regularization}, where the Tikhonov regularization method is
discussed.

It is also important to note that the minimization of the generalized functional
$\mathcal{L}_{\rm G}[\bm{q}]$ can be reduced to the selective SVD method (SSVD).
In the SSVD method, the regularized solution $\bm{q}_{\rm SSVD}$ is calculated
according to the formula:
\begin{equation}
  \label{eq:ssvd-solution}
  \bm{q}_{\rm SSVD} = \sum\limits_{i=0}^{n}\phi_i\frac{(\bm{u}_i,\bm{f})}{\sigma_i}\bm{v}_i,
\end{equation}
where $\phi_i$ are some weight coefficients. In the simple case, when the matrix
$\hat{\mathcal{B}}$ in eq.~\eqref{eq:tikhonov-functional-general} is equal to
the identity matrix (i.e. in the case of eq.~\eqref{eq:tikhonov-functional}),
the weight coefficients can be found using the following formula~\footnote{The
eq.~\eqref{eq:tikhonov-ssvd-corefficients} is correct only if values of the
function $f(t)$ are measured at equidistant points and $l^2$ vector norms
are used in eq.~\eqref{eq:tikhonov-functional}. In the case of non-equidistant
points, this equation becomes more complicated. In this case, we have to write
down norms in eq.~\eqref{eq:tikhonov-functional}, taking into account the
weights of contributions from different $x$ and $t$ points to the dot products.
This can be easily done by using the $L^2$ function norms in
eq.~\eqref{eq:tikhonov-functional}. The $L^2$ norm of an arbitrary function
$g(x)$ is $\|g\|_2=\sqrt{\int\limits^{x_{\text{max}}}_{x_{\text{min}}}
  |g(x)|^2dx}$, where $x\in [x_{\text{min}}, x_{\text{max}}]$. Since the problem
is discrete, instead of the functions $f(t)$ and $q(x)$, one can use their
interpolations.}:
\begin{equation}
  \label{eq:tikhonov-ssvd-corefficients}
  \phi_i = \frac{\sigma^2_i}{\sigma^2_i + \lambda}.
\end{equation}
More details about the methods for solving ill-posed problems, as well as the
methods of optimal choice of the regularization parameter, can be found in the
works~\cite{Tikhonov:1963,Hansen1993,Tikhonov:1995,Hansen2010DiscreteIP,Kabanikhin2011}.

\subsection{\label{sec:problem-formulation}Proposed method}
The condition number depends on the properties of a particular integral operator
and can indeed be large in many applications such as image
processing~\cite{Dahl2010AlgorithmIR,Sadek2012,Hearn2014,Tirer2019},
geophysics~\cite{Ballani2002,Deidda2003,ABDELAZEEM2013,Brufati2016} and high
energy physics~\cite{Hocker1996,Kuusela2012,Spano2013,Kuusela2015,Kuusela2017}.
However, as it is shown in section~\ref{sec:discretization}, the condition
number of the matrix of the integral operator from
eq.~\eqref{eq:visible-Born-relationship-kuraev-fadin-2} is of order unity due to
the properties of the function $F(x,s)$. Therefore, one can hope that the
ill-posedness of problem given by
eq.~\eqref{eq:visible-Born-relationship-kuraev-fadin-2} can be neglected. Since
the matrix of the system of linear equations corresponding to integral
equation~\eqref{eq:visible-Born-relationship-kuraev-fadin-2} is well
conditioned, the numerical solution can be obtained directly by inverting this
matrix. This is the new method for obtaining a Born cross section, proposed in
this paper. Further, this method will be referred to as the naive method. The
conditionality of the matrices corresponding to integral
equations~\eqref{eq:visible-Born-relationship-kuraev-fadin-2} and
\eqref{eq:visible-Born-relationship-kuraev-fadin-2-blured} are discussed in
section~\ref{sec:discretization}. Section~\ref{sec:naive-method} presents the
results of applying the naive method obtained using a number of numerical
experiments.

Note, that when eq.~\eqref{eq:visible-Born-relationship-kuraev-fadin-2-blured}
is reduced to a system of linear equations, the condition number can be large if
the c.m.\ energy spread is comparable to or greater than the distance between
c.m.\ energy points. Therefore, the numerical solution to this equation cannot
always be found using the naive method, i.e.\ without using regularization
techniques.

On the other hand, it is known~\cite{Kuusela2012} that when using various
regularization techniques, such as Tikhonov regularization, the covariance
matrix of the regularized solution is incorrect, since this solution is biased.
For this reason, regularization techniques can hardly be used for precise
obtaining of a Born cross section.

\section{\label{sec:discretization}Discretization of the problem}
\subsection{\label{sec:discretization-eq-kf}Discretization of equation~\eqref{eq:visible-Born-relationship-kuraev-fadin-2}}
By the phrase ``discretization of the problem'' in this work we mean the
reduction of an integral equation to a system of linear equations. In order to
reduce the integral eq.~\eqref{eq:visible-Born-relationship-kuraev-fadin-2} to a
system of linear equations, we first interpolate an unknown Born cross section,
taking into account the fact that the values of a Born cross section at the
c.m.\ energies below the threshold energy are equal to zero. The next step is to
linearly express the interpolation coefficients in terms of the unknown values
of a Born cross section at points with the c.m.\ energies
$\sqrt{s_1},\sqrt{s_2},\ldots,\sqrt{s_{\rm N}}$. Assuming that the measurements
of visible cross section and detection efficiency were carried out at points
with these c.m.\ energies, we can write the following system of equations:
\begin{equation}
  \label{eq:KF-system}
  \begin{cases}
    \int\limits^{1 - s_{\rm T}/s_1}_{0}
    F(x, s_1)\varepsilon(x, s_1)\sigma^{\rm interp}_{\rm Born}(s_1(1-x))dx &= \sigma_{\rm vis}(s_1),\\
    \int\limits^{1 - s_{\rm T}/s_2}_{0}
    F(x, s_2)\varepsilon(x, s_2)\sigma^{\rm interp}_{\rm Born}(s_2(1-x))dx &= \sigma_{\rm vis}(s_2),\\
    &\ldots\\
    \int\limits^{1 - s_{\rm T}/s_{\rm N}}_{0} F(x, s_{\rm N})\varepsilon(x,
    s_{\rm N})\sigma^{\rm interp}_{\rm Born}(s_{\rm N}(1-x))dx &=
    \sigma_{\rm vis}(s_{\rm N}),
\end{cases}
\end{equation}
where $\sigma^{\rm interp}_{\rm Born}(s)$ is the interpolation function of a
Born section, which linearly depends on the unknown values of a Born cross
section at the points with the c.m.\ energies
$\sqrt{s_1},\sqrt{s_2},\ldots,\sqrt{s_{\rm N}}$. After taking the integrals in
eq.~\eqref{eq:KF-system}, we obtain the following system of linear equations:
\begin{equation}
  \label{eq:KF-SLAE}
  \hat{\mathcal{F}}\bm{\sigma}_{\rm Born} = \bm{\sigma}_{\rm vis},
\end{equation}
where $\hat{\mathcal{F}}$ is the matrix of this system of linear equations,
$\bm{\sigma}_{\rm Born}$ is the vector of the Born cross section unknown values
and $\bm{\sigma}_{\rm vis}$ is the vector of the visible cross section values at
the considered c.m.\ energies.

To obtain the system of linear equations~\eqref{eq:KF-SLAE} in the above way, it is
necessary that the used interpolation $\hat{L}$ has the linearity property. i.e.\ 
the interpolation coefficients should linearly depend on the unknown values of
a Born cross section or, which is the same, the following equality should be
satisfied:
\begin{equation}
  \label{eq:linearity-property}
  \hat{L}[\psi(x)] = \sum\limits_{i}c_i\hat{L}[h_i(x)],
\end{equation}
where $\psi(x)$ is the function linearly expressed in terms of the functions
$h_i(x)$:
\begin{equation}
  \psi(x) = \sum\limits_{i}c_ih_i(x)
\end{equation}
and $c_i$ are arbitrary real coefficients. In this work, piecewise linear
interpolation is used, as well as cubic spline interpolation. Both of these
interpolation kinds satisfy the linearity property~\eqref{eq:linearity-property}
with respect to the values of a Born cross section at different c.m.\ energy
points. A package used in this work to obtain numerical solutions to
equations~\eqref{eq:visible-Born-relationship-kuraev-fadin-2} and
\eqref{eq:visible-Born-relationship-kuraev-fadin-2-blured} also includes the
ability to alternate these kinds of interpolation at different c.m.\ energy
intervals. This package is called ISRSolver and is discussed in more detail in
section~\ref{sec::ISRSolver}.

\begin{figure}[tbp]
  \centering
  \includegraphics[width=\textwidth]{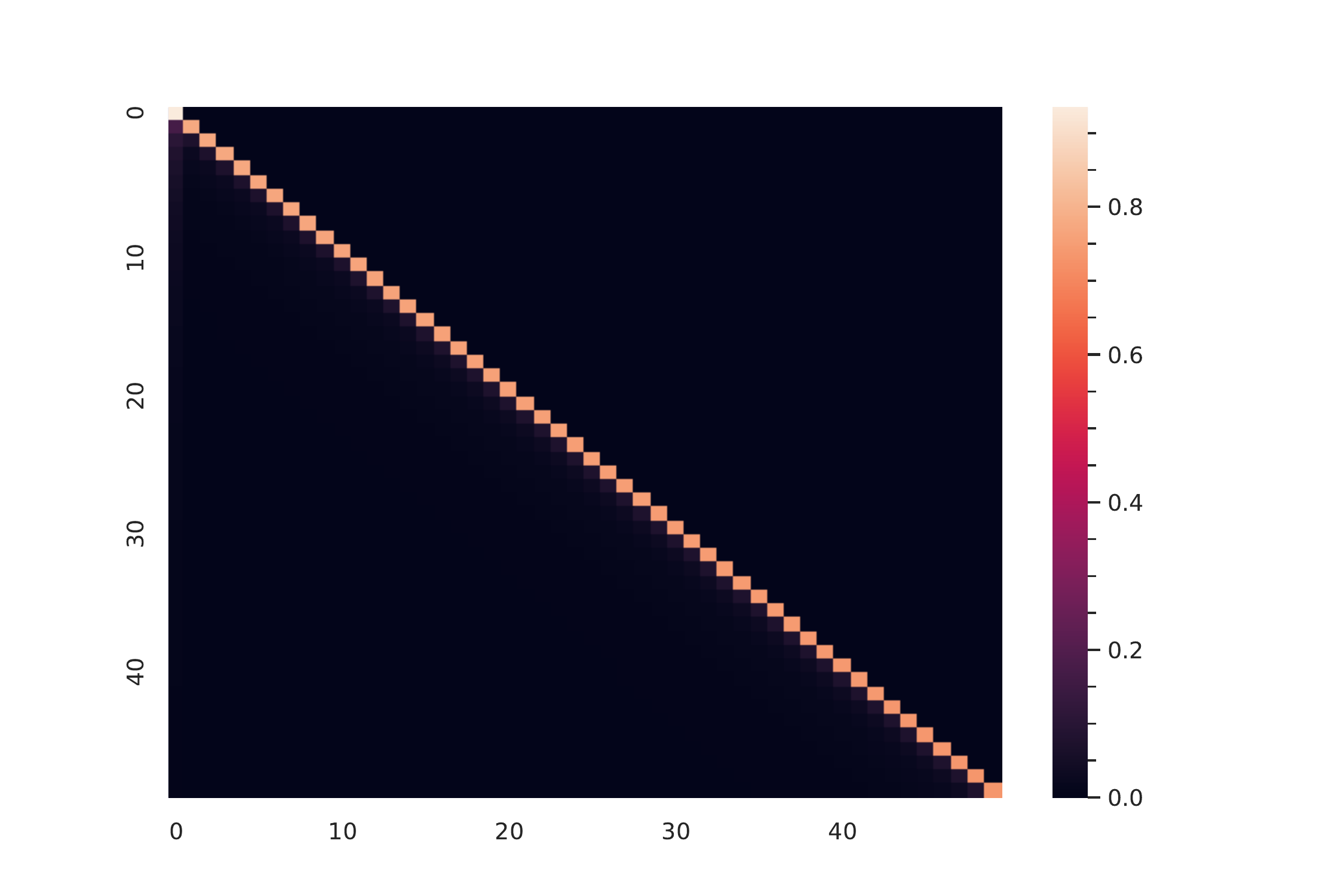}
  \caption{\label{fig:int-op-etapipi-linterp-no-energy-spread} The integral
    operator matrix $\hat{\mathcal{F}}$ in the case of piecewise linear interpolation
    of the Born cross section. The matrix corresponds to the $\eta\pi^+\pi^-$
    threshold energy.}
\end{figure}
It should be noted that when the piecewise linear interpolation is used, the
matrix $\hat{\mathcal{F}}$ is a lower triangular matrix with small off-diagonal
elements. An example of the matrix $\hat{\mathcal{F}}$ obtained using piecewise
linear interpolation is shown in
figure~\ref{fig:int-op-etapipi-linterp-no-energy-spread}. This matrix
corresponds to the threshold energy of the $e^+e^-\rightarrow\eta\pi^+\pi^-$
process and is obtained using $50$ equally spaced c.m.\ energy points belonging
the range from $1.18$ to $2.00$ GeV. The condition number of the
matrix $\hat{\mathcal{F}}$ in this case is approximately equal to $1.6$.

\begin{figure}[tbp]
  \centering
  \includegraphics[width=\textwidth]{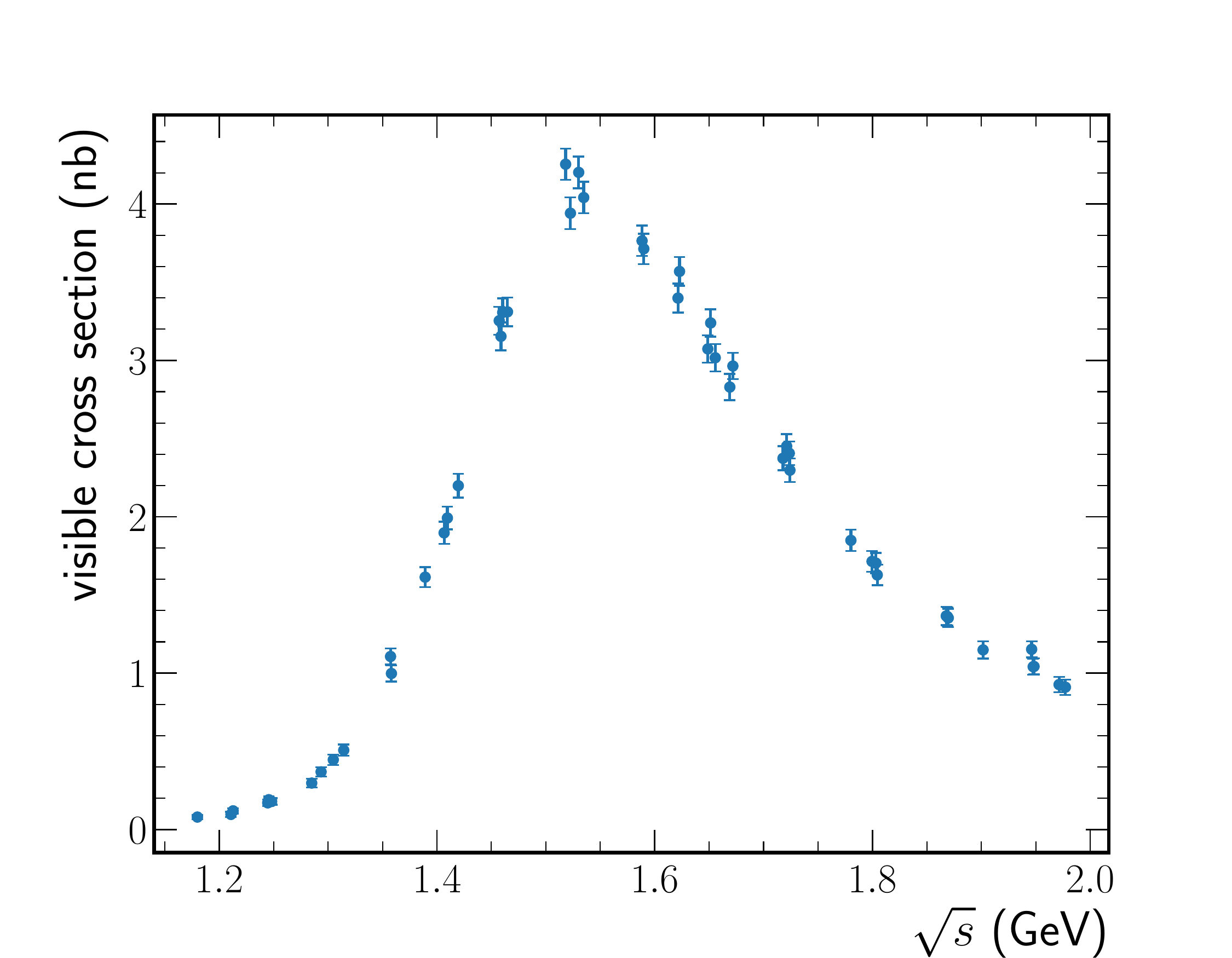}
  \caption{\label{fig:vcs-etapipi-nonuniform} An example of the dependence of
    the $e^+e^-\rightarrow\eta\pi^+\pi^-$ visible cross section on the
    c.m.\ energy. The distribution of the c.m.\ energy points is highly
    non-uniform. The cross section shown in the figure is not the result of an
    experiment. The cross section points are generated randomly using the
    vector meson dominance model.}
\end{figure}

Let us consider an example similar to the previous one, but corresponding to an
extremely non-uniform distribution of c.m.\ energy points. The positions of the
c.m.\ energy points for this example are shown in
figure~\ref{fig:vcs-etapipi-nonuniform}. The number of points, as in the
previous case, is $50$. Despite the non-uniform distribution of the
c.m.\ energies, the condition number is still comparable to unity. In this case,
the condition number is approximately $1.9$.

In the case of cubic spline interpolation or mixed type
interpolation, the matrix $\hat{\mathcal{F}}$ is not a lower triangular matrix. If
the interpolation function describes a Born cross section well, that is, this
function does not have oscillatory outliers between the points of this cross
section, then the condition number is approximately the same as in the examples
described above.

\subsection{Calculation and estimation of the condition number}
The condition number from the previous and subsequent examples is calculated
using the formula $cond(\hat{\mathcal{F}})=\sigma_{\rm max} / \sigma_{\rm min}$.
To find the minimum ($\sigma_{\rm min}$) and maximum ($\sigma_{\rm max}$)
singular values, the singular value decomposition of the matrix
$\hat{\mathcal{F}}$ is performed numerically. In the ISRsolver package, the
singular value decomposition of matrices is carried out using the Eigen 3 linear
algebra library~\cite{eigenweb}.

Finding an analytical expression for the condition number even in the case of
piecewise linear interpolation of a Born cross section is a quite complicated
task due to the bulky kernel function $F(x, s)$ and the large size of the matrix
$\hat{\mathcal{F}}$. However, in the case of piecewise linear interpolation of
a Born cross section, a rough estimation of the condition number can be
obtained analytically, which makes it possible to qualitatively understand the
reason that the condition number of the matrix $\hat{\mathcal{F}}$ is of the
order of unity. It is known that in the case when the matrix $\hat{\mathcal{C}}$
is a normal matrix, i.e.\ $[\hat{\mathcal{C}}, \hat{\mathcal{C}}^{\dag}] = 0$,
the condition number can be found as the ratio of the largest eigenvalue to the
smallest one. A special case of a normal matrix is a diagonal matrix. Since the
off-diagonal elements of the matrix $\hat{\mathcal{F}}$ are small, this matrix
is close to a diagonal matrix, and as a consequence to a normal matrix.
Based on this assumption, it is possible to obtain a rough estimation of the
condition number using the formula $cond(\hat{\mathcal{F}}) = \lambda_{\rm max}
/ \lambda_{\rm min}$, where $\lambda_{\rm max}$ is the maximum eigenvalue of the
matrix $\hat{\mathcal{F}}$ and $\lambda_{\rm min}$ is the minimum one. If the
matrix $\hat{\mathcal{F}}$ is lower triangular matrix, its eigenvalues coincide
with the diagonal matrix elements. Diagonal matrix elements in the case of
piecewise linear interpolation are given by the following formula:
\begin{equation}
  \hat{\mathcal{F}}_{jj} = \int\limits^{1 - \frac{s_{j - 1}}{s_j}}_0F(x, s_j)\frac{\sqrt{s_j (1 - x)} - \sqrt{s_{j - 1}}}{\sqrt{s_j} - \sqrt{s_{j - 1}}}dx,
\end{equation}
where the index $j$ takes values from one to $N$ and $s_{0}=s_{\rm T}$. For
simplicity, the last equation is obtained under the assumption that the
detection efficiency is equal to unity. Further, assuming that the
c.m.\ energies are located equidistantly, it can be
easily shown that the diagonal elements of the matrix $\hat{\mathcal{F}}$
decrease while the index $j$ increases. Thus, taking into account the above
assumptions, we can roughly estimate the condition number by the following
formula:
\begin{equation}
  \label{eq:condnum-rough-estimation-1}
  cond(\hat{\mathcal{F}})\sim\frac{\hat{\mathcal{F}}_{11}}{\hat{\mathcal{F}}_{NN}}.
\end{equation}

The last formula in the case of the matrix shown in
figure~\ref{fig:int-op-etapipi-linterp-no-energy-spread} gives an estimation of
the condition number equal to $1.3$, which is $0.3$ less than the exact value of
the condition number in this case.

\subsection{Closeness of equation~\eqref{eq:visible-Born-relationship-kuraev-fadin-2} to a well-posed problem}
It can be shown in an alternative way that the ill-posed problem given by
eq.~\eqref{eq:visible-Born-relationship-kuraev-fadin-2} is close to some
well-posed problem. To begin with, let us introduce a typical c.m. energy scale
$\Gamma$ at which a cross section changes and assume that this scale is greater
than or of the order of several MeV. Let us also introduce a condition on the
range $0\div\Delta{x}$ of the argument $x$, at which a cross section changes
weakly: $\sqrt{s} - \sqrt{s(1-\Delta{x})}\ll\Gamma$. This condition leads us to
the following inequality for the upper bound of the corresponding $x$-range:
\begin{equation}
  \label{eq:upper-x-limit-inequality}
  \Delta{x}\ll\frac{2\Gamma}{\sqrt{s}}.
\end{equation}
Taking into account the last inequality,
eq.~\eqref{eq:visible-Born-relationship-kuraev-fadin-2} can be written as
follows:
\begin{equation}
  \label{eq:volterra-second-kind}
  \begin{split}
    \sigma_{\rm vis}(s)  &\approx\sigma_{\rm Born}(s)\int\limits^{\Delta{x}}_{0}F(x, s)\varepsilon(x, s)dx +
    \int\limits^{1 - s_{\rm T}/s}_{\Delta{x}}
    F(x, s)\varepsilon(x, s)\sigma_{\rm Born}(s(1-x))dx\\
    &= g(s,\Delta{x})\sigma_{\rm Born}(s) +
    \int\limits^{1 - s_{\rm T}/s}_{\Delta{x}}
    F(x, s)\varepsilon(x, s)\sigma_{\rm Born}(s(1-x))dx,
\end{split}
\end{equation}
where $g(s, \Delta{x})=\int\limits^{\Delta{x}}_{0}F(x, s)\varepsilon(x, s)dx$.
Let us consider an example where $\Gamma=\Gamma_{\phi}\approx4.3\text{ MeV}$ and
$\sqrt{s}=m_{\phi}\approx1020\text{ MeV}$. In this case the parameter
$\Delta{x}$ must satisfy the following inequality: $\Delta{x}\ll
8.4\times10^{-3}$. When the parameter $\Delta{x}=2\times10^{-5}$ is set, the
function $g(m^2_{\phi}, \Delta{x})$ is approximately equal to $0.52$, i.e.\ has
value comparable to one. Since division by a number of the order of unity is a
well-posed problem~\cite{Kabanikhin2011}, let us divide
eq.~\eqref{eq:volterra-second-kind} by the function $g(s, \Delta{x})$, where
$s\sim m^2_{\phi}$ ($\Gamma_{\phi}\ll m_{\phi}$). The equation obtained as the result of division can be
written as the Volterra integral equation of the second kind. It is well known
fact~\cite{Kabanikhin2011} that the Volterra integral equation of the second
kind is a well-posed problem. It is also worth noting that the parameter
$\Delta{x}$ can be always chosen in a such way for any reasonable parameters
$\Gamma$ and $\sqrt{s}$ that the coefficient $g(s, \Delta{x})$ is comparable to
one and the inequality~\eqref{eq:upper-x-limit-inequality} is satisfied. Thus,
the problem given by eq.~\eqref{eq:visible-Born-relationship-kuraev-fadin-2} is
close to some well-posed problem.

\subsection{\label{sec:discretization-eq-kf-blured}Discretization of equation~\eqref{eq:visible-Born-relationship-kuraev-fadin-2-blured}}
Equation~\eqref{eq:visible-Born-relationship-kuraev-fadin-2-blured}
can be reduced to a system of linear equations in the same way as
eq.~\eqref{eq:visible-Born-relationship-kuraev-fadin-2}. The system of linear
equations in this case has the following form:
\begin{equation}
  \hat{\mathcal{G}}\hat{\mathcal{F}}\bm{\sigma}_{\rm Born}=\bm{\sigma}_{\rm vis},
\end{equation}
where matrix $\hat{\mathcal{G}}$ corresponds to the external integral operator
from eq.~\eqref{eq:visible-Born-relationship-kuraev-fadin-2-blured}, while
matrix $\hat{\mathcal{F}}$ corresponds to the internal integral operator and in
fact is the same as the analogous matrix from eq.~\eqref{eq:KF-SLAE}.

\begin{figure}[tbp]
  \centering
  \begin{subfigure}[t]{0.47\textwidth}
    \centering
    \includegraphics[width=\textwidth]{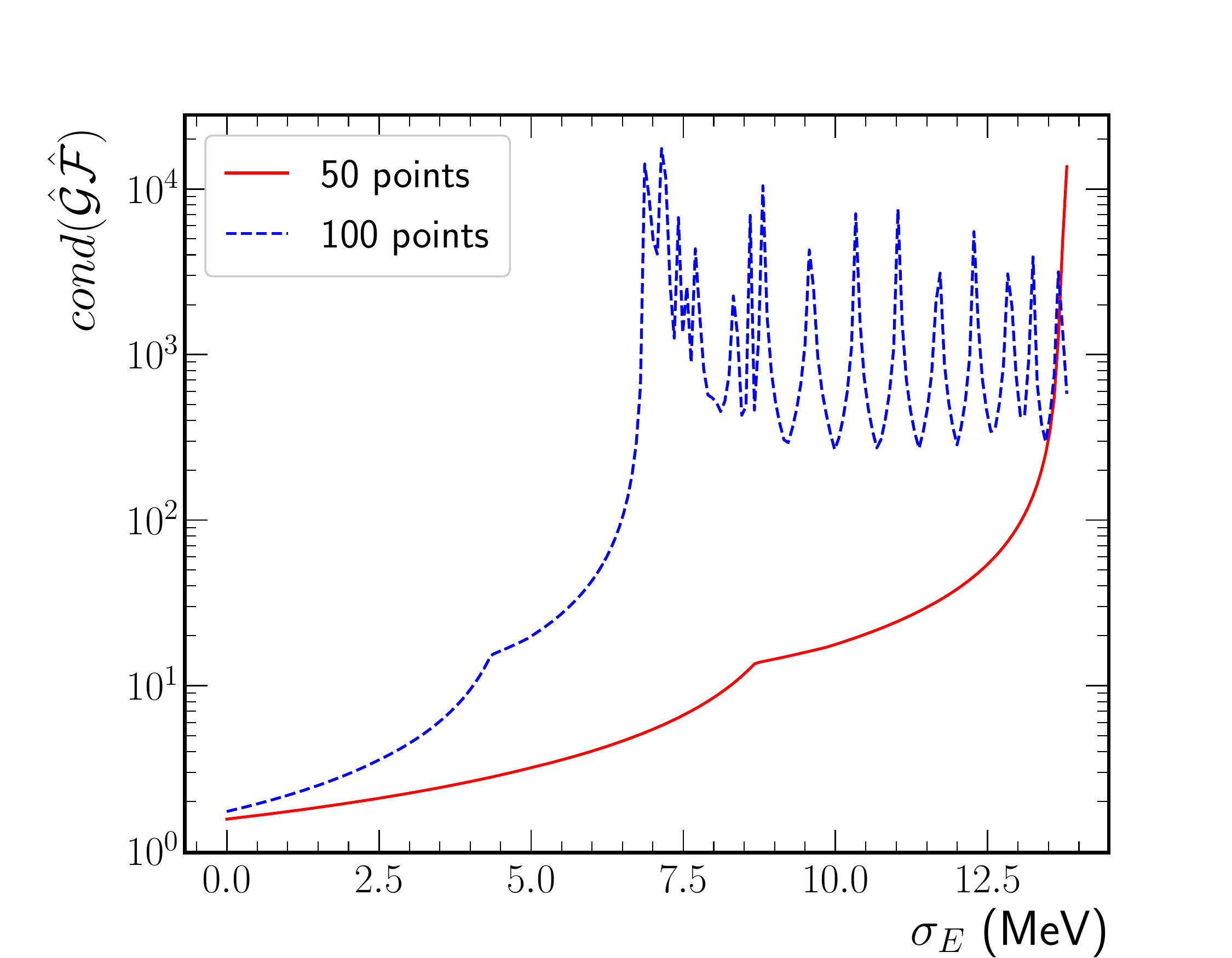}
    \caption{\label{fig:cond_sigma1} The case of $50$ (solid curve) and $200$
      (dashed curve) c.m.\ energy points equally spaced in the range
      $1.18$-$2.00$ GeV.}
  \end{subfigure}
  \hfill
  \begin{subfigure}[t]{0.47\textwidth}
    \centering
    \includegraphics[width=\textwidth]{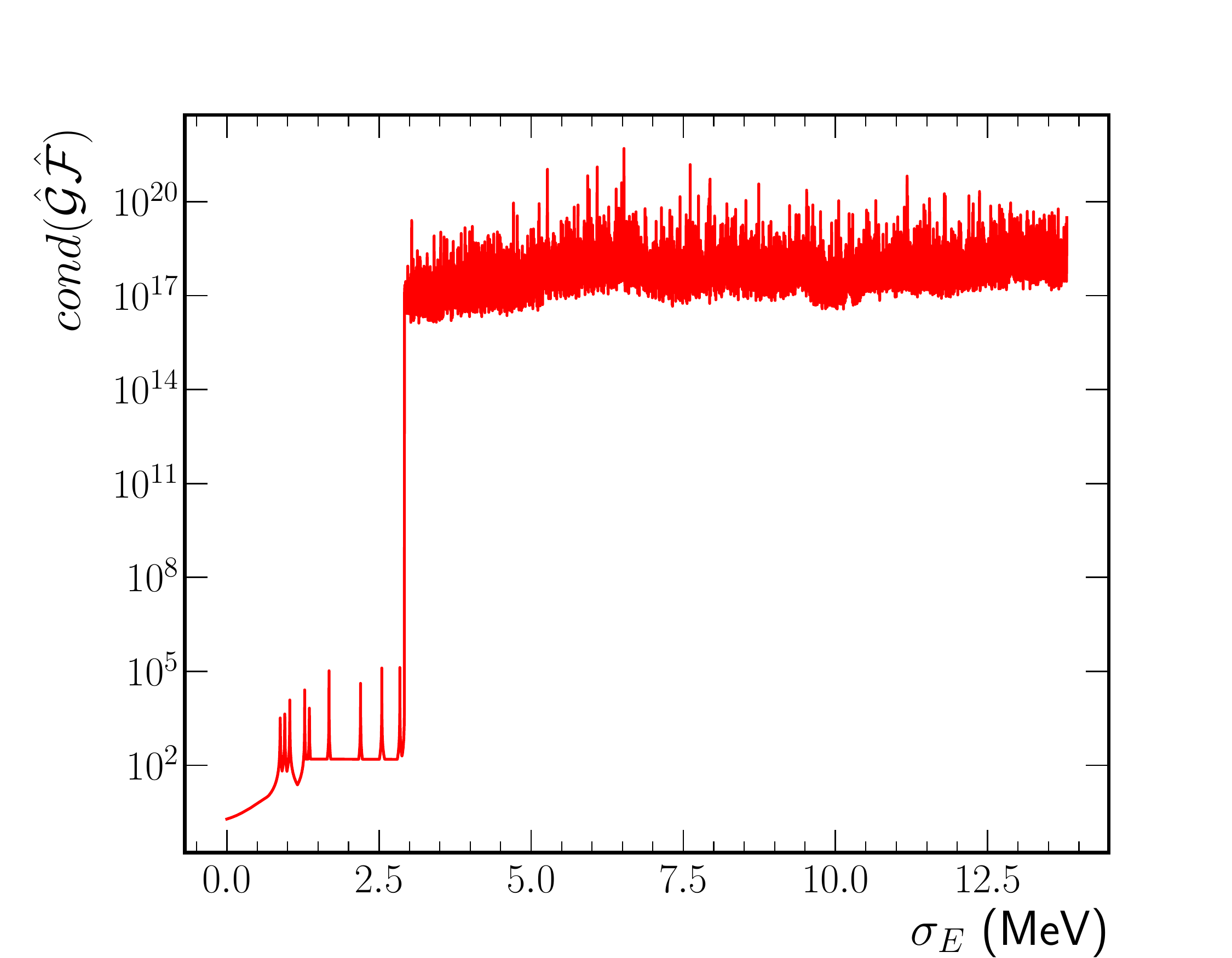}
    \caption{\label{fig:cond_sigma2} The case of $50$ points, which are highly
      non-uniformly distributed in the c.m.\ energy range from $1.18$
      to $2.00$ GeV. The c.m.\ energy points used in this case
      correspond to figure~\ref{fig:vcs-etapipi-nonuniform}.}
  \end{subfigure}
  \caption{\label{fig:cond_sigma} The dependence of the condition number
    $cond(\hat{\mathcal{G}}\hat{\mathcal{F}})$ on the c.m.\ energy spread.}
\end{figure}

However, due to the limits in the outer integral from
eq.~\eqref{eq:visible-Born-relationship-kuraev-fadin-2-blured}, the matrix
$\hat{\mathcal{G}}\hat{\mathcal{F}}$ of the full integral operator is not a
lower triangular matrix in this case even if piecewise linear interpolation is
used. Since the kernel $\frac{1}{\sqrt{2\pi\sigma^2_{\rm
      E}}}\exp{\left(-\frac{(E - \sqrt{s})^2}{2\sigma^2_{\rm E}}\right)}$ of the
external integral operator can change quite smoothly for large values of the
c.m.\ energy spread, the condition number of the matrix
$\hat{\mathcal{G}}\hat{\mathcal{F}}$ in this case can be quite large. Typical
dependencies of the condition number of the matrix
$\hat{\mathcal{G}}\hat{\mathcal{F}}$ on the value of the parameter $\sigma_E$
are shown in figure~\ref{fig:cond_sigma1}. The solid curve in this figure is
obtained with the same c.m.\ energy spread at each point. The integral operator
matrices used to obtain this curve correspond to $50$ equally spaced c.m. energy
points in range from $1.18$ to $2.00$ and the $\eta\pi^+\pi^-$ threshold energy.
It is seen from this figure that the condition number remains of the order of
unity for values of the c.m.\ energy spread of the order of several MeV:
$\sigma_E\lessapprox3\text{ MeV}$. The reason is that the distance ($16.4\text{
  MeV}$) between the c.m.\ energy points in this case is much greater than the
parameter $\sigma_E$. As the density of c.m.\ energy points increases, the
dependence of the condition number on the parameter becomes steeper. An example
of this is the dashed curve shown in the same figure. This curve also represents
the dependence of the condition number on the c.m.\ energy spread, but obtained
at twice the density of c.m.\ energy points than the solid curve. Some
oscillations are observed on the dashed curve when the energy spread exceeds
$6\text{ MeV}$. Similar oscillations are observed for the solid curve too, but
they appear outside the figure starting at about $14\text{ MeV}$.

In the case when the distance between some c.m.\ energy points is less than or
of the order of the c.m.\ energy spread, the condition number increases
dramatically. A typical dependence of the condition number on the c.m.\ energy
spread in this case is shown in figure~\ref{fig:cond_sigma2}. This dependence
is obtained with the highly non-uniform distribution of c.m.\ energy points,
which is shown in figure~\ref{fig:vcs-etapipi-nonuniform}.
From figures~\ref{fig:cond_sigma2} and \ref{fig:vcs-etapipi-nonuniform} it
follows that if some c.m.\ energy points are located at a distance much less
than the c.m.\ energy spread from each other, then the condition number grows
rapidly with an increase of the c.m.\ energy spread and can be large even when
the c.m.\ energy spread is of the order of several MeV. In this case, the
numerical solution of
eq.~\eqref{eq:visible-Born-relationship-kuraev-fadin-2-blured} at such points is
strongly scattered. If the number of close energy points is small, then a large
scatter is observed only at these points.

\section{\label{sec:naive-method}Naive method}
\subsection{Covariance matrix}
Before considering examples of obtaining a Born cross section using the naive
method, let us derive the relationship between the covariance matrices of Born
and visible cross sections. In this section, we denote the matrix of the
integral operator as $\hat{\mathcal{A}}$. Therefore, in the case of
eq.~\eqref{eq:visible-Born-relationship-kuraev-fadin-2},
$\hat{\mathcal{A}}=\hat{\mathcal{F}}$, while in the case of
eq.~\eqref{eq:visible-Born-relationship-kuraev-fadin-2-blured},
$\hat{\mathcal{A}}=\hat{\mathcal{G}}\hat{\mathcal{F}}$. The covariance matrix
$\hat{\Lambda}$ of a visible cross section can be written as follows:
\begin{equation}
  \begin{split}
    \hat{\Lambda}_{ij} &= {\rm Cov}[\bm{\sigma}_{\rm vis}]_{ij}= {\rm Cov}[\hat{\mathcal{A}}\bm{\sigma}_{\rm Born}]_{ij}\\
    &=  {\rm E}[(\hat{\mathcal{A}}\bm{\sigma}_{\rm Born} - {\rm E}[\hat{\mathcal{A}}\bm{\sigma}_{\rm Born}])_i(\hat{\mathcal{A}}\bm{\sigma}_{\rm Born} - {\rm E}[\hat{\mathcal{A}}\bm{\sigma}_{\rm Born}])_j]\\
    &= {\rm E}[\hat{\mathcal{A}}_{ik}(\bm{\sigma}_{\rm Born} - {\rm E}[\bm{\sigma}_{\rm Born}])_k(\bm{\sigma}_{\rm Born} - {\rm E}[\bm{\sigma}_{\rm Born}])_l\hat{\mathcal{A}}_{jl}] \\
    &= \hat{\mathcal{A}}_{ik}{\rm E}[(\bm{\sigma}_{\rm Born} - {\rm E}[\bm{\sigma}_{\rm Born}])_k(\bm{\sigma}_{\rm Born} - {\rm E}[\bm{\sigma}_{\rm Born}])_l]\hat{\mathcal{A}}_{jl}\\
    &= \hat{\mathcal{A}}_{ik}\hat{\mathcal{M}}_{kl}\hat{\mathcal{A}}_{jl} = (\hat{\mathcal{A}}\hat{\mathcal{M}}\hat{\mathcal{A}}^{\intercal})_{ij},
  \end{split}
\end{equation}
where $\hat{\mathcal{M}}$ is the covariance matrix of a Born cross section,
the functional ${\rm E}[.]$ denotes the expectation of a value in square brackets
and the functional ${\rm Cov}[.]$ denotes the covariance matrix of a vector in
square brackets, it is also assumed that the summation is taken over the
repeated indices. Thus, the covariance matrix $\hat{\mathcal{M}}$ of a Born
cross section is expressed in terms of the covariance matrix $\hat{\Lambda}$ of
a visible cross section and the integral operator matrix $\hat{\mathcal{A}}$
by means of the following formula:
\begin{equation}
  \label{eq:naive-method-covmat}
  \hat{\mathcal{M}} = \hat{\mathcal{A}}^{-1}\hat{\Lambda}\left(\hat{\mathcal{A}}^{-1}\right)^{\intercal}.
\end{equation}
In this paper, we assume that both covariance matrices $\hat{\Lambda}$ and
$\hat{\mathcal{M}}$ correspond only to statistical uncertainties. It is assumed
that a visible cross section at different c.m.\ energy points is measured
independently, so that its covariance matrix $\hat{\Lambda}$ is a diagonal
matrix. In principle it is possible to use a non-diagonal matrix
$\hat{\Lambda}$. All the formulas will be the same in this case.

\subsection{\label{sec:numerical-experiments}Numerical experiments}
\subsubsection{\label{sec:numerical-experiments-without-enspread}Numerical experiments with equation~\eqref{eq:visible-Born-relationship-kuraev-fadin-2}}
Next, let us consider a few numerical experiments that demonstrate how the naive
method works. The idea behind these numerical experiments is to test this method
using c.m.\ energy dependencies of some known model cross sections. At the first
stage of each numerical experiment considered below, a model visible cross
section is calculated by substituting a model Born cross section into
eq.~\eqref{eq:visible-Born-relationship-kuraev-fadin-2}. Further, the points of
a visible cross section are generated according to the normal distribution using
the covariance matrix $\hat{\Lambda}$. As noted in the previous paragraph, the
off-diagonal elements of this matrix are equal to zero, while the diagonal
elements are taken proportional to a visible cross section in the examples
considered below. The last stage of each numerical experiment consists in
obtaining a numerical solution to
eq.~\eqref{eq:visible-Born-relationship-kuraev-fadin-2} using the naive method,
i.e.\ by solving the system of linear equations~\eqref{eq:KF-SLAE}, where the
vector of a generated visible cross section is substituted as the right-hand
side. Finally, the accuracy of the naive method can be estimated by comparing
a model Born cross section with a numerical solution obtained using this method.

\begin{figure}[tbp]
  \centering
  \begin{subfigure}[t]{0.47\textwidth}
    \centering
    \includegraphics[width=\textwidth]{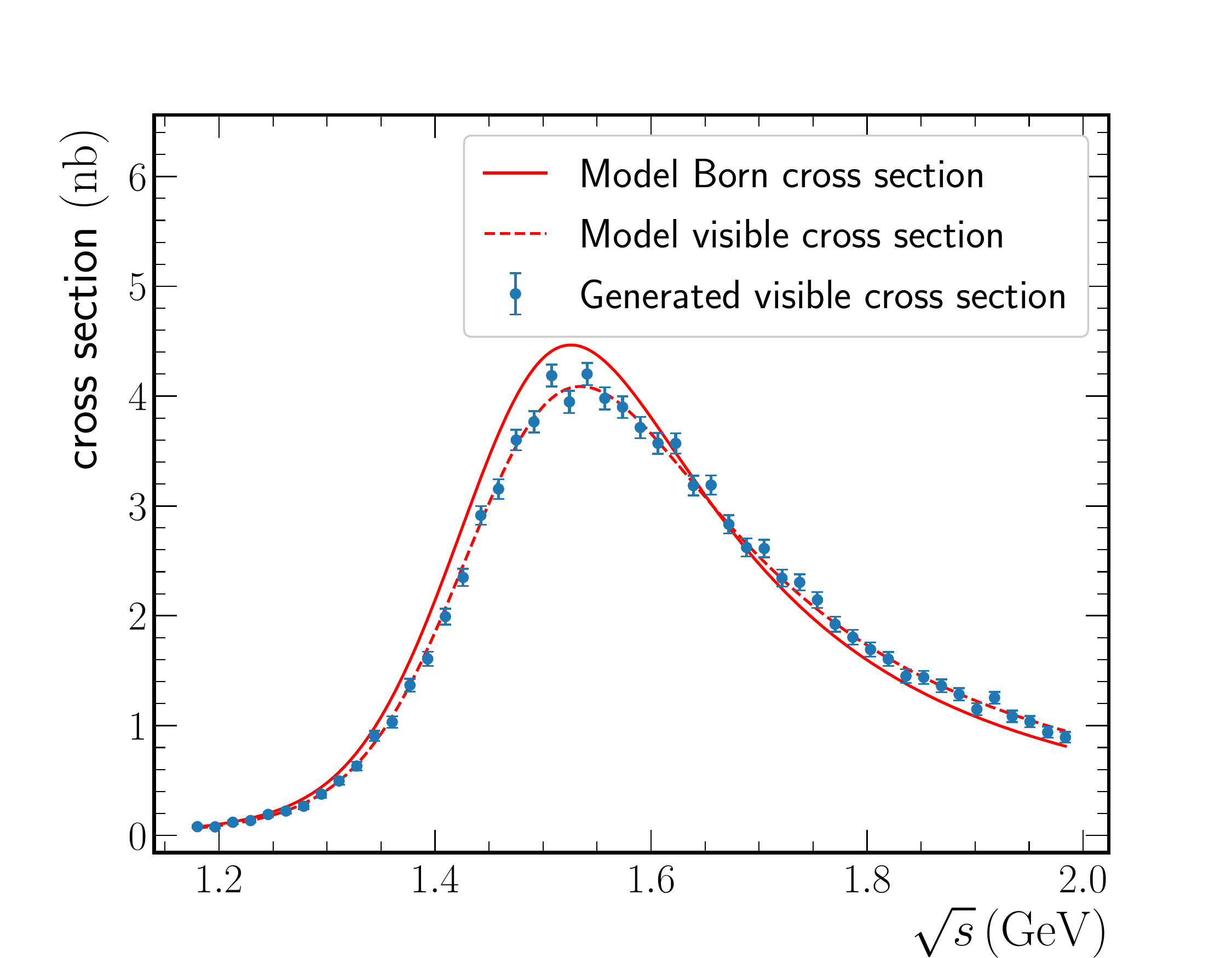}
    \caption{\label{fig:model-bcs-vcs-etapipi-linterp-no-energy-spread-errkoeff5em2}
      Model Born, model visible and generated visible cross sections.}
  \end{subfigure}
  \hfill
  \begin{subfigure}[t]{0.47\textwidth}
    \centering
    \includegraphics[width=\textwidth]{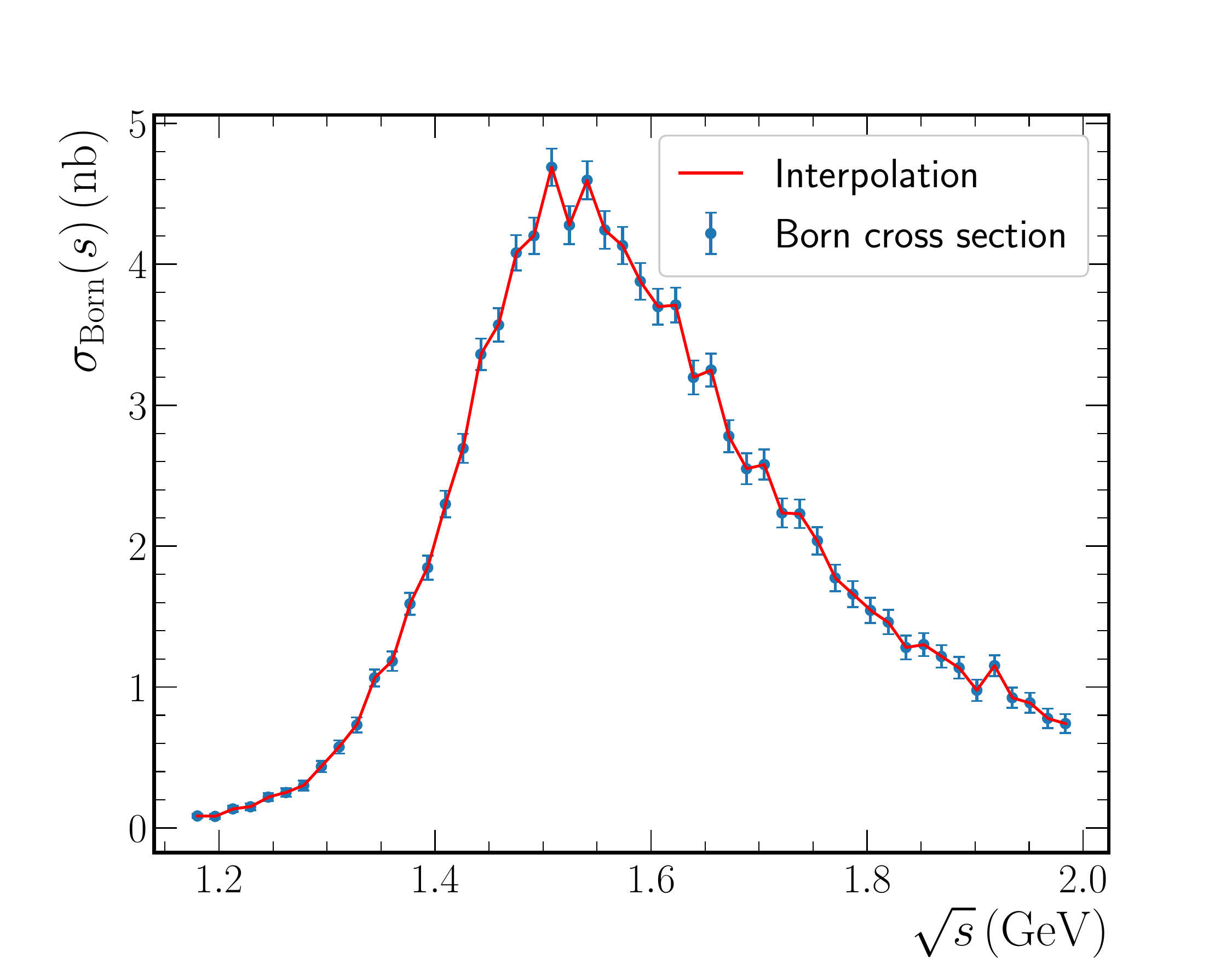}
    \caption{\label{fig:bcs-interp-etapipi-linterp-no-energy-spread-errkoeff5em2}
      Numerical solution of eq.~\eqref{eq:visible-Born-relationship-kuraev-fadin-2} obtained using the naive method.}
  \end{subfigure}
  \vfill
  \begin{subfigure}[t]{0.47\textwidth}
    \centering
    \includegraphics[width=\textwidth]{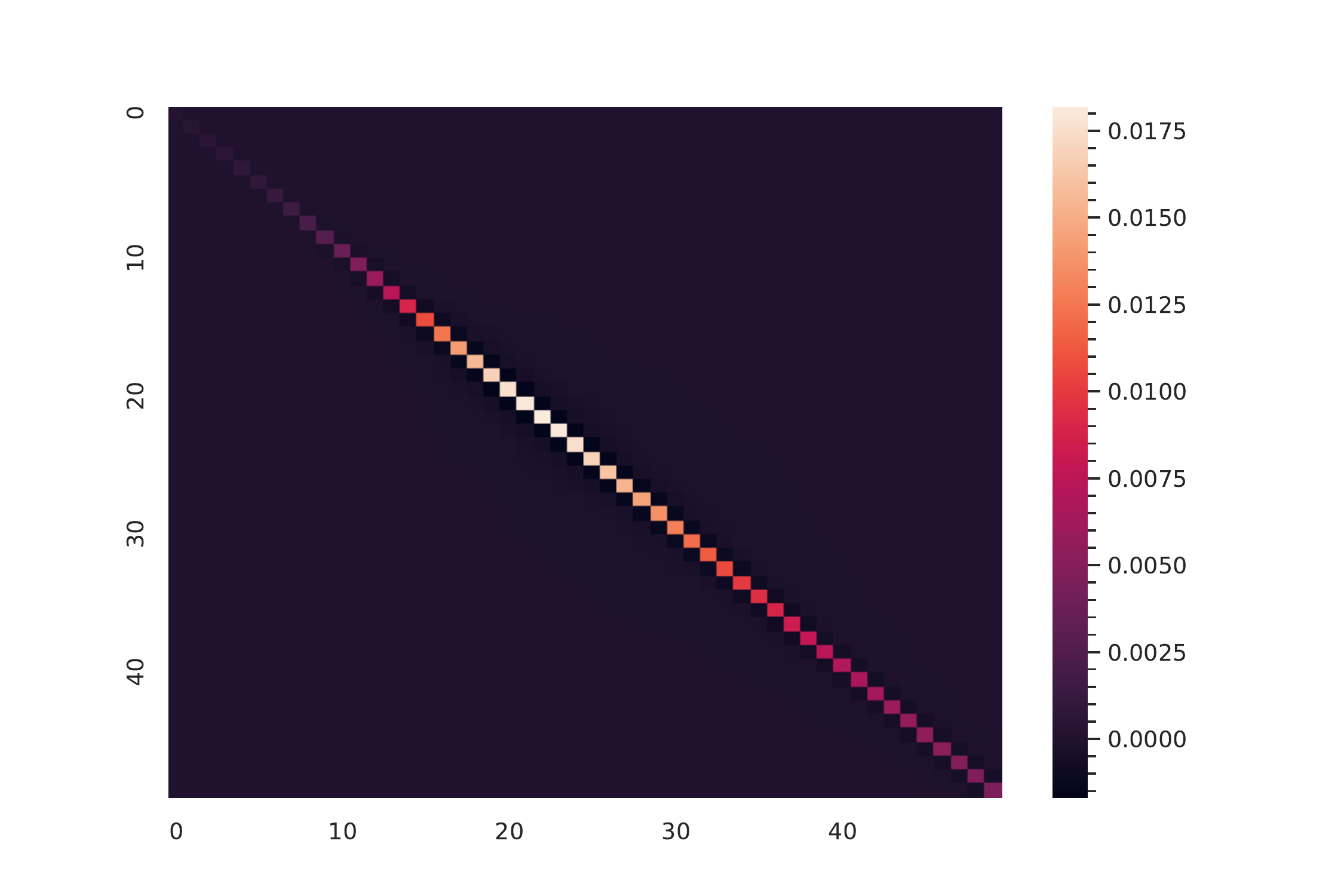}
    \caption{\label{fig:inv-error-matrix-etapipi-linterp-no-energy-spread}
      Covariance matrix $\mathcal{M}$ of the numerical solution of
      eq.~\eqref{eq:visible-Born-relationship-kuraev-fadin-2} in the case of
      piecewise linear interpolation.}
  \end{subfigure}
  \hfill
  \begin{subfigure}[t]{0.47\textwidth}
    \centering
    \includegraphics[width=\textwidth]{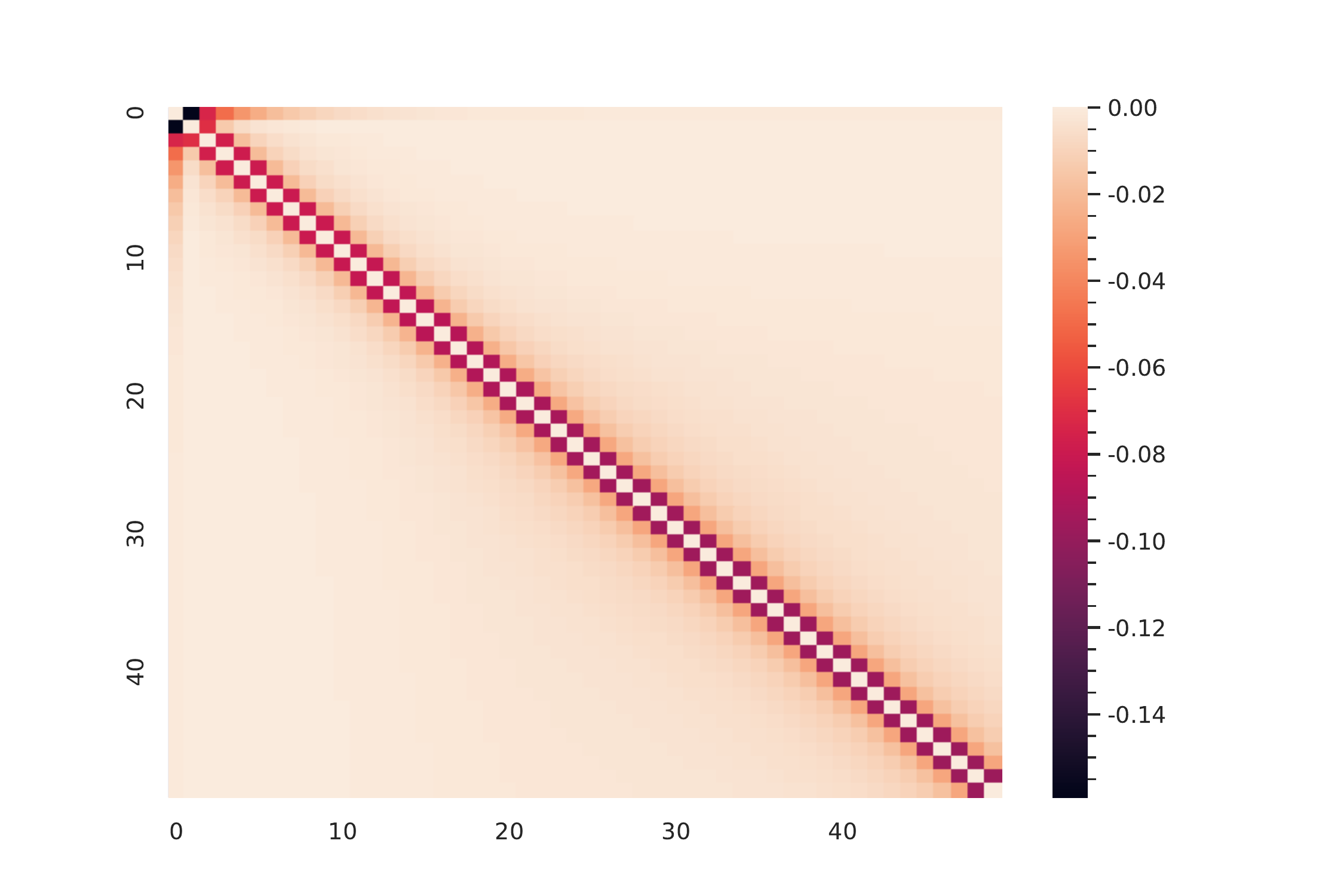}
    \caption{\label{fig:pearson-correlation-etapipi} Pearson correlation
      coefficient with subtracted diagonal elements.}
  \end{subfigure}
  \vfill
  \begin{subfigure}[t]{0.47\textwidth}
    \centering
    \includegraphics[width=\textwidth]{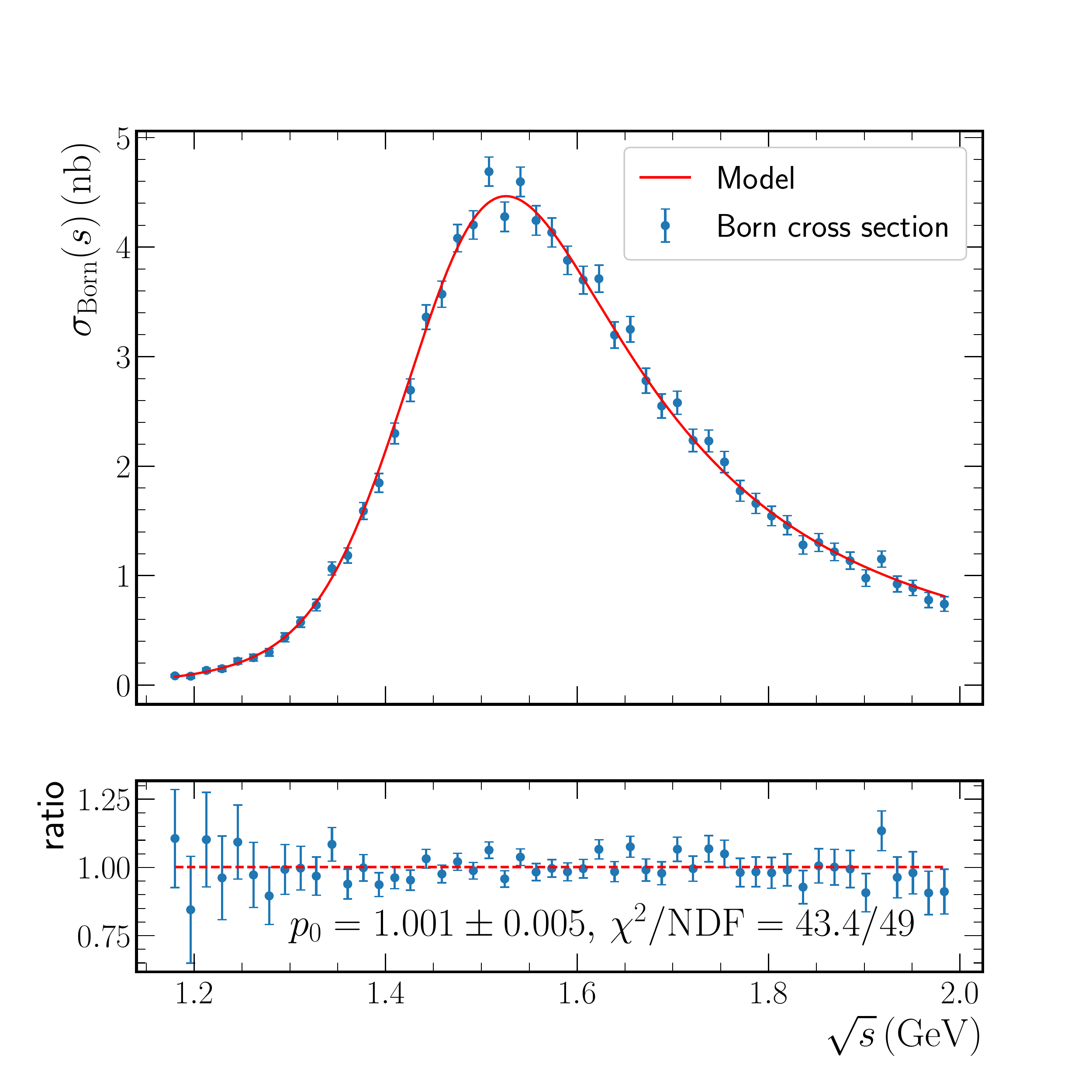}
    \caption{\label{fig:bcs-ratio-etapipi-linterp-no-energy-spread-errkoeff5em2}
      Comparison of the $e^+e^-\rightarrow\eta\pi^+\pi^-$ model Born cross
      section and the numerical solution of
      eq.~\eqref{eq:visible-Born-relationship-kuraev-fadin-2} obtained using the
      naive method with the generated visible cross section.}
  \end{subfigure}
  \hfill
  \begin{subfigure}[t]{0.47\textwidth}
    \centering
    \includegraphics[width=\textwidth]{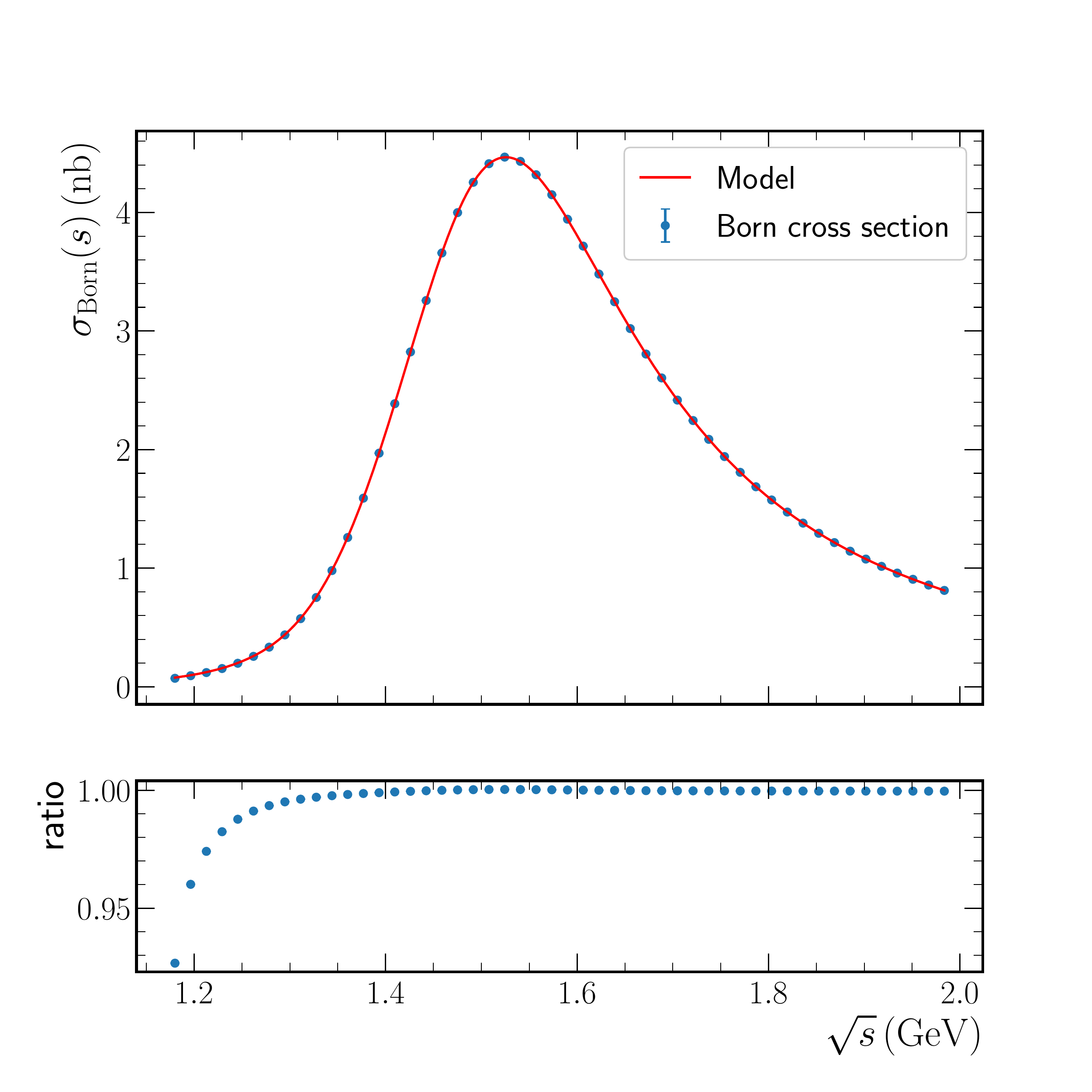}
    \caption{\label{fig:bcs-ratio-etapipi-linterp-no-energy-spread-errkoeff1em6}
      Comparison of the $e^+e^-\rightarrow\eta\pi^+\pi^-$ model Born cross
      section and the numerical solution of
      eq.~\eqref{eq:visible-Born-relationship-kuraev-fadin-2} obtained using the
      naive method with the model visible cross section.}
  \end{subfigure}
  \caption{\label{fig:numexp-etapipi-results1} Results of the numerical
    experiment with the $e^+e^-\rightarrow\eta\pi^+\pi^-$ cross
    section.}
\end{figure}
In the first example, let us consider the Born cross section of the
$e^+e^-\rightarrow\eta\pi^+\pi^-$ process within the framework of the vector
meson dominance model (VMD). Since we are only interested in the test of the
naive method for obtaining the Born cross section, we consider only a simple
model containing only $\rho\rightarrow\rho\eta$ and
$\rho^{\prime}\rightarrow\rho\eta$ intermediate states. To describe the
dependence of the $e^+e^-\rightarrow\eta\pi^+\pi^-$ Born cross section on the
c.m. energy, the function and parameters from the work~\cite{Gribanov:2019qgw}
are used. The corresponding model Born cross section is shown in
figure~\ref{fig:model-bcs-vcs-etapipi-linterp-no-energy-spread-errkoeff5em2} as
a solid curve. The model visible cross section obtained by substituting the
model Born cross section in
eq.~\eqref{eq:visible-Born-relationship-kuraev-fadin-2} is shown in this figure
as a dashed curve. The detection efficiency in this example is taken equal to
one. The visible cross section values are generated randomly according to the
model visible cross section at $50$ points equally spaced in the c.m. energy
range from $1.18$ to $2.00$ GeV. The generated visible cross section is shown in
figure~\ref{fig:model-bcs-vcs-etapipi-linterp-no-energy-spread-errkoeff5em2}
as points with error bars. The Born cross section obtained using the naive
method is shown in
figure~\ref{fig:bcs-interp-etapipi-linterp-no-energy-spread-errkoeff5em2}. The
solid curve in this figure is the corresponding piecewise linear interpolation.
The corresponding integral operator matrix $\hat{\mathcal{F}}$ and the Born
cross section covariance matrix $\hat{\mathcal{M}}$ are shown in
figure~\ref{fig:int-op-etapipi-linterp-no-energy-spread} and
figure~\ref{fig:inv-error-matrix-etapipi-linterp-no-energy-spread},
respectively. Figure~\ref{fig:inv-error-matrix-etapipi-linterp-no-energy-spread}
shows that the absolute values of the off-diagonal elements of the covariance
matrix $\hat{\mathcal{M}}$ are small in comparison with the diagonal elements of
this matrix. This statement is valid for the covariance matrix of the numerical
solution of eq.~\eqref{eq:visible-Born-relationship-kuraev-fadin-2}, since the
kernel function $F(x, s)$ decreases rapidly as the argument $x$ increases.
Moreover, all off-diagonal elements of the covariance matrix are non-positive.
This fact can be seen from figure~\ref{fig:pearson-correlation-etapipi}. This
figure shows the matrix of the Pearson correlation coefficient with subtracted
diagonal elements:
$\frac{\hat{\mathcal{M}}_{ij}}{\sqrt{\hat{\mathcal{M}}_{ii}\hat{\mathcal{M}}_{jj}}}
- \delta_{ij}$. The nature of the negative correlation between the points of the
Born cross section is simple. Since the value of the integral from
eq.~\eqref{eq:visible-Born-relationship-kuraev-fadin-2} at each c.m.\ energy
point is fixed at the value of the visible cross section at that point, an
increase in the Born cross section at one c.m.\ energy point leads to its
decrease in other c.m.\ energy points. Comparison of the model Born cross
section and the Born cross section obtained using the naive method is shown in
figures~\ref{fig:bcs-ratio-etapipi-linterp-no-energy-spread-errkoeff5em2} and
\ref{fig:bcs-ratio-etapipi-linterp-no-energy-spread-errkoeff1em6}. The model
Born cross section is shown as a solid curve, and the Born cross section
obtained using the naive method is shown as points with error bars. At the
bottom of each of these two figures, the ratio of the numerical solution to the
model Born cross section is additionally shown. The Born cross section shown in
figure~\ref{fig:bcs-ratio-etapipi-linterp-no-energy-spread-errkoeff5em2} is
obtained using the generated visible cross section as the right side of
eq.~\eqref{eq:KF-system}. The Born cross section shown in
figure~\ref{fig:bcs-ratio-etapipi-linterp-no-energy-spread-errkoeff1em6} is
obtained using the exact model visible cross section as the right side of
eq.~\eqref{eq:KF-system}.
Figure~\ref{fig:bcs-ratio-etapipi-linterp-no-energy-spread-errkoeff5em2} shows
that there is agreement between the model Born cross section and the numerical
solution within a given statistical uncertainty. However, it can be seen from
figure~\ref{fig:bcs-ratio-etapipi-linterp-no-energy-spread-errkoeff1em6} that
there is a systematic discrepancy of the numerical solution relative to the
model Born cross section at the c.m.\ energies close to the threshold energy.
This discrepancy is caused by the fact that interpolation poorly describes the
threshold behavior of the cross section. Despite the fact that at the point with
the lowest c.m.\ energy the relative value of the discrepancy reaches $7\%$, the
absolute value of this discrepancy is small due to the smallness of the Born
cross section near the threshold.
Figure~\ref{fig:bcs-ratio-etapipi-linterp-no-energy-spread-errkoeff1em6} also
shows that the relative discrepancy rapidly decreases with increasing c.m.
energy.

\begin{figure}[tbp]
  \centering
  \begin{subfigure}[t]{0.47\textwidth}
    \centering
    \includegraphics[width=\textwidth]{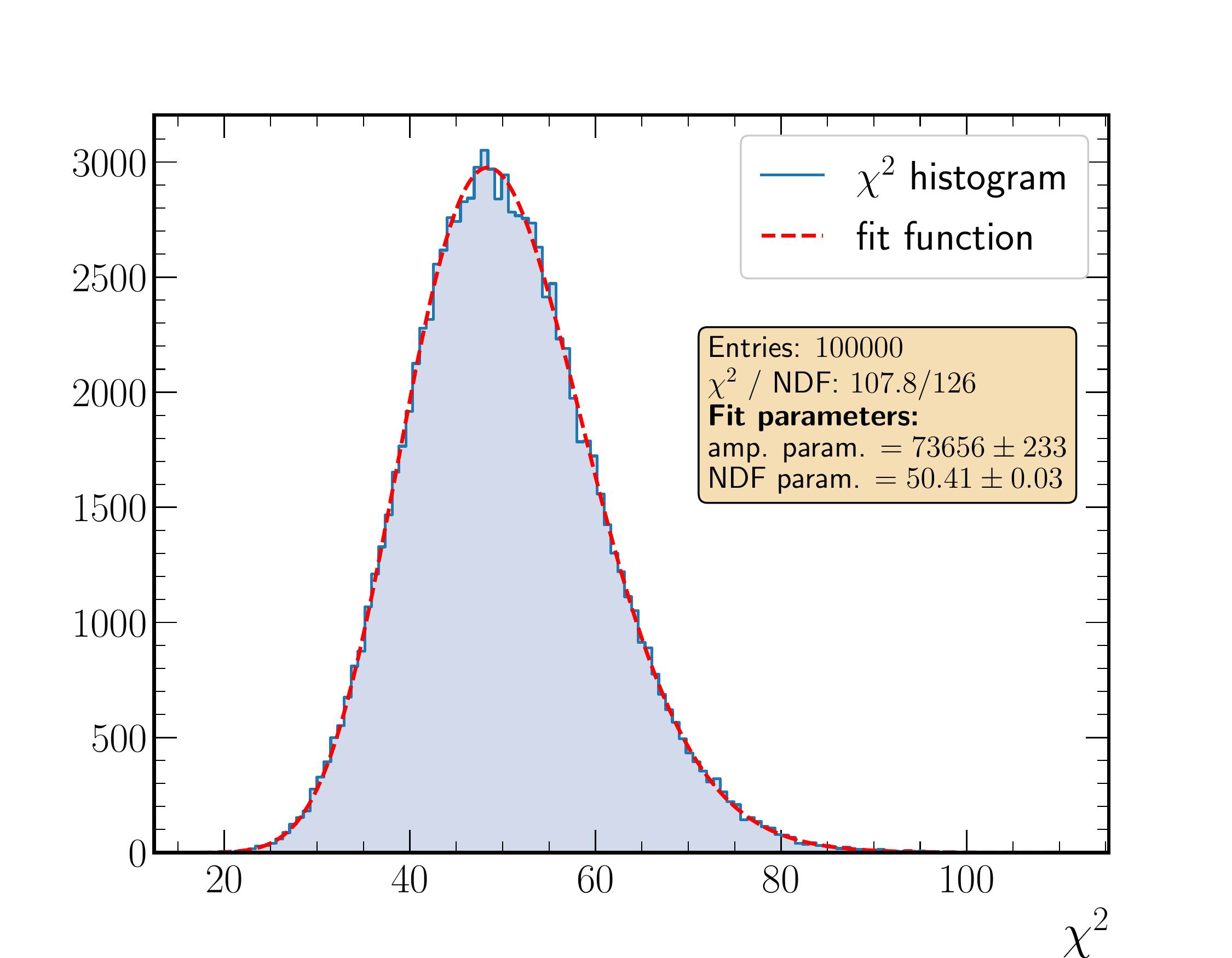}
    \caption{\label{fig:chi2_etapipi_model_fcn} Chi-square distribution with
      respect to the model Born cross section.}
  \end{subfigure}
  \hfill
  \begin{subfigure}[t]{0.47\textwidth}
    \centering
    \includegraphics[width=\textwidth]{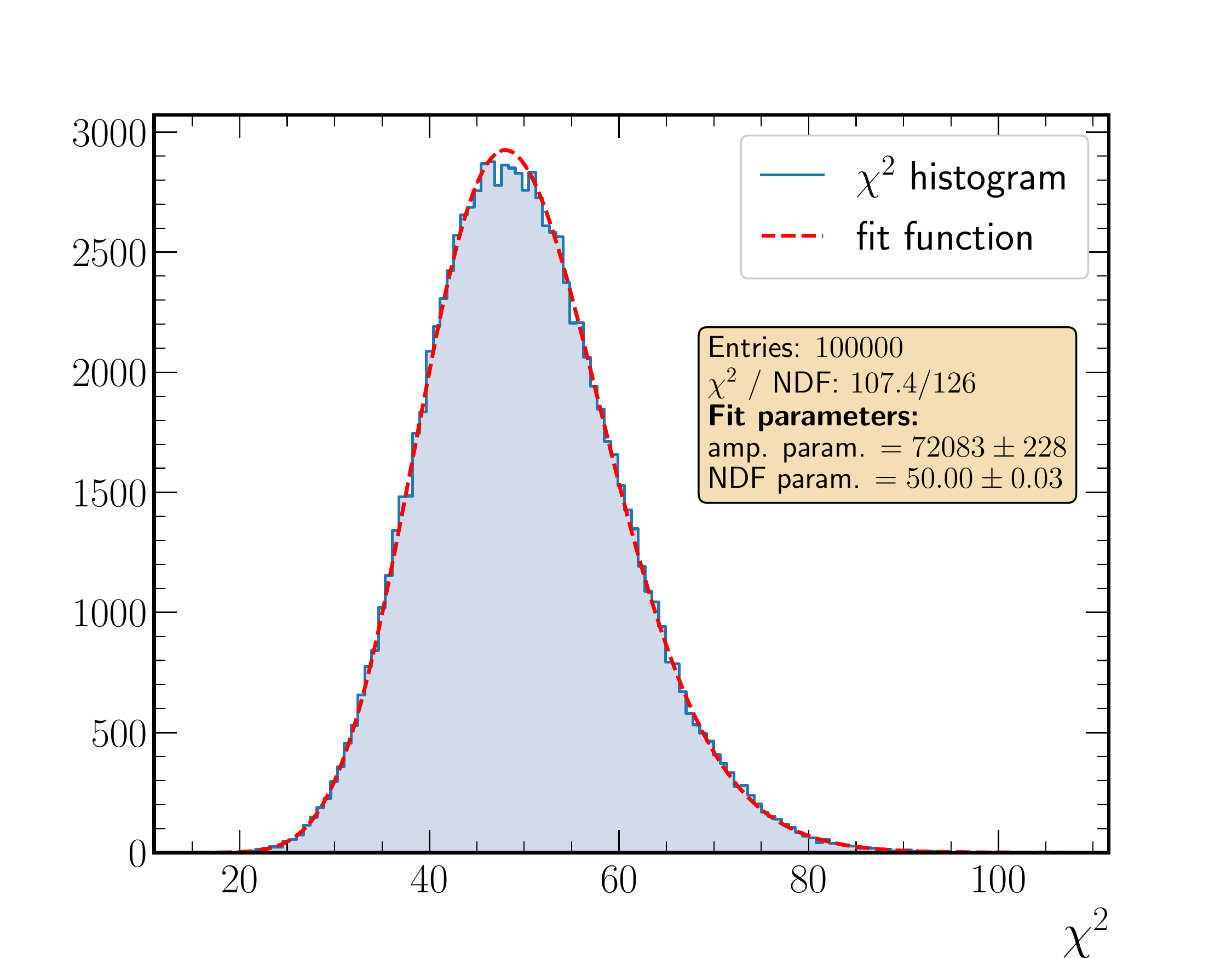}
    \caption{\label{fig:chi2_etapipi_data} Chi-square distribution with respect
      to the numerical solution averaged over all numerical experiments.}
  \end{subfigure}
  \caption{\label{fig:chi2_etapipi} Chi-square distribution for the
    $e^+e^-\rightarrow\eta\pi^+\pi^-$ Born cross section obtained using the
    naive method.}
\end{figure}

The discrepancy between the numerical solution and the model Born cross section,
as well as the fact that the covariance matrix describes the fluctuations of the
numerical solution in various numerical experiments, can be tested by
considering a chi-square histogram. Let us consider the chi-square $\chi^2_{\rm
  model}$ of the numerical solution calculated with respect to the model Born
cross section. In this case, the chi-square is given as follows:
\begin{equation}
  \label{eq:chi2_etapipi_model}
  \chi^2_{\rm model} = \left(\bm{\sigma}^{\rm naive}_{\rm Born} - \bm{\sigma}^{\rm model}_{\rm Born}\right)^{\intercal}\hat{\mathcal{M}}^{-1}\left(\bm{\sigma}^{\rm naive}_{\rm Born} - \bm{\sigma}^{\rm model}_{\rm Born}\right),
\end{equation}
where $\bm{\sigma}^{\rm naive}_{\rm Born}$ is the numerical solution obtained
using the naive method, $\bm{\sigma}^{\rm model}_{\rm Born}$ is the model Born
cross section. The chi-square~\eqref{eq:chi2_etapipi_model} histogram obtained
as a result of $10^5$ numerical experiments with the cross section of the
$e^+e^-\rightarrow\eta\pi^+\pi^-$ process is shown in
figure~\ref{fig:chi2_etapipi_model_fcn}. In each numerical experiment, the
visible cross section is generated according to the same covariance matrix
$\hat{\Lambda}$. The chi-square histogram is fitted with
a chi-square probability density function (PDF). The normalization factor (amp.\
param.\ in figure~\ref{fig:chi2_etapipi_model_fcn}) and the number of degrees of
freedom (NDF param.\ in figure~\ref{fig:chi2_etapipi_model_fcn}) are free fit
parameters. The dashed curve in figure~\ref{fig:chi2_etapipi_model_fcn}
corresponds to the fitting function. The fit parameters are also shown in this
figure. It can be seen from the figure that the fitting function describes well
the chi-square histogram. The value of the NDF parameter obtained as a result of
the fit is $50.41$, which is slightly different from the expected NDF value. The
expected value of NDF is $50$ because the cross section is obtained at $50$
points. The difference between the NDF parameter in fit and its expected value
is associated with the accuracy of the interpolation of the Born cross section.
Indeed, if we plot the chi-square distribution with respect to the numerical
solution averaged over all numerical experiments, then the deviation of the NDF
parameter from the expected value disappears, because the discrepancies
associated with interpolation are canceled out in the difference between the
numerical solution and its mean value. The corresponding chi-square distribution
is shown in figure~\ref{fig:chi2_etapipi_data}.

\begin{figure}[tbp]
   \centering
   \begin{subfigure}[t]{0.47\textwidth}
     \centering
     \includegraphics[width=\textwidth]{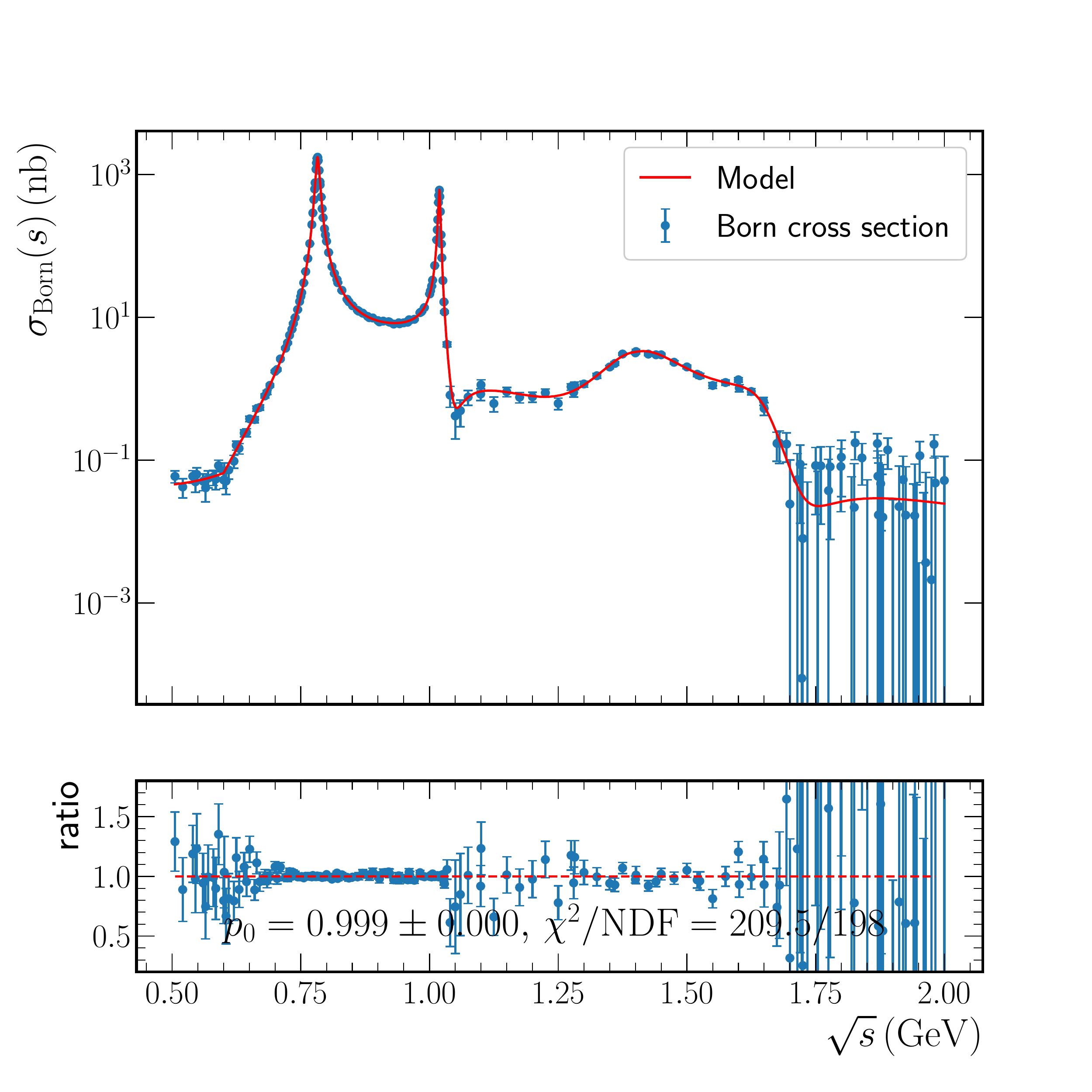}
     \caption{\label{fig:bcs_ratio_3pi_noenspread_errkoeff5em2_numpt200_method_SLAE_interp_mixed}
       Comparison of the $e^+e^-\rightarrow\pi^+\pi^-\pi^0$ model Born cross
       section and the Born cross section obtained using the naive method with
       the generated visible cross section.}
   \end{subfigure}
   \hfill
   \begin{subfigure}[t]{0.47\textwidth}
     \centering
     \includegraphics[width=\textwidth]{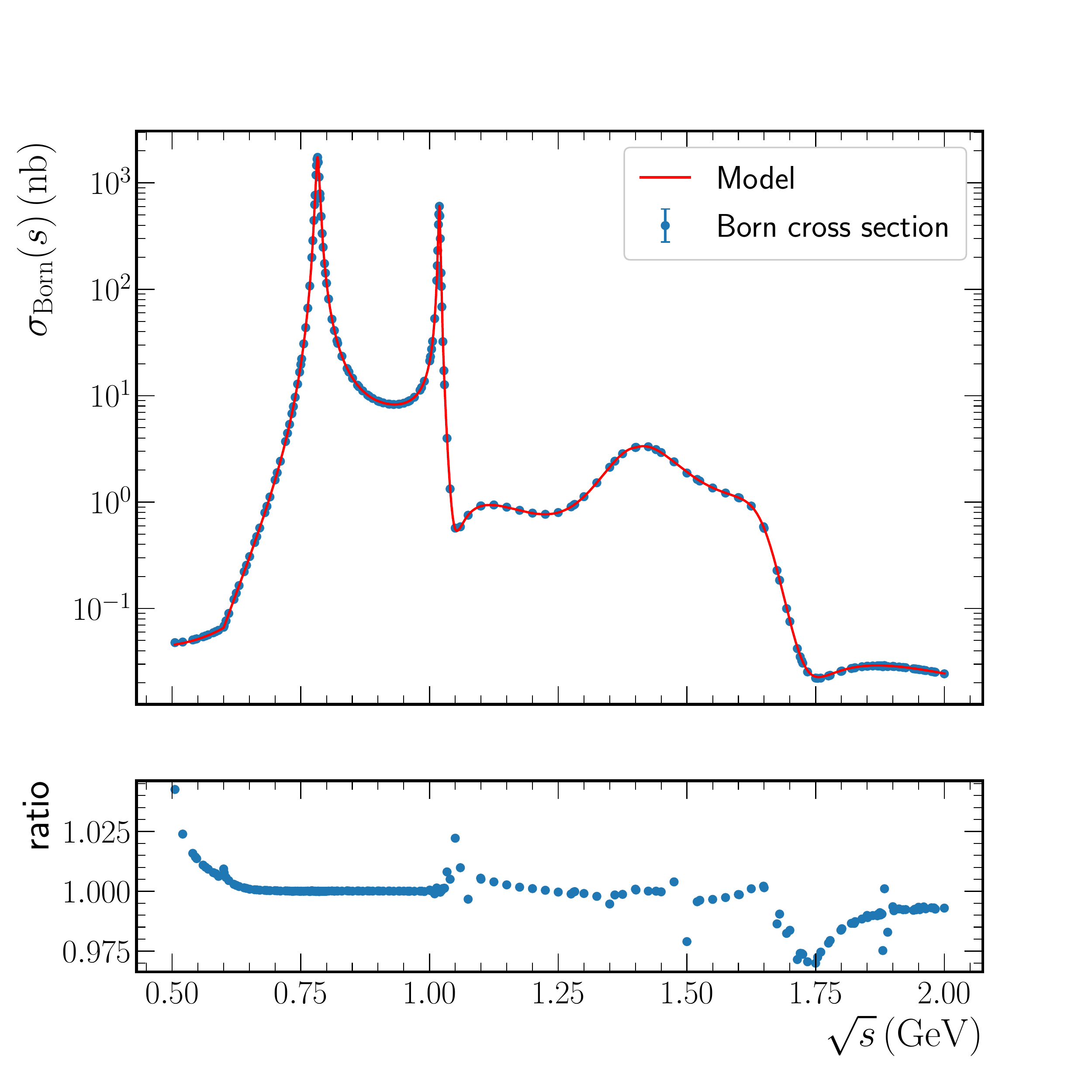}
     \caption{\label{fig:bcs_ratio_3pi_noenspread_errkoeff0_numpt200_method_SLAE_interp_mixed}
       Comparison of the $e^+e^-\rightarrow\pi^+\pi^-\pi^0$ model Born cross
       section and the Born cross section obtained using the naive method with
       the model visible cross section.}
   \end{subfigure}
   \caption{\label{fig:bcs_ratio_3pi_noenspread} Comparison of the
     $e^+e^-\rightarrow\pi^+\pi^-\pi^0$ model Born cross section with the Born
     cross section obtained using the naive method. The results shown in this
     figure are obtained using mixed interpolation.}
 \end{figure}

As the second example, let us consider the cross section of the
$e^+e^-\rightarrow\pi^+\pi^-\pi^0$ process. To describe the dependence of the
Born cross section of this process on the c.m.\ energy, the vector-meson
dominance model is also used. In this work, to describe the energy dependence of
the Born cross section, we use the function and parameters given in the
work~\cite{Epifanov2006}. However, we also introduced nonzero $\omega(1420)$ and
$\omega(1650)$ contributions in order to make the behavior of the cross section
more complicated at energies above the $\phi$-meson production energy. The model
Born cross section of the $e^+e^-\rightarrow\pi^+\pi^-\pi^0$ is shown in
figure~\ref{fig:bcs_ratio_3pi_noenspread_errkoeff5em2_numpt200_method_SLAE_interp_mixed}
as a solid curve. The considered cross section has two sharp peaks at the
$\omega$ and $\phi$ production energies. In order for interpolation to describe
them well, it is necessary to provide a high density of c.m.\ energy points in
the regions of these peaks. The density of points outside the peaks may be
lower. To test the naive method and to what extent this method is suitable under
the conditions of real experiment and the complex behavior of the cross section,
in this case, such c.m.\ energy points are used that are close to the points
obtained during the operation of the VEPP-2000 collider for several seasons of
data acquisition. A comparison of the model Born cross section and the numerical
solution of eq.~\eqref{eq:visible-Born-relationship-kuraev-fadin-2} obtained
using the naive method in this case is shown in
figure~\ref{fig:bcs_ratio_3pi_noenspread}. The model Born cross section is shown
in
figures~\ref{fig:bcs_ratio_3pi_noenspread_errkoeff5em2_numpt200_method_SLAE_interp_mixed}
and
\ref{fig:bcs_ratio_3pi_noenspread_errkoeff0_numpt200_method_SLAE_interp_mixed}
as a solid curve, and the numerical solution is shown as points with error bars.
The numerical solution shown in
figure~\ref{fig:bcs_ratio_3pi_noenspread_errkoeff5em2_numpt200_method_SLAE_interp_mixed}
is obtained using the generated visible cross section as the right part of
eq.~\eqref{eq:KF-SLAE}, while the numerical solution shown in
figure~\ref{fig:bcs_ratio_3pi_noenspread_errkoeff0_numpt200_method_SLAE_interp_mixed}
is obtained using the exact model visible cross section as the right side of
this equation.
Figure~\ref{fig:bcs_ratio_3pi_noenspread_errkoeff0_numpt200_method_SLAE_interp_mixed}
shows that, as in the case with the $e^+e^-\rightarrow\eta\pi^+\pi^-$ process,
there is a discrepancy between the numerical solution and the model Born cross
section near the c.m.\ energies close to the threshold energy. As noted above,
this discrepancy is due to the fact that interpolation poorly describes the
threshold behavior of the Born cross section. In addition, a significant
relative discrepancy between the numerical solution and the model Born cross
section is observed at the c.m.\ energy of $1.75$ GeV. This discrepancy is due
to the fact that interpolation does not adequately describe the sharp behavior
of the Born cross section at this energy. However, it should be noted that the
absolute value of this discrepancy small as well as in the case of the
discrepancy near the threshold energy.

\begin{figure}[tbp]
   \centering
   \begin{subfigure}[t]{0.47\textwidth}
     \centering
     \includegraphics[width=\textwidth]{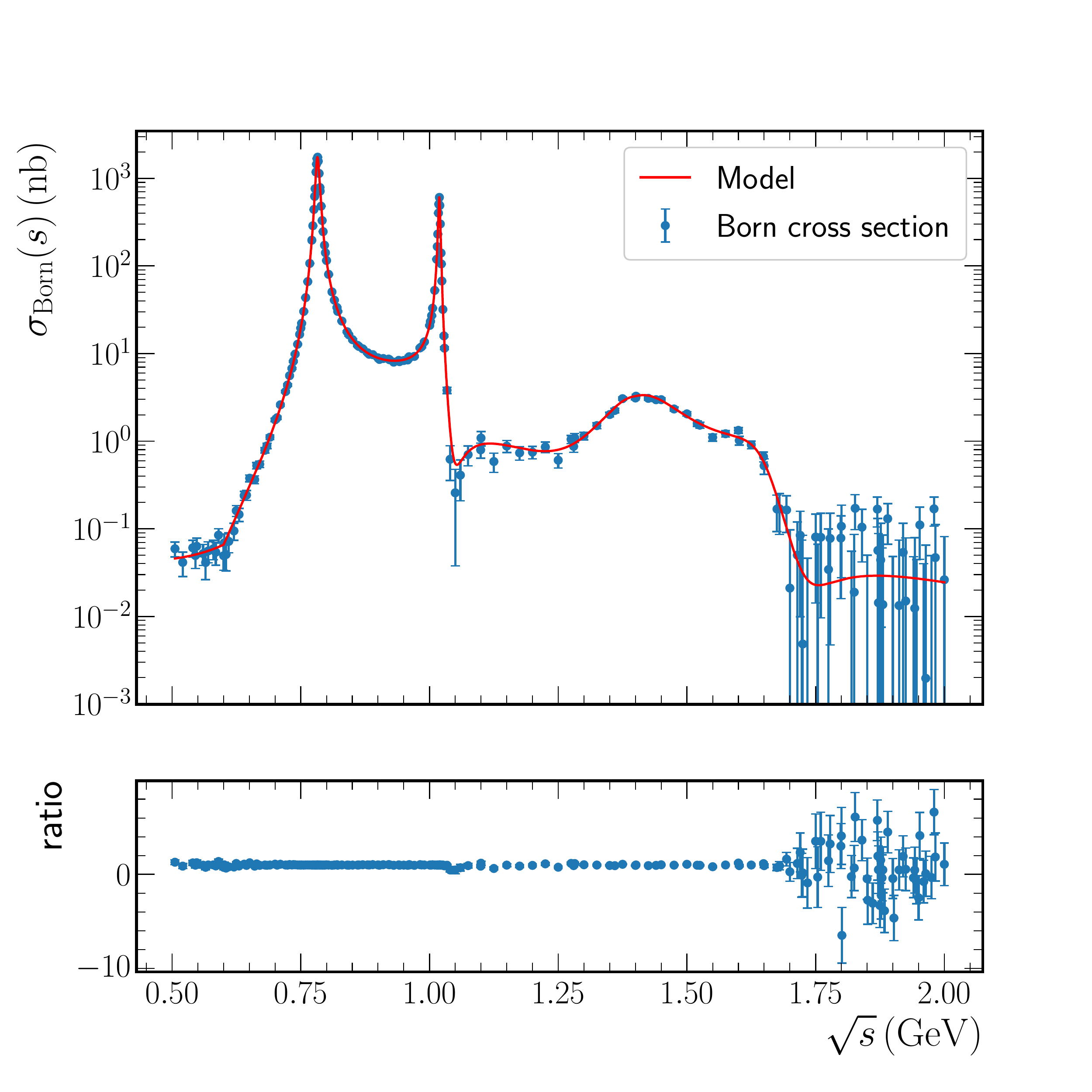}
     \caption{\label{fig:ratio_bcs_3pi_errkoeff5em2_200pt_noenspread_lininterp}
       Comparison of the $e^+e^-\rightarrow\pi^+\pi^-\pi^0$ model Born cross
       section and the Born cross section obtained using the naive method with
       the generated visible cross section.}
   \end{subfigure}
   \hfill
   \begin{subfigure}[t]{0.47\textwidth}
     \centering
     \includegraphics[width=\textwidth]{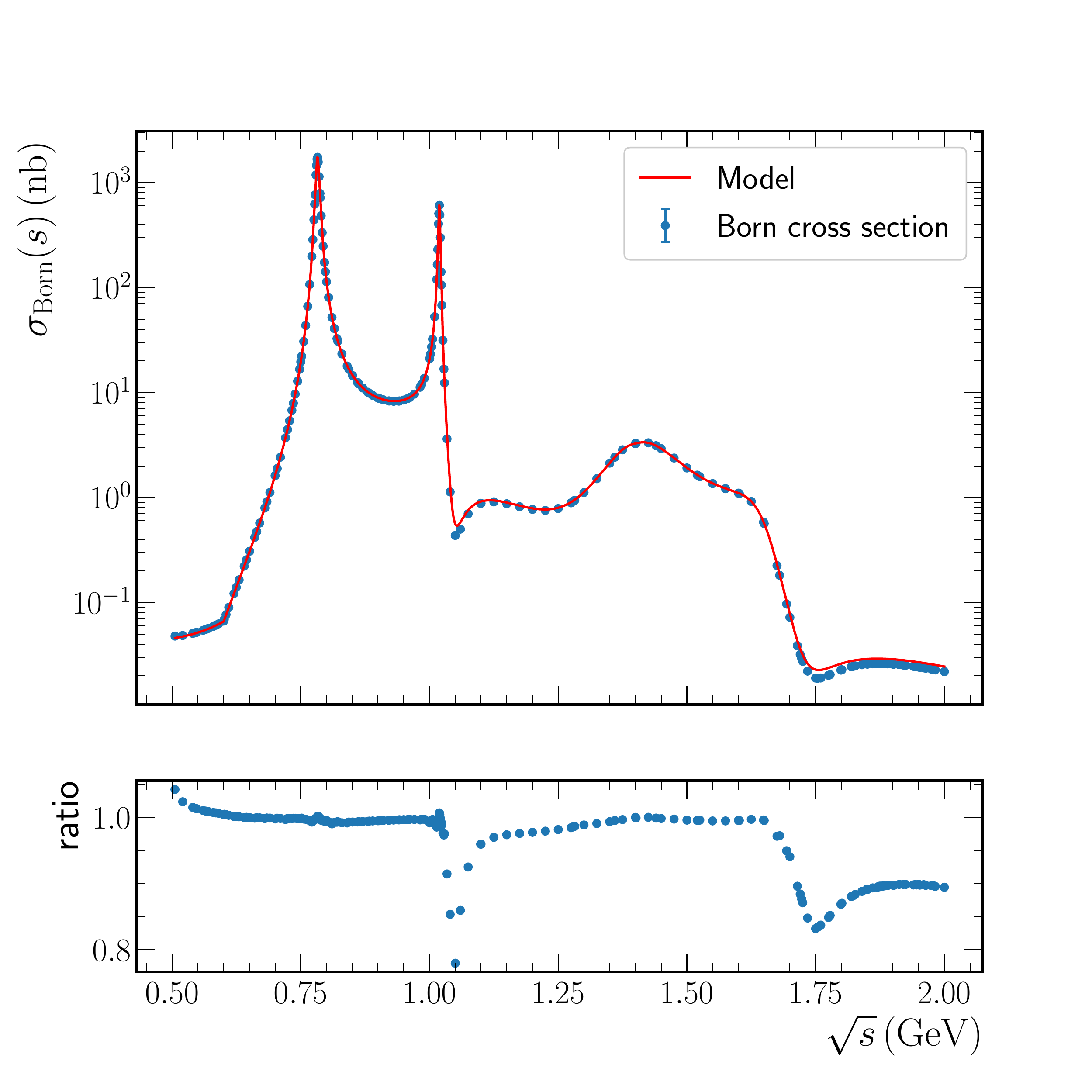}
     \caption{\label{fig:ratio_bcs_3pi_errkoeff0_200pt_noenspread_lininterp}
       Comparison of the $e^+e^-\rightarrow\pi^+\pi^-\pi^0$ model Born cross
       section and the Born cross section obtained using the naive method with
       the model visible cross section.}
   \end{subfigure}
   \caption{\label{fig:ratio_bcs_3pi_200pt_noenspread_lininterp} Comparison of
     the $e^+e^-\rightarrow\pi^+\pi^-\pi^0$ model Born cross section with the
     Born cross section obtained using the naive method. The results shown in
     the figure are obtained using piecewise linear interpolation.}
 \end{figure}

The results shown in figure~\ref{fig:bcs_ratio_3pi_noenspread} are obtained
using mixed interpolation, i.e.\ piecewise linear interpolation of the Born
cross section is used on some c.m.\ energy ranges, while cubic spline
interpolation is used on other ranges. Mixed interpolation is used for the
reason that in the energy ranges with abrupt changes in the Born cross section,
the accuracy of piecewise linear interpolation decreases significantly. A
comparison between the numerical solution and the model Born cross section,
similar to that shown in figure~\ref{fig:bcs_ratio_3pi_noenspread}, is shown in
figure~\ref{fig:ratio_bcs_3pi_200pt_noenspread_lininterp}, but in the case of
using piecewise linear interpolation. It can be seen from this figure that in
areas with abrupt changes in the cross-section, the interpolation accuracy is
indeed significantly lower than in the case of using mixed interpolation.
 
\begin{figure}[tbp]
   \centering
   \begin{subfigure}[t]{0.47\textwidth}
     \centering
     \includegraphics[width=\textwidth]{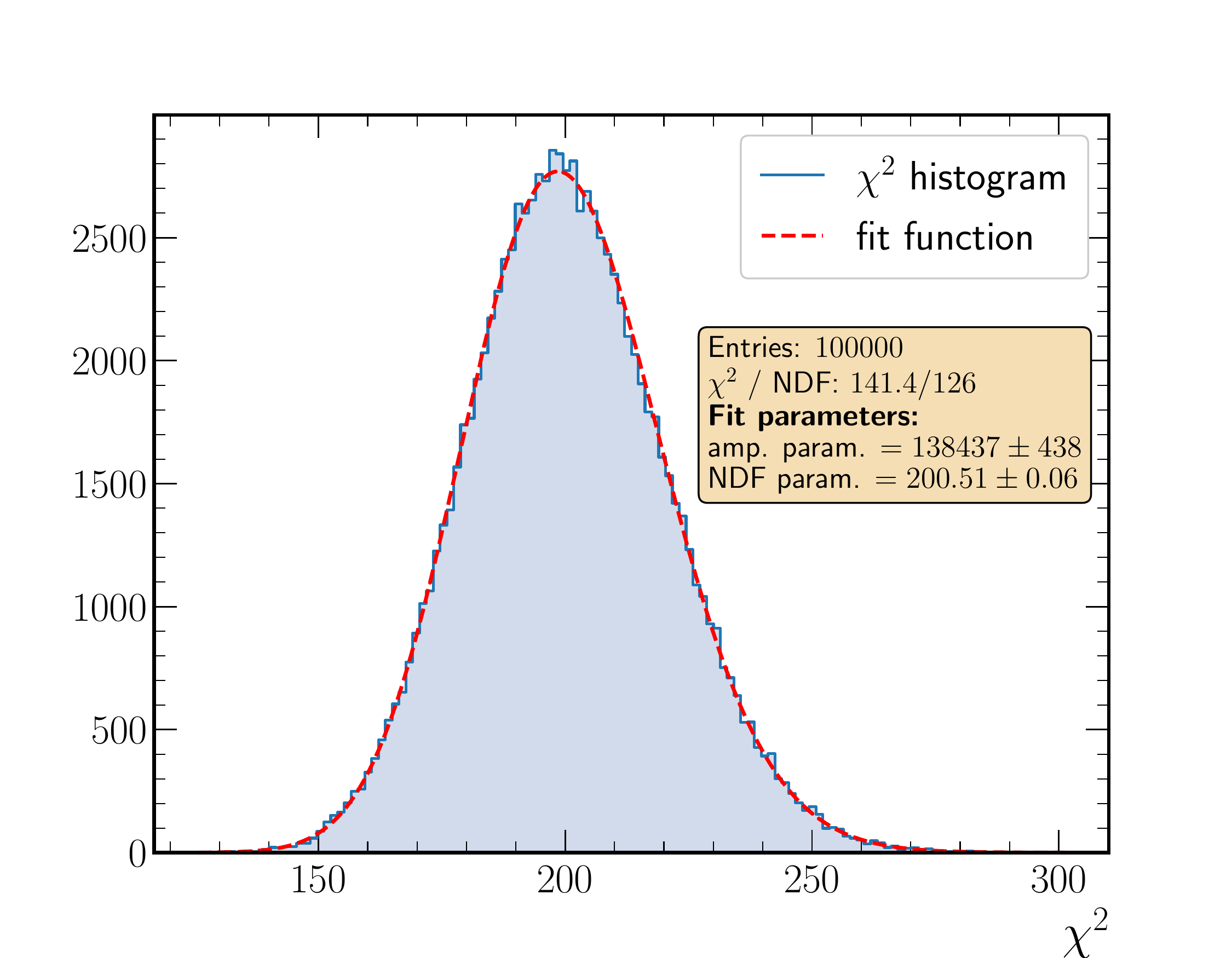}
     \caption{\label{fig:chi2_3pi_model_mixed} Chi-square distribution with
       respect to the model Born cross section.}
   \end{subfigure}
   \hfill
   \begin{subfigure}[t]{0.47\textwidth}
     \centering
     \includegraphics[width=\textwidth]{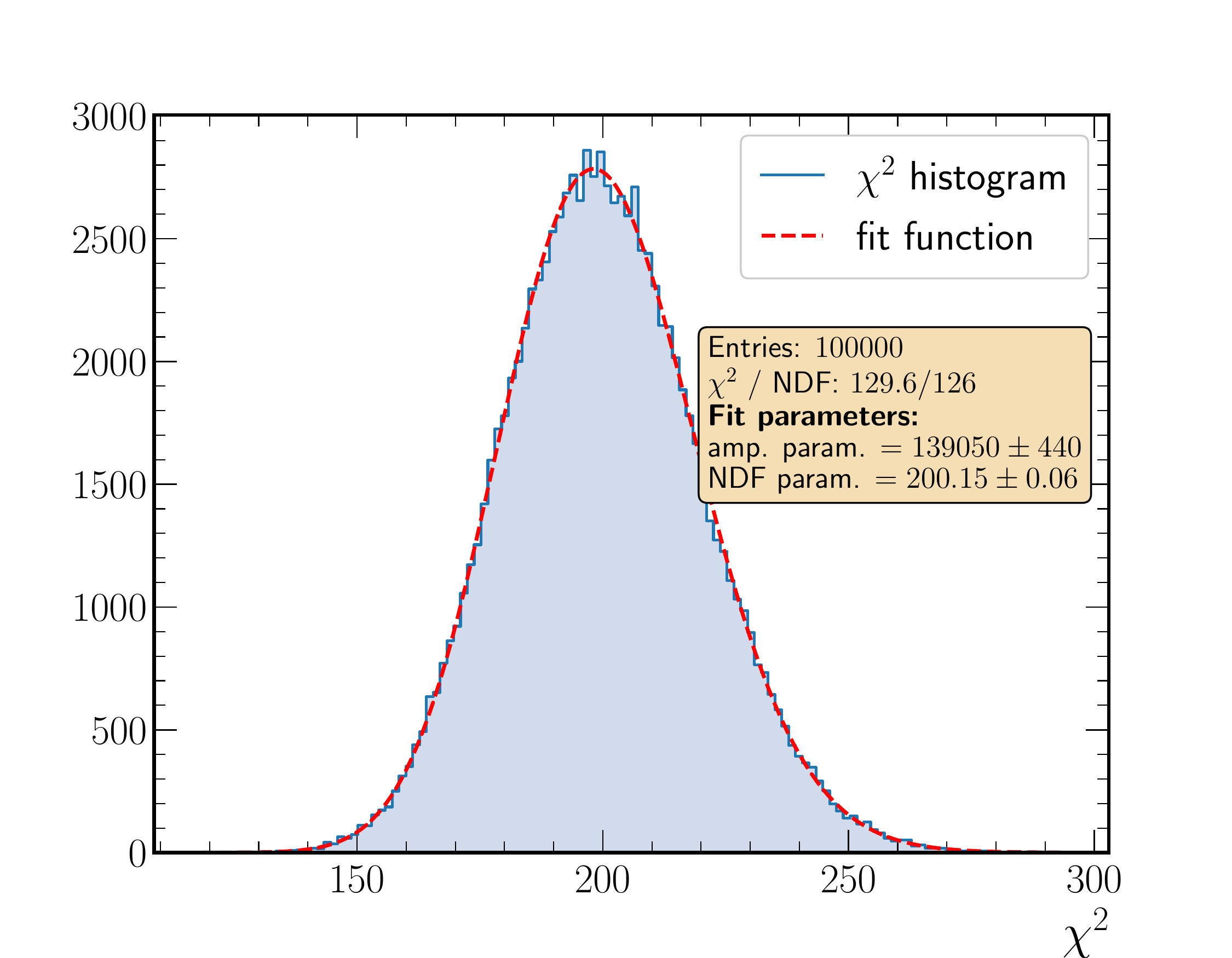}
     \caption{\label{fig:chi2_3pi_data_mixed} Chi-square distribution with
       respect to the numerical solution averaged over all numerical
       experiments.}
   \end{subfigure}
   \caption{\label{fig:chi2_3pi_mixed} Chi-square distribution for the
     $e^+e^-\rightarrow\pi^+\pi^-\pi^0$ Born cross section obtained using the
     naive method. The numerical solution is obtained using mixed interpolation.}
 \end{figure}

\begin{figure}[tbp]
   \centering
   \begin{subfigure}[t]{0.47\textwidth}
     \centering
     \includegraphics[width=\textwidth]{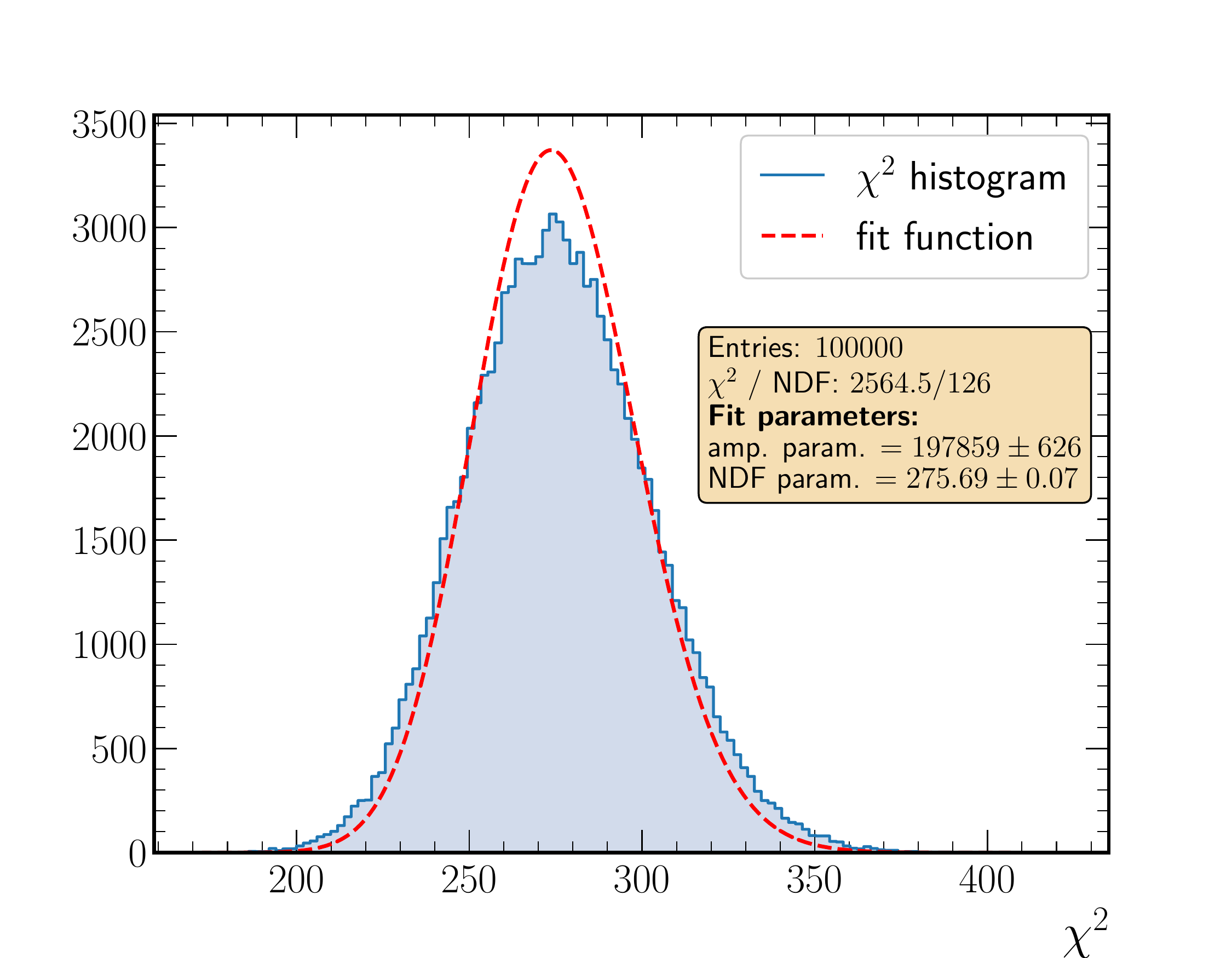}
     \caption{\label{fig:chi2_3pi_model_linear} Chi-square distribution with
       respect to the model Born cross section.}
   \end{subfigure}
   \hfill
   \begin{subfigure}[t]{0.47\textwidth}
     \centering
     \includegraphics[width=\textwidth]{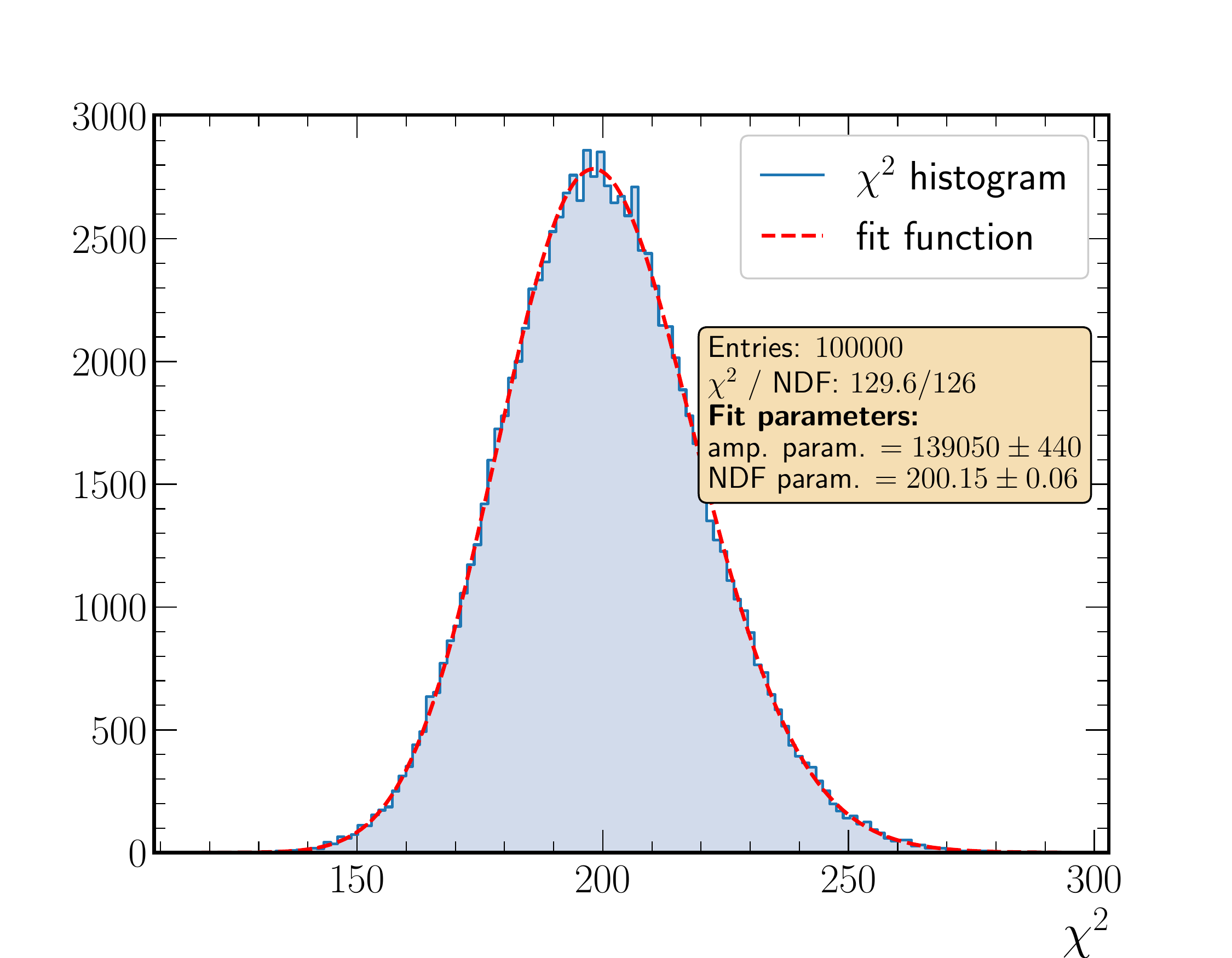}
     \caption{\label{fig:chi2_3pi_data_linear} Chi-square distribution with
       respect to the numerical solution averaged over all numerical
       experiments.}
   \end{subfigure}
   \caption{\label{fig:chi2_3pi_linear} Chi-square distribution for the
     $e^+e^-\rightarrow\pi^+\pi^-\pi^0$ Born cross section obtained using the
     naive method. The numerical solution is obtained using piecewise linear
     interpolation.}
 \end{figure}

 Similarly, as in the case of the process $e^+e^-\rightarrow\eta\pi^+\pi^-$, the
 chi-square histograms of the numerical solution for the
 $e^+e^-\rightarrow\pi^+\pi^-\pi^0$ is obtained in the cases of mixed
 interpolation and piecewise linear interpolation.
 Figure~\ref{fig:chi2_3pi_mixed} shows the chi-square histograms obtained using
 mixed interpolation of the Born cross section.
 Figure~\ref{fig:chi2_3pi_model_mixed} shows the chi-square histogram of the
 numerical solution with respect to the model Born cross section, while
 figure~\ref{fig:chi2_3pi_data_mixed} shows the chi-square histogram of the
 numerical solution with respect to the numerical solution averaged over all
 numerical experiments. The histograms shown in
 figures~\ref{fig:chi2_3pi_model_mixed} and \ref{fig:chi2_3pi_data_mixed} are
 fitted with a chi-squared distribution. The amplitude of this distribution and
 the number of degrees of freedom are free parameters of the fit. The fit curve
 is shown as a solid curve in each of these two figures. The parameters obtained
 as a result of the corresponding fit are shown in each of
 figures~\ref{fig:chi2_3pi_model_mixed} and \ref{fig:chi2_3pi_data_mixed}. It
 can be seen from figure~\ref{fig:chi2_3pi_mixed} that there is good agreement
 between the histograms and the fitting curves. The number of degrees of
 freedom for the chi-square distribution in each of
 figures~\ref{fig:chi2_3pi_model_mixed} and \ref{fig:chi2_3pi_data_mixed} is
 consistent with the expected value of this parameter equal to $200$, i.e.\ the
 number of c.m.\ energy points. Figure~\ref{fig:chi2_3pi_linear} is similar to
 figure~\ref{fig:chi2_3pi_mixed}, but is obtained using linear interpolation of
 the Born cross section. The histogram shown in
 figure~\ref{fig:chi2_3pi_model_linear} (chi-square with respect to the model
 Born cross section) does not correspond to the expected chi-square distribution
 with $200$ degrees of freedom. This discrepancy is associated with a
 significant discrepancy between the numerical solution and the model Born cross
 section, caused by the fact that piecewise linear interpolation has low
 accuracy in the case of sharply varying cross sections. In contrast, the
 histogram shown in figure~\ref{fig:chi2_3pi_data_linear} (chi-square with
 respect to averaged numerical solution) corresponds to a chi-square
 distribution with 200 degrees of freedom.

 \subsubsection{Numerical experiments with equation~\eqref{eq:visible-Born-relationship-kuraev-fadin-2-blured}}
\begin{figure}[tbp]
   \centering
   \begin{subfigure}[t]{0.47\textwidth}
     \centering
     \includegraphics[width=\textwidth]{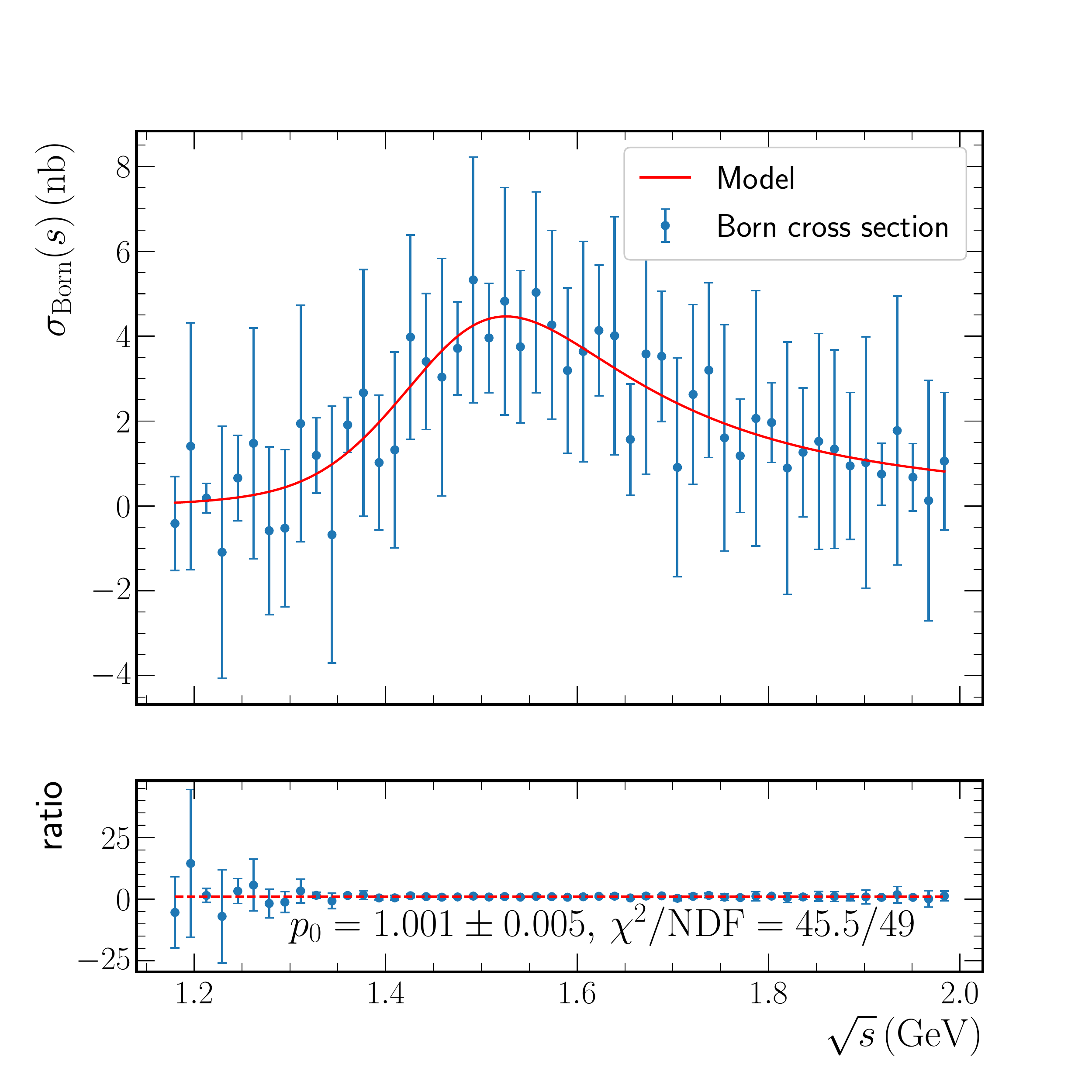}
     \caption{\label{fig:bcs_ratio_etapipi_errkoeff5em2_enspread1p0em2_method_SLAE}
       Numerical solution of
       eq.~\eqref{eq:visible-Born-relationship-kuraev-fadin-2-blured}
       obtained using the naive method.}
   \end{subfigure}
   \hfill
   \begin{subfigure}[t]{0.47\textwidth}
     \centering
     \includegraphics[width=\textwidth]{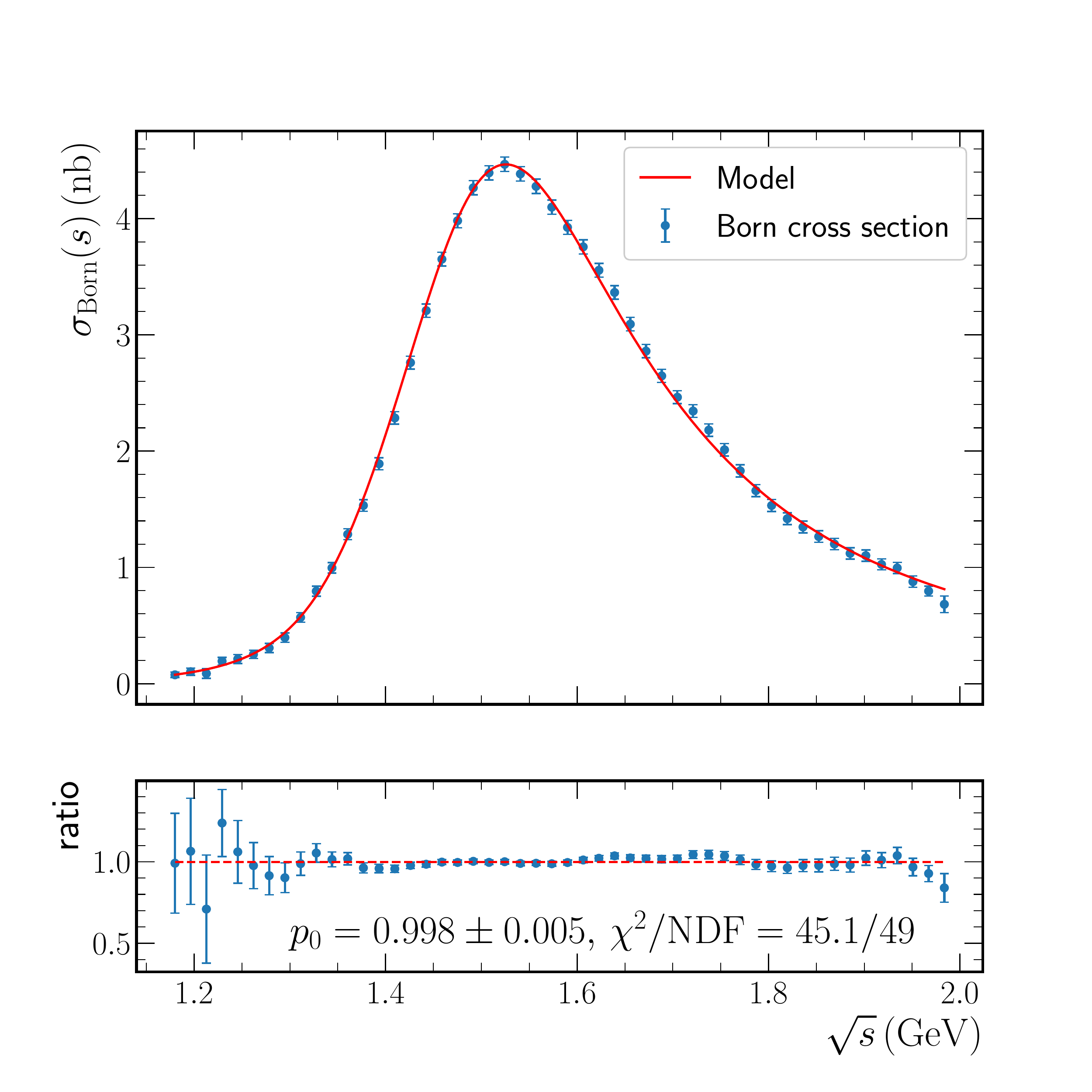}
     \caption{\label{fig:bcs_ratio_etapipi_errkoeff5em2_enspread1p0em2_method_tikhonov_optalpha0p721}
       Numerical solution of
       eq.~\eqref{eq:visible-Born-relationship-kuraev-fadin-2-blured} obtained
       using the Tikhonov regularization method. The value ($\lambda\approx 0.78\text{
         GeV}^2\text{nb}^{-2}$) of the regularization parameter is chosen using
       the L-curve criterion.}
   \end{subfigure}
   \caption{\label{fig:bcs_ratio_etapipi_enspread1p0em2} Comparison of the
     $e^+e^-\rightarrow\eta\pi^+\pi^-$ model Born cross and the numerical
     solution of eq.~\eqref{eq:visible-Born-relationship-kuraev-fadin-2-blured}.
     Values of the parameter $\sigma_E$ at each c.m.\ energy
     point are same and equal to $20\text{ MeV}$.}
 \end{figure}

 Next, let us consider obtaining a numerical solution to
 eq.~\eqref{eq:visible-Born-relationship-kuraev-fadin-2-blured} using the naive
 method. As noted in section~\ref{sec:discretization-eq-kf-blured}, the
 condition number of the matrix $\hat{\mathcal{G}}\hat{\mathcal{F}}$ in this
 case can be large, and, therefore, small perturbations of the visible cross
 section can lead to large perturbations in the numerical solution. Let us also
 consider a numerical experiment similar to the first numerical experiment from
 section~\ref{sec:numerical-experiments-without-enspread} with the only
 difference that the c.m.\ energy spread $\sigma_E=20\text{ MeV}$ is used. This
 spread is extremely large compared to the typical c.m.\ energy spread at
 VEPP-$2000$ and is used for demonstration purposes only. The condition number
 of the matrix $\hat{\mathcal{G}}\hat{\mathcal{F}}$ in this case is
 approximately equal to $201.6$. The numerical experiment consists in the fact
 that the visible cross section is generated using the model Born cross section
 of the $e^+e^-\rightarrow\eta\pi^+\pi^-$ process.
 Eq.~\eqref{eq:visible-Born-relationship-kuraev-fadin-2-blured} is then
 numerically solved using the naive method. An example of a numerical solution
 of eq.~\eqref{eq:visible-Born-relationship-kuraev-fadin-2-blured} is shown in
 figure~\ref{fig:bcs_ratio_etapipi_errkoeff5em2_enspread1p0em2_method_SLAE} as
 points with error bars. The solid curve in this figure denotes the model Born
 cross section. The lower part of
 figure~\ref{fig:bcs_ratio_etapipi_errkoeff5em2_enspread1p0em2_method_SLAE}
 shows the ratio of the numerical solution to the model Born cross section.
 Since the value of the parameter $\sigma_E$ ($20\text{ MeV}$) is comparable to
 the distance ($16.4\text{ MeV}$) between the c.m.\ energy points, there is a
 larger scatter of the points of the numerical solution than in
 figure~\ref{fig:bcs-ratio-etapipi-linterp-no-energy-spread-errkoeff5em2} due to
 the ill-posedness of the problem.
 Figure~\ref{fig:bcs_ratio_etapipi_errkoeff5em2_enspread1p0em2_method_SLAE} also
 shows that error bars of the numerical solution (square roots of the diagonal
 elements of the corresponding covariance matrix) are also significantly larger
 than in
 figure~\ref{fig:bcs-ratio-etapipi-linterp-no-energy-spread-errkoeff5em2}. Since
 the matrix $\hat{\mathcal{G}}\hat{\mathcal{F}}$ is a full rank matrix, the
 covariance matrix of the numerical solution obtained using the naive method is
 correct~\cite{Kuusela2012} in the sense that the square roots of its diagonal
 elements are correct $68.27\%$ confidence intervals for the Born cross section
 in the absence of any model knowledge about its behavior. It should also be
 noted that with an increase in the value of the parameter $\sigma_E$, the
 scatter of the points of the numerical solution increases. As shown in
 section~\ref{sec:discretization-eq-kf-blured} (see
 figure~\ref{fig:cond_sigma2}), for sufficiently large values of the
 c.m.\ energy spread, the condition number of the matrix
 $\hat{\mathcal{G}}\hat{\mathcal{F}}$ becomes too large
 ($cond(\hat{\mathcal{G}}\hat{\mathcal{F}})\sim10^{18}$) and the calculation of
 the covariance matrix of the numerical solution using
 eq.~\eqref{eq:naive-method-covmat} leads to incorrect results due to
 insufficient machine accuracy.

 We can reduce the scatter of the points of the numerical solution, assuming
 that it should be more or less smooth. This idea is a core of various
 regularization methods such as Tikhonov regularization method. However, the
 assumption of smoothness leads to an additional systematic uncertainty
 associated with the fact that the numerical solution in this case is biased
 (regularization error).
 Figure~\ref{fig:bcs_ratio_etapipi_errkoeff5em2_enspread1p0em2_method_tikhonov_optalpha0p721}
 shows a comparison of the model Born cross section and the regularized
 numerical solution of
 eq.~\eqref{eq:visible-Born-relationship-kuraev-fadin-2-blured} obtained using
 the Tikhonov regularization method under the conditions described above. The
 numerical solution is presented in this figure as points with error bars. The
 model Born cross section is shown in
 figure~\ref{fig:bcs_ratio_etapipi_errkoeff5em2_enspread1p0em2_method_tikhonov_optalpha0p721}
 as a solid curve.
 Figure~\ref{fig:bcs_ratio_etapipi_errkoeff5em2_enspread1p0em2_method_tikhonov_optalpha0p721}
 shows that the points of the numerical solution are close to the curve of the
 model Born cross section, and their scatter is significantly less than in
 figure~\ref{fig:bcs_ratio_etapipi_errkoeff5em2_enspread1p0em2_method_SLAE}.
 However, a comparison of
 figures~\ref{fig:bcs_ratio_etapipi_errkoeff5em2_enspread1p0em2_method_tikhonov_optalpha0p721}
 and \ref{fig:bcs-ratio-etapipi-linterp-no-energy-spread-errkoeff5em2} shows
 that the error bars shown in
 figure~\ref{fig:bcs_ratio_etapipi_errkoeff5em2_enspread1p0em2_method_tikhonov_optalpha0p721}
 are significantly less than the error bars shown in
 figure~\ref{fig:bcs-ratio-etapipi-linterp-no-energy-spread-errkoeff5em2}. The
 reason is that since the regularized numerical solution is biased, the square
 roots of the diagonal elements of its covariance matrix do not represent the
 correct $68.27\%$ confidence intervals for the Born cross section points. As
 already noted in section~\ref{sec:problem-formulation}, this fact significantly
 limits the applicability of regularization methods for precise obtaining a Born
 cross section.

\begin{figure}[tbp]
   \centering
   \begin{subfigure}[t]{0.47\textwidth}
     \centering
     \includegraphics[width=\textwidth]{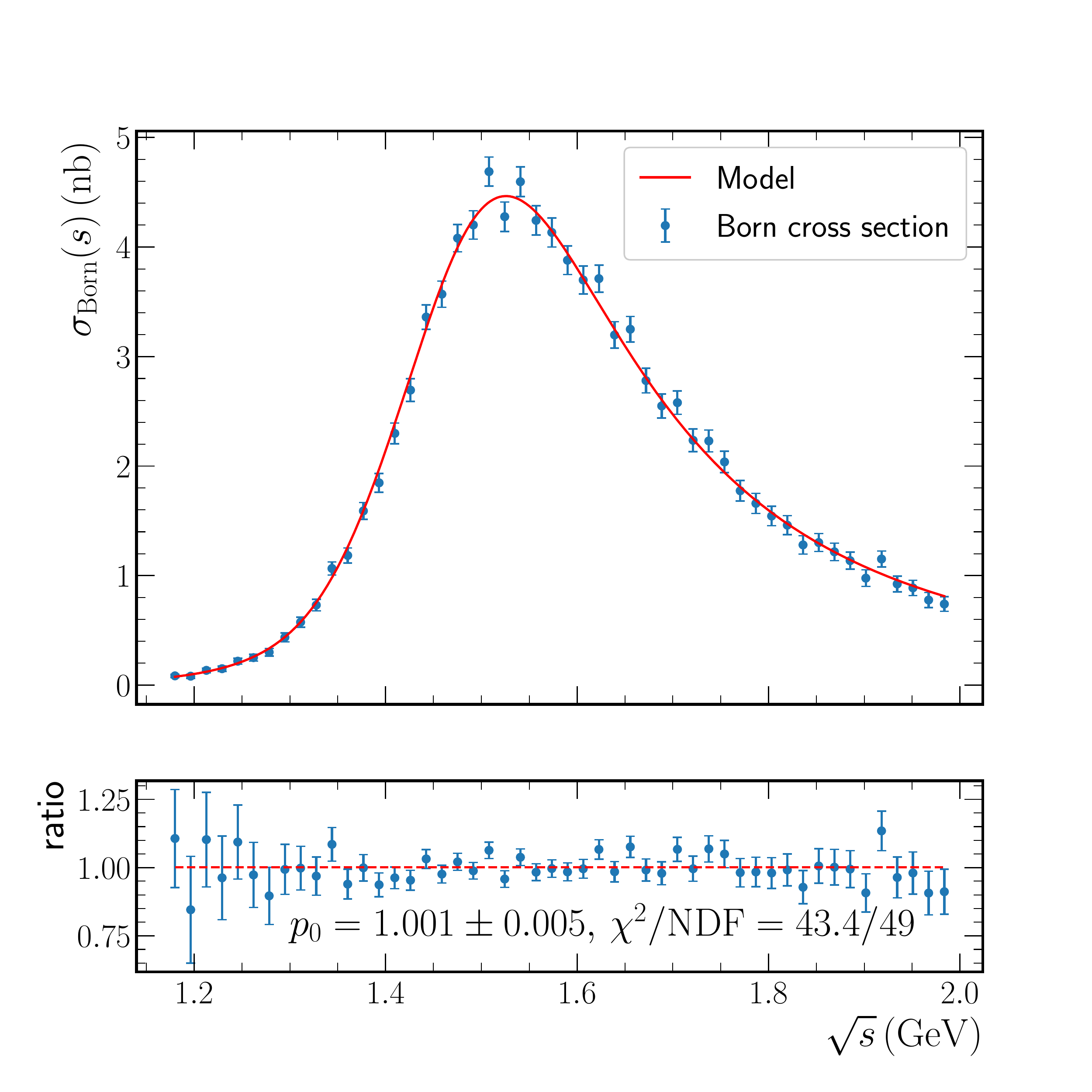}
     \caption{\label{fig:bcs_ratio_etapipi_errkoeff5em2_enspread2p0em3_method_SLAE}
       Numerical solution of
       eq.~\eqref{eq:visible-Born-relationship-kuraev-fadin-2-blured} obtained
       using the naive method.}
   \end{subfigure}
   \hfill
   \begin{subfigure}[t]{0.47\textwidth}
     \centering
     \includegraphics[width=\textwidth]{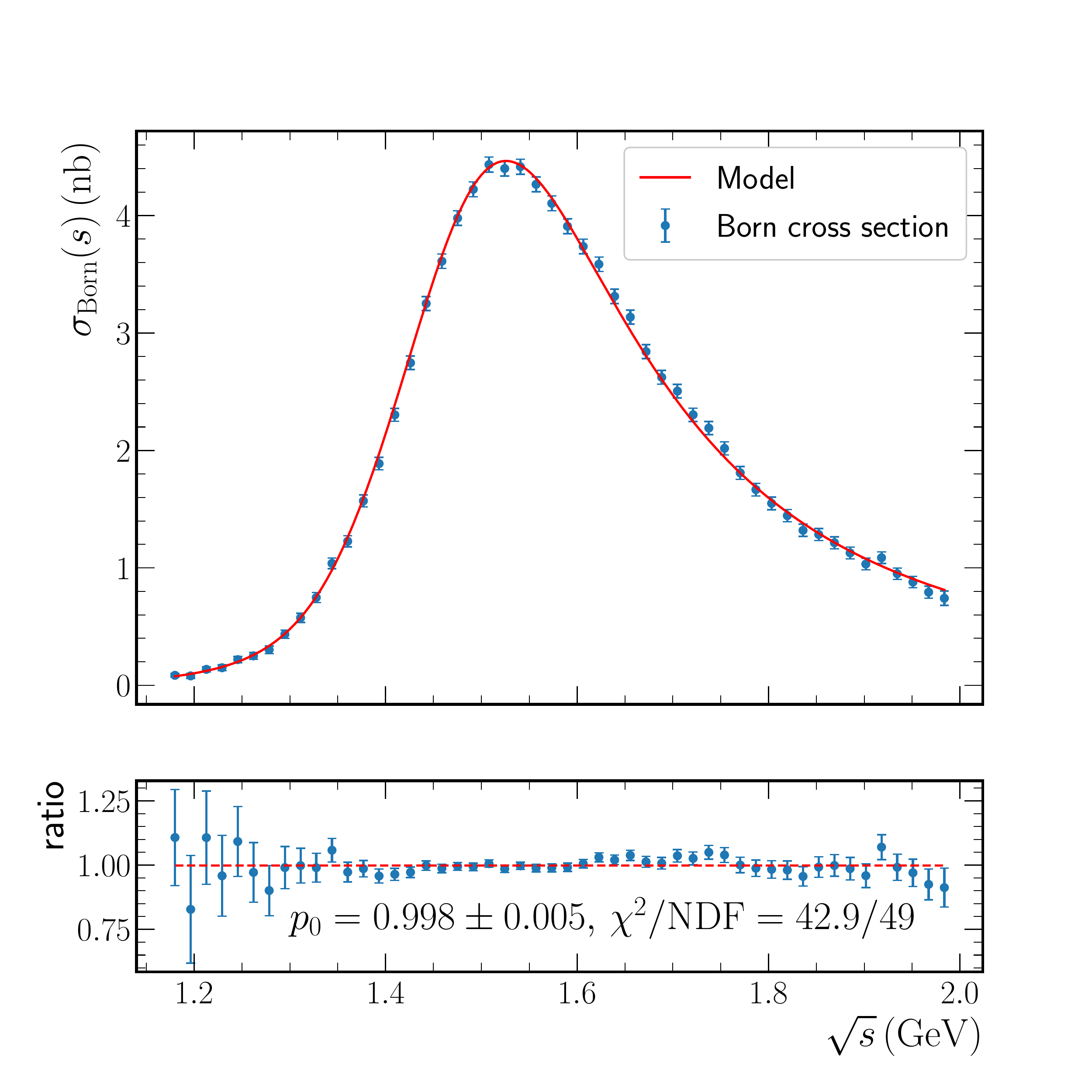}
     \caption{\label{fig:bcs_ratio_etapipi_errkoeff5em2_enspread2p0em3_method_tikhonov_optalpha0p9016}
       Numerical solution of
       eq.~\eqref{eq:visible-Born-relationship-kuraev-fadin-2-blured} obtained
       using the Tikhonov regularization method. The value ($\lambda\approx
       0.90\text{ GeV}^2\text{nb}^{-2}$) of the regularization parameter is
       chosen using the L-curve criterion.}
   \end{subfigure}
   \caption{\label{fig:bcs_ratio_etapipi_enspread2p0em3} Comparison of the
     $e^+e^-\rightarrow\eta\pi^+\pi^-$ model Born cross and the numerical
     solution of eq.~\eqref{eq:visible-Born-relationship-kuraev-fadin-2-blured}.
     Values of the parameter $\sigma_E$ at each c.m.\ energy
     point are same and equal to $2\text{ MeV}$.}
 \end{figure}

 In section~\ref{sec:discretization-eq-kf-blured}, it is shown that if the
 c.m.\ energy spread is much less than the distance between c.m. energy points,
 then the condition number of the matrix $\hat{\mathcal{G}}\hat{\mathcal{F}}$ is
 of the order of unity. This fact allows us to hope that in the case of small
 values of the c.m. energy spread the effects associated with the
 ill-posedness of the problem given by
 eq.~\eqref{eq:visible-Born-relationship-kuraev-fadin-2-blured} will be small,
 and the considered problem can be solved using the naive method. Let us
 consider a numerical experiment similar to the previous one, but with the
 parameter $\sigma_E$ equal to $2\text{ MeV}$. The condition number of the
 matrix $\hat{\mathcal{G}}\hat{\mathcal{F}}$ in this case is approximately equal
 to $2.0$. A comparison of the model Born cross section and the numerical
 solution of eq.~\eqref{eq:visible-Born-relationship-kuraev-fadin-2-blured},
 obtained using the naive method, is shown in
 figure~\ref{fig:bcs_ratio_etapipi_errkoeff5em2_enspread2p0em3_method_SLAE}. It
 can be seen from this figure that the numerical solution and the model Born
 cross section are in good agreement, and the error bars of the numerical
 solution are of the same order of magnitude as in
 figure~\ref{fig:bcs-ratio-etapipi-linterp-no-energy-spread-errkoeff5em2}.

 Figure~\ref{fig:bcs_ratio_etapipi_errkoeff5em2_enspread2p0em3_method_tikhonov_optalpha0p9016}
 shows a comparison of the model Born cross section and the numerical solution
 obtained using the Tikhonov regularization method. This numerical solution,
 like the numerical solution discussed in the previous paragraph, corresponds to
 the parameter $\sigma_E$ equal to $2\text{ MeV}$. As in the example shown in
 figure~\ref{fig:bcs_ratio_etapipi_errkoeff5em2_enspread1p0em2_method_tikhonov_optalpha0p721},
 the error bars of the numerical solution are small compared to the similar
 error bars shown in
 figure~\ref{fig:bcs-ratio-etapipi-linterp-no-energy-spread-errkoeff5em2}. It
 should be noted that the value of the elements of the covariance matrix depends
 on the choice of the regularization parameter. In the case of the
 figures~\ref{fig:bcs_ratio_etapipi_errkoeff5em2_enspread1p0em2_method_tikhonov_optalpha0p721}
 and
 \ref{fig:bcs_ratio_etapipi_errkoeff5em2_enspread2p0em3_method_tikhonov_optalpha0p9016},
 the value of the regularization parameter is chosen using the L-curve
 criterion. The L-curve criterion, as well as the Tikhonov regularization
 method, is considered in detail in section~\ref{sec:regularization}.

\begin{figure}[tbp]
   \centering
   \begin{minipage}[t]{0.47\textwidth}
     \centering
     \includegraphics[width=\textwidth]{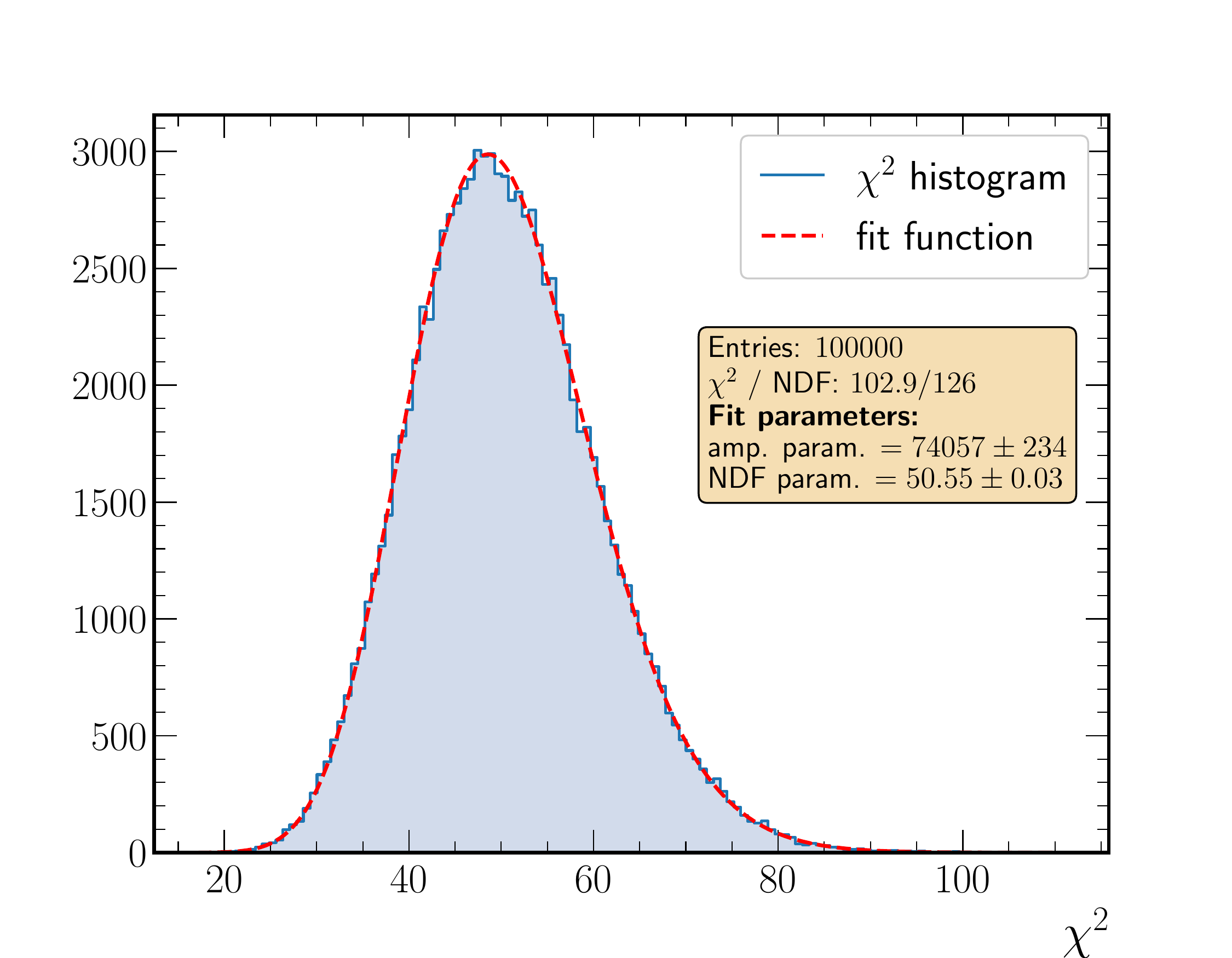}
     \caption{\label{fig:chi2_etapipi_enspread2em3} An example of the chi-square
       distribution for the numerical solution of
       eq.~\eqref{eq:visible-Born-relationship-kuraev-fadin-2-blured} with
       respect to the model Born cross section. The parameter $\sigma_E$ is
       small compared to the distance between the c.m.\ energy points and is
       equal to $2\text{ MeV}$. The numerical solution is obtained using the
       naive method.}
   \end{minipage}
   \hfill
   \begin{minipage}[t]{0.47\textwidth}
     \centering
     \includegraphics[width=\textwidth]{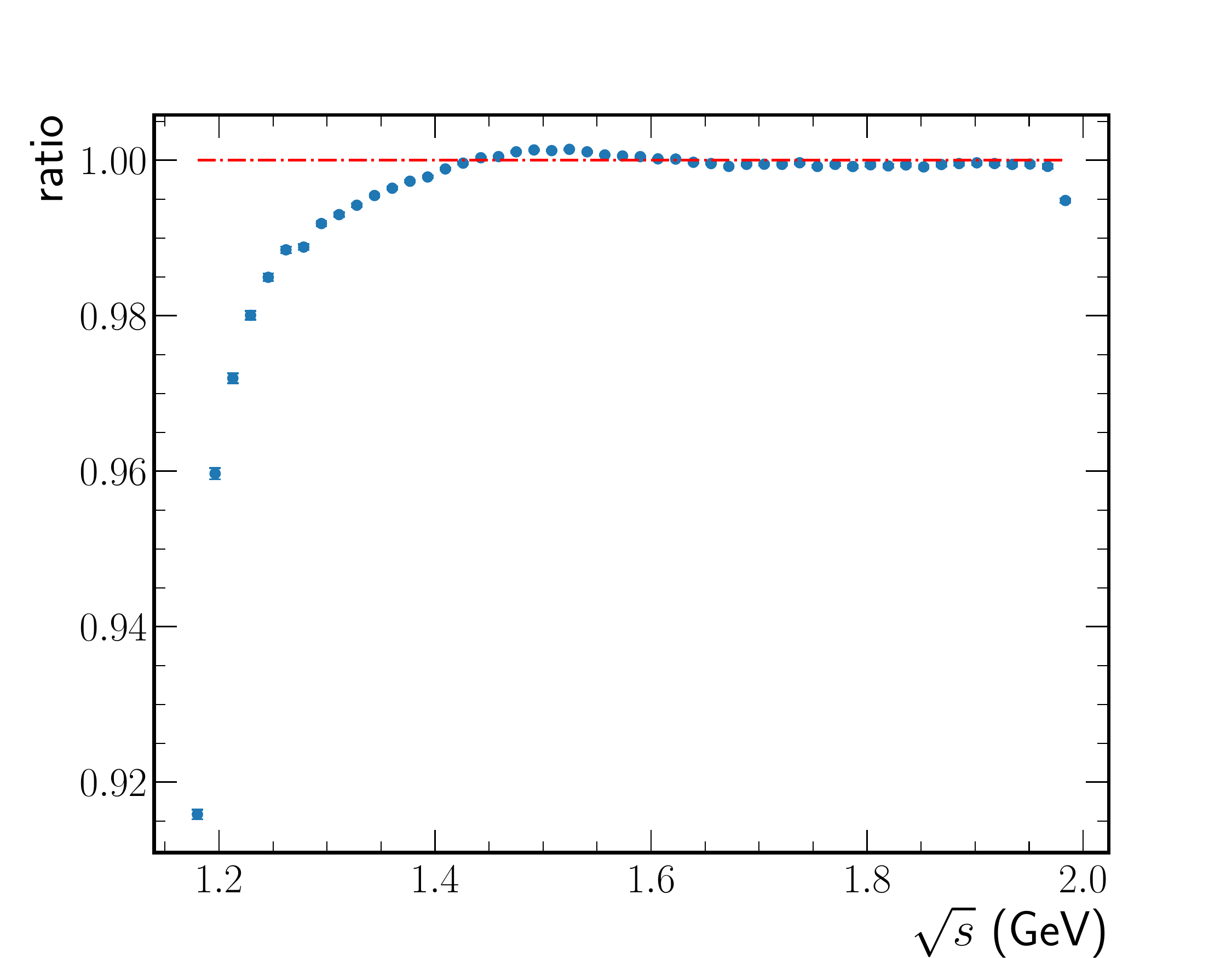}
     \caption{\label{fig:ratio_etapipi_enspread2em3} An example of the ratio of
       the numerical solution of
       eq.~\eqref{eq:visible-Born-relationship-kuraev-fadin-2-blured} to the
       model Born cross section, averaged over $10^5$ numerical experiments. The
       parameter $\sigma_E$ is small compared to the distance between the
       c.m.\ energy points and is equal to $2\text{ MeV}$. The numerical
       solution is obtained using the naive method.}
   \end{minipage}
 \end{figure}

 Let us consider again the naive method for solving
 eq.~\eqref{eq:visible-Born-relationship-kuraev-fadin-2-blured} in the case when
 the parameter $\sigma_E$ is equal to $2\text{ MeV}$. As in the examples from
 section~\ref{sec:numerical-experiments-without-enspread}, it is possible to
 plot the chi-square histogram of the numerical solution obtained using the
 naive method. This histogram is shown in
 figure~\ref{fig:chi2_etapipi_enspread2em3}. As in the previous examples, the
 histogram is fitted using the chi-square distribution function. The NDF
 parameter obtained with fit is $50.55\pm0.03$, which is consistent with the
 expected number of degrees of freedom equal to $50$. However, as in the case of
 the example of the chi-square distribution from
 figure~\ref{fig:chi2_etapipi_model_fcn}, there is a slight systematic deviation
 of this parameter from the expected number of degrees of freedom, associated
 mainly with interpolation accuracy.
 
 Figure~\ref{fig:ratio_etapipi_enspread2em3} shows the ratio
 between the numerical solution ($\sigma_E=2\text{ MeV}$) obtained using the
 naive method and the model Born cross section, averaged over $10^{5}$ numerical
 experiments. In each numerical experiment, the visible cross section is
 generated using the same covariance matrix $\hat{\Lambda}$. It can be seen from
 the figure that at points with the c.m.\ energies close to the minimum c.m.\
 energy, there is a significant relative deviation of the numerical solution
 from the model Born cross section. As in the examples described in
 section~\ref{sec:numerical-experiments-without-enspread}, this deviation is due
 to the fact that interpolation poorly describes the threshold behavior of the
 Born cross section. Since the cross section at these c.m.\ energy points is
 small, the absolute values of the deviation of the numerical solution from the
 model Born cross section are also small.
 
 Figure~\ref{fig:ratio_etapipi_enspread2em3} also shows that at the point with
 the maximum c.m.\ energy, a relative deviation of the numerical solution with
 respect to the model cross section is observed. This deviation is approximately
 equal to $0.5\%$ and is associated with the lack of experimental information on
 the visible cross section at energies above the maximum. The cross section at
 the c.m.\ energies above the maximum must be extrapolated to perform its
 convolution with the normal distribution when finding a numerical solution to
 eq.~\eqref{eq:visible-Born-relationship-kuraev-fadin-2-blured}. For instance, a
 constant equal to the value of the cross section at the point with the maximum
 c.m.\ energy can be used as an extrapolation function. Thus, this relative
 difference of the numerical solution in comparison with the model Born cross
 section is caused by the extrapolation accuracy.

 It should be noted that the naive method works better with smooth cross
 sections. In the case of sharp cross sections, it is necessary to have a
 sufficient density of c.m.\ energy points in order to provide the desired
 interpolation accuracy. On the other hand, as the density of c.m.\ energy
 points increases, the distance between them decreases. At a certain density of
 c.m.\ energy points, the distance between them can become of the order of the
 c.m.\ energy spread. As discussed above, this leads to a significant
 scatter in a numerical solution, which is due to the
 ill-posedness of the problem.

 In the example with the process $e^+e^-\rightarrow\pi^+\pi^-\pi^0$ from
 section~\ref{sec:numerical-experiments-without-enspread}, the density of
 c.m.\ energy points is such that the distance between some of these points is
 comparable to $1\text{ MeV}$. On the other hand, the authors were unable to
 provide good interpolation accuracy when using a lower density of c.m.\ energy
 points. Thus, the naive method is not applicable for obtaining the cross
 section of the process $e^+e^-\rightarrow\pi^+\pi^-\pi^0$ with the c.m.\ energy
 spread of the order of $1\text{ MeV}$ or more.

 \subsubsection{Numerical experiments with $\varepsilon(x,s)\neq{1}$}
\begin{figure}[tbp]
   \centering
   \begin{subfigure}[t]{0.47\textwidth}
     \centering
     \includegraphics[width=\textwidth]{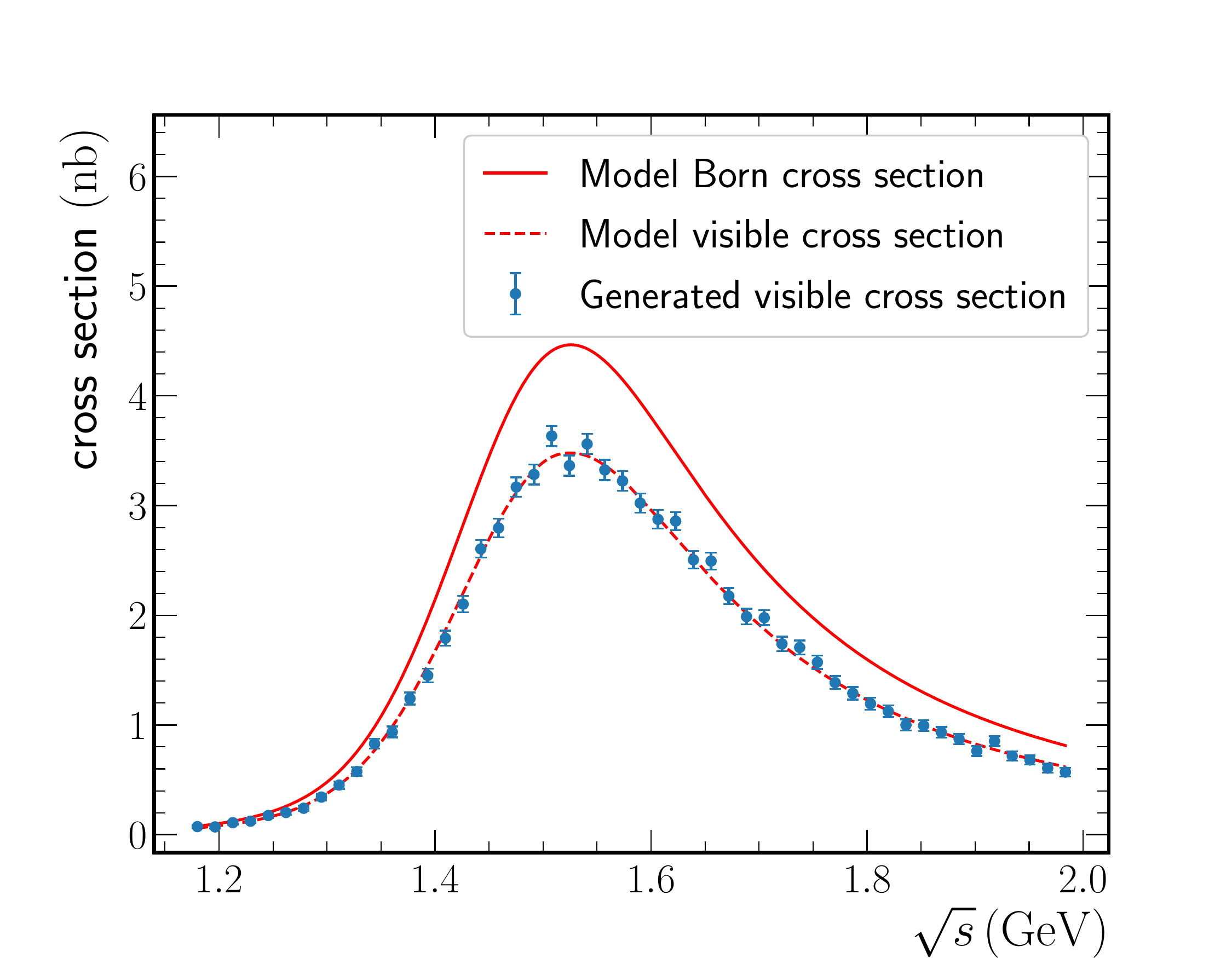}
     \caption{\label{fig:model_bcs_vcs_etapipi_efficiency} Model Born, model
       visible and generated visible cross sections.}
   \end{subfigure}
   \hfill
   \begin{subfigure}[t]{0.47\textwidth}
     \centering
     \includegraphics[width=\textwidth]{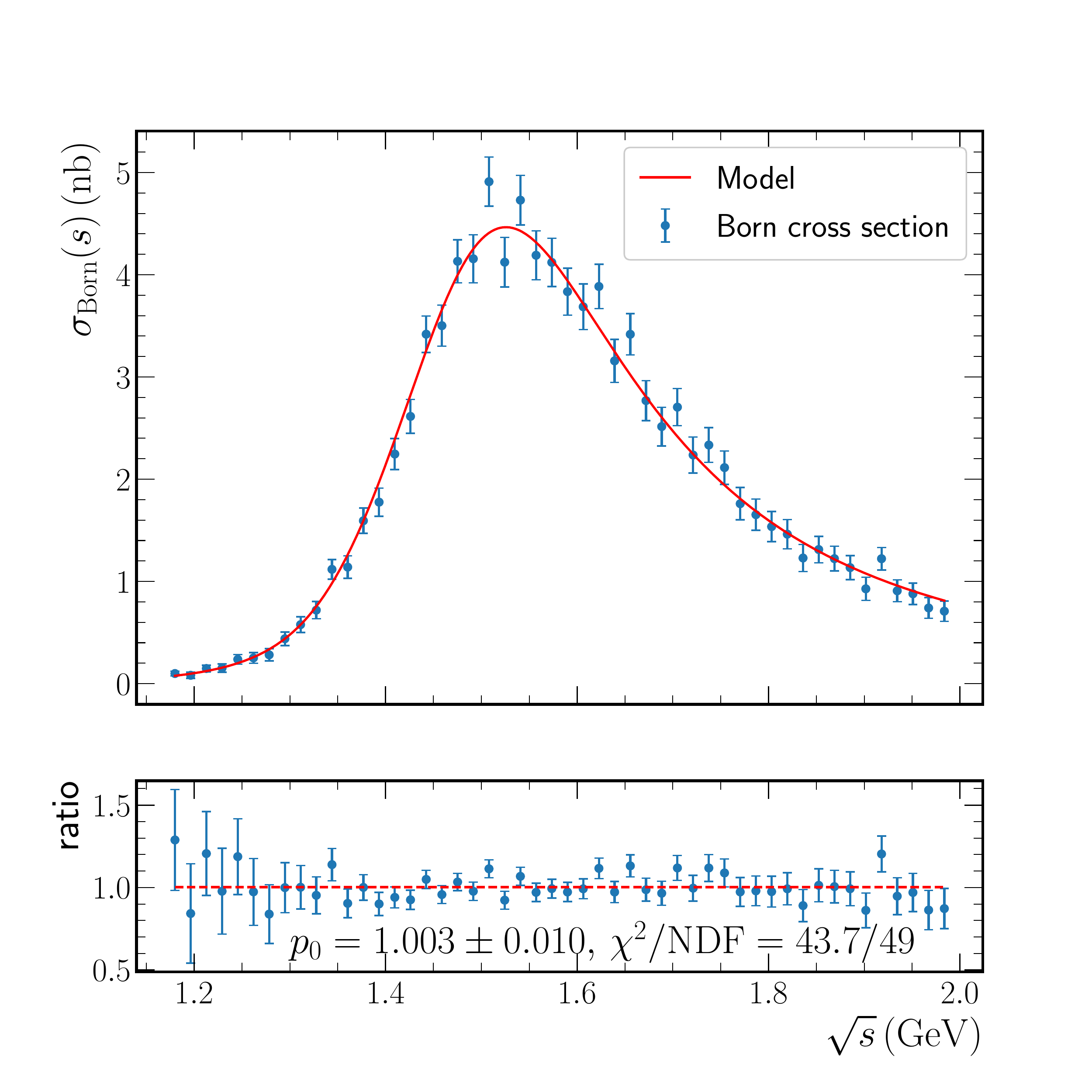}
     \caption{\label{fig:bcs_etapipi_ratio_efficiency}Comparison of the
       $e^+e^-\rightarrow\eta\pi^+\pi^-$ model Born cross section and the
       numerical solution obtained using the naive method. The generated visible
       cross section is used as the right-hand side of
       eq.~\eqref{eq:visible-Born-relationship-kuraev-fadin-2}.}
   \end{subfigure}
   \vfill
   \begin{subfigure}[t]{0.47\textwidth}
     \centering
     \includegraphics[width=\textwidth]{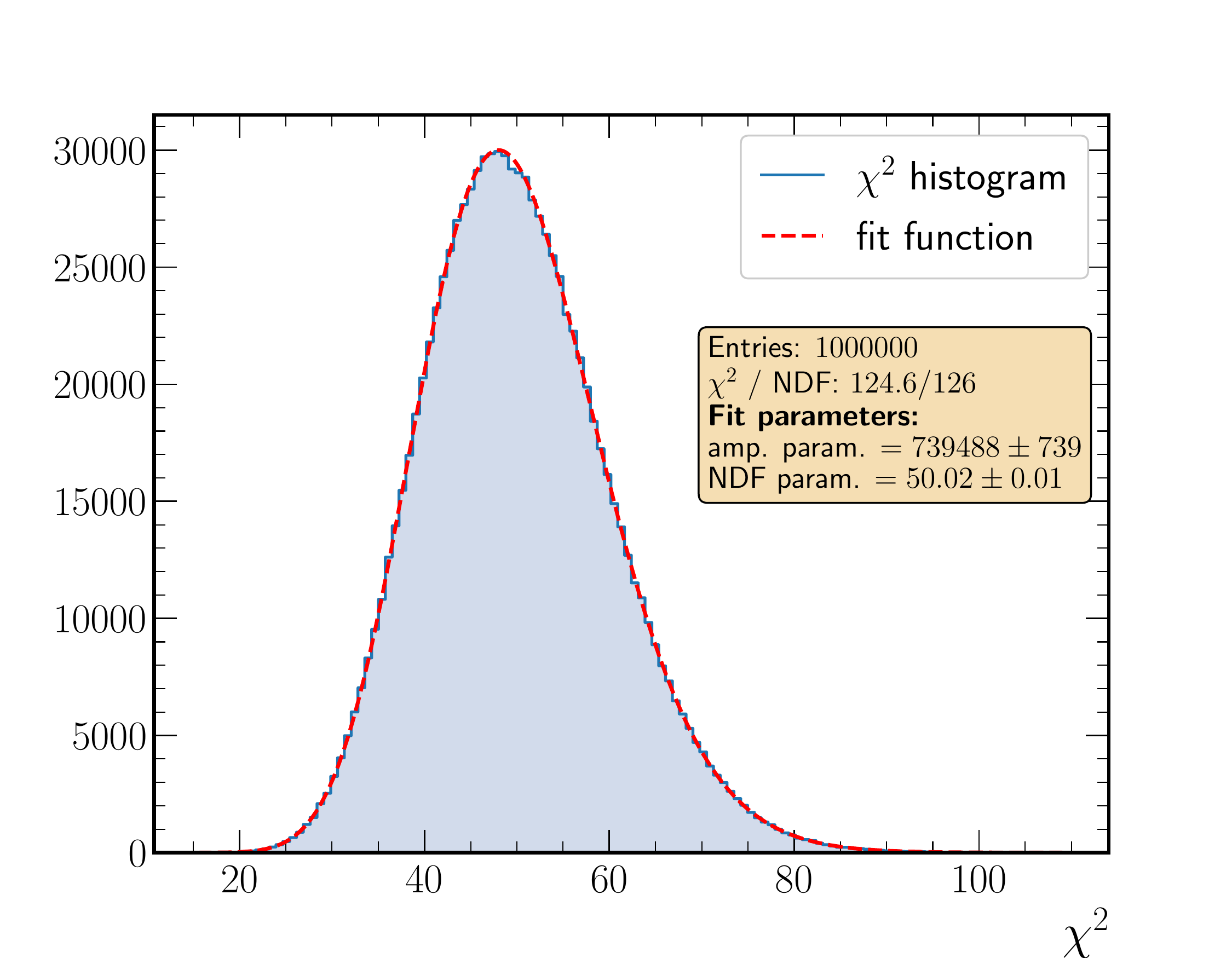}
     \caption{\label{eq:chi2_etapipi_efficiency}Chi-square distribution with
       respect to the model Born cross section.}
   \end{subfigure}
   \hfill
   \begin{subfigure}[t]{0.47\textwidth}
     \centering
     \includegraphics[width=\textwidth]{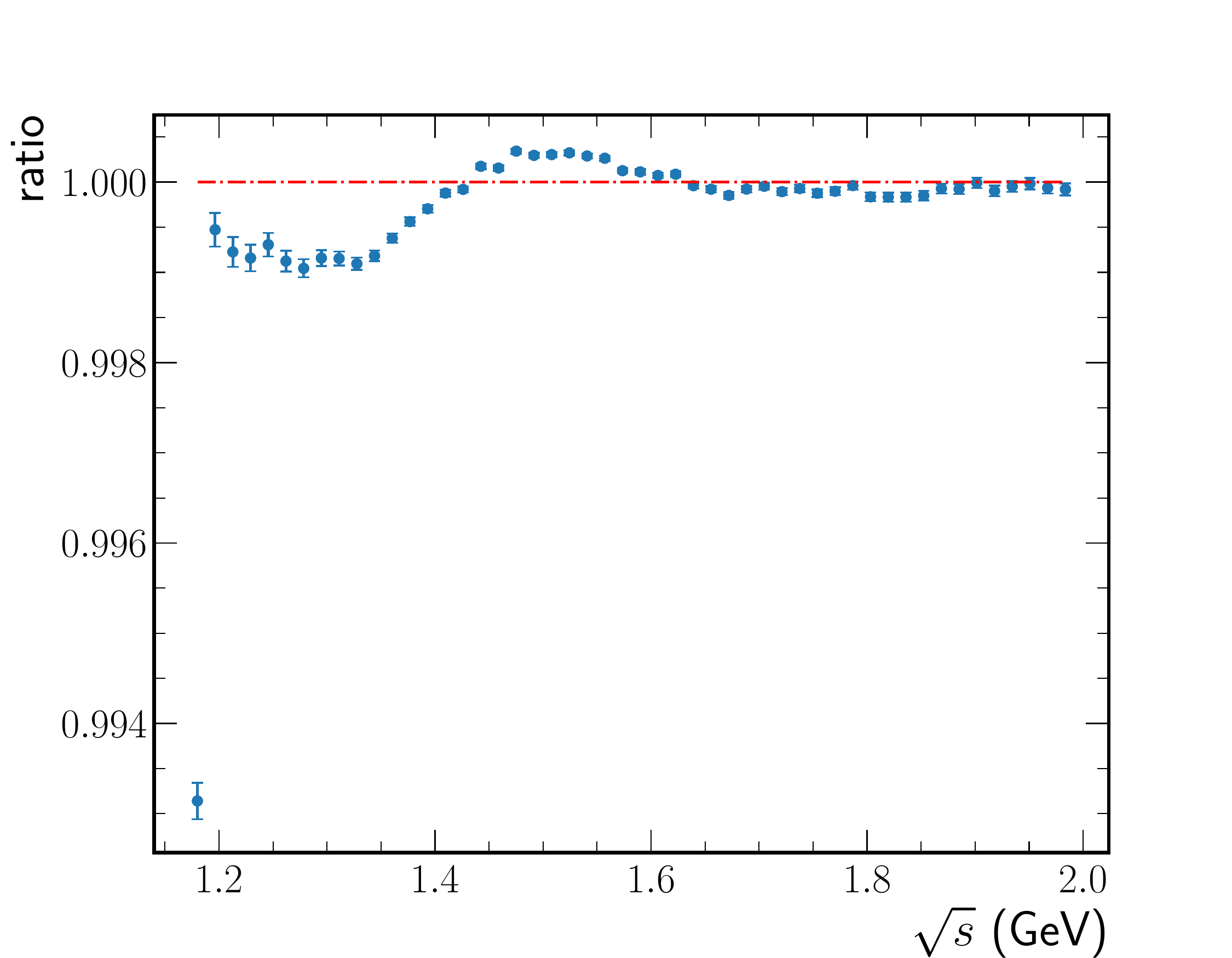}
     \caption{\label{eq:ratio_etapipi_efficiency} Ratio of the numerical
       solution of eq.~\eqref{eq:visible-Born-relationship-kuraev-fadin-2} to
       the model Born cross section, averaged over $10^6$ numerical
       experiments.}
   \end{subfigure}
   \caption{Results of the numerical experiment with the
     $e^+e^-\rightarrow\eta\pi^+\pi^-$ cross section obtained using the naive
     method with a detection efficiency equal to unity.}
 \end{figure}

 Up to this point, we have considered finding a numerical solution to
 equations~\eqref{eq:visible-Born-relationship-kuraev-fadin-2} and
 \eqref{eq:visible-Born-relationship-kuraev-fadin-2-blured} with a detection
 efficiency equal to unity. Let us consider an example of finding a numerical
 solution to eq.~\eqref{eq:visible-Born-relationship-kuraev-fadin-2} with the
 non-trivial detection efficiency:
 \begin{equation}
   \label{eq:test-efficiency}
   \varepsilon(x, s) =
   \begin{cases}
     1, & \sqrt{s} - \sqrt{s(1-x)} < 10\text{ MeV},\\
     0, & \text{otherwise.}
   \end{cases}
 \end{equation}
 In this example, we consider a numerical experiment in which the Born cross
 sections of the $e^+e^-\rightarrow\eta\pi^+\pi^-$ process are found as a
 numerical solution to eq.~\eqref{eq:visible-Born-relationship-kuraev-fadin-2}
 using the naive method at $50$ equally spaced points in the c.m.\ energy range
 from $1.18$ to $2.00$ GeV. Figure~\ref{fig:model_bcs_vcs_etapipi_efficiency}
 shows the corresponding model Born and model visible cross sections. The
 generated visible cross section is also shown in this figure. The model Born
 cross section is shown as a solid curve, and the model visible cross section is
 shown as a dashed curve. The generated visible cross section is drawn as points
 with error bars. It can be seen from the figure that the visible cross-section
 is smaller than the visible cross-section obtained in the similar example with
 a detection efficiency equal to one
 (figure~\ref{fig:model-bcs-vcs-etapipi-linterp-no-energy-spread-errkoeff5em2}).
 Figure~\ref{fig:bcs_etapipi_ratio_efficiency} shows a comparison of the
 numerical solution of eq.~\eqref{eq:visible-Born-relationship-kuraev-fadin-2}
 with the model Born cross section. It can be seen from this figure that the
 numerical solution and the model Born cross section are in agreement.
 Figure~\ref{eq:chi2_etapipi_efficiency} shows the chi-square histogram of the
 numerical solution with respect to the model Born cross section. As in the
 previous examples, the histogram is fitted with a chi-square distribution. Fit
 parameters are shown in the figure. The number of degrees of freedom of the
 chi-square distribution is in good agreement with the expected value of this
 parameter, equal to $50$. Figure~\ref{eq:ratio_etapipi_efficiency} shows the
 ratio of the numerical solution to the model Born cross section, averaged over
 $10^6$ numerical experiments. It can be seen from the figure that the relative
 deviation at low energies became less than in the similar example with the
 detection efficiency equal to unity
 (figure~\ref{fig:bcs-ratio-etapipi-linterp-no-energy-spread-errkoeff1em6}). A
 significant relative deviation is observed only at the point with the minimum
 c.m.\ energy and is approximately equal to $0.7\%$. At the rest of the energy
 points, the relative deviation is small. In the case of using the detection
 efficiency equal to one, the relative deviation in the points with the lowest
 energy is approximately equal to $7\%$. The reason for the decrease in the
 relative difference between the numerical solution and the model Born cross
 section is that when the function $F(x, s)$ is multiplied by the detection
 efficiency, which decreases with increasing $x$, the kernel of the integral
 operator decreases faster than in the case with the
 detection efficiency equal to unity.

 \section{\label{sec:regularization} Regularization}
 \subsection{Tikhonov regularization}
 In this section, we will briefly consider the application of the Tikhonov
 regularization method to
 equations~\eqref{eq:visible-Born-relationship-kuraev-fadin-2} and
 \eqref{eq:visible-Born-relationship-kuraev-fadin-2-blured}. The Tikhonov
 regularization method, unlike the naive method, can be used to find a solution
 to eq.~\eqref{eq:visible-Born-relationship-kuraev-fadin-2-blured} in the case
 when the parameter $\sigma_E$ is greater or comparable to the distances between
 c.m.\ energy points. However, the application of the Tikhonov regularization
 method is essentially limited by the fact that the diagonal elements of the
 covariance matrix for the numerical solution of
 eq.~\eqref{eq:visible-Born-relationship-kuraev-fadin-2-blured} (or
 eq.~\eqref{eq:visible-Born-relationship-kuraev-fadin-2}) do not represent the
 correct $68.27\%$ confidence intervals for the Born cross section points. More
 details about this fact can be found in work~\cite{Kuusela2012}.

 In this work, the Tikhonov regularization method is applied to
 eq.~\eqref{eq:visible-Born-relationship-kuraev-fadin-2-blured} by means of
 minimizing the functional of the following form:
 \begin{equation}
   \label{eq:chi2-tikhonov-functional}
   \mathcal{L}_G[\bm{\sigma}_{\rm Born}]=\left(\hat{\mathcal{A}}\bm{\sigma}_{\rm Born} - \bm{\sigma}_{\rm vis}\right)^{\intercal}\hat{\Lambda}^{-1}\left(\hat{\mathcal{A}}\bm{\sigma}_{\rm Born} - \bm{\sigma}_{\rm vis}\right) + \lambda\|\hat{\mathcal{D}}\bm{\sigma}_{\rm Born}\|^2.
 \end{equation}
 Comparing equations~\eqref{eq:chi2-tikhonov-functional} and
 \eqref{eq:tikhonov-functional-general}, we see that in the second term of the
 functional $\mathcal{L}_G[\bm{\sigma}_{\rm Born}]$, the matrix
 $\hat{\mathcal{B}}$ has been replaced by the matrix $\hat{\mathcal{D}}$, which
 in this work denotes the matrix of the c.m.\ energy derivative operator. The
 regularization functional $\mathcal{L}_G[\bm{\sigma}_{\rm Born}]$ given by
 eq.~\eqref{eq:chi2-tikhonov-functional} also differs from the standard
 functional of the Tikhonov regularization by the presence of the
 chi-square term instead of the term $\|\hat{\mathcal{A}}\bm{\sigma}_{\rm Born}
 - \bm{\sigma}_{\rm vis}\|^2$. The replacement of the term
 $\|\hat{\mathcal{A}}\bm{\sigma}_{\rm Born} - \bm{\sigma}_{\rm vis}\|^2$ by the
 chi-square term is performed to identify the minimization of the functional
 $\mathcal{L}_G$ with the minimization of the chi-square under the condition
 $\lambda\|\hat{\mathcal{D}}\bm{\sigma}_{\rm Born}\|^2$ on the smoothness of the
 numerical solution. In fact, the results obtained using term
 $\|\hat{\mathcal{A}}\bm{\sigma}_{\rm Born} - \bm{\sigma}_{\rm vis}\|^2$ are
 very similar to those obtained using the chi-square term and can be obtained in
 a similar manner.

 Minimization of the functional $\mathcal{L}_G[\bm{\sigma}_{\rm Born}]$ with
 respect to unknown values of the numerical solution leads to the following
 system of linear equations:
 \begin{equation}
   \label{eq:tikhonov-regularized-equation}
   \left( \hat{\mathcal{A}}^{\intercal}\hat{\Lambda}^{-1}\hat{\mathcal{A}}
     + \lambda\hat{\mathcal{Q}}
   \right)\bm{\sigma}^{\rm Tikh}_{\rm Born} = \hat{\mathcal{A}}^{\intercal}\hat{\Lambda}^{-1}\bm{\sigma}_{\rm vis},
 \end{equation}
 where $\bm{\sigma}^{\rm Tikh}_{\rm Born}$ is the regularized numerical solution
 of eq.~\eqref{eq:visible-Born-relationship-kuraev-fadin-2-blured} (or
 eq.~\eqref{eq:visible-Born-relationship-kuraev-fadin-2}), and the matrix
 $\hat{\mathcal{Q}}$ is defined as follows:
 \begin{equation}
   \label{eq:Q-matrix}
   \hat{\mathcal{Q}}_{ij} = \sum\limits_kJ_k\hat{\mathcal{D}}_{ki}\hat{\mathcal{D}}_{kj}.  
 \end{equation}
 The coefficients $J_k$ in eq.~\eqref{eq:Q-matrix} are the weights of the
 c.m.\ energy points in a dot product. The values of these coefficients
 depend on the distance between the c.m.\ energy points and the type of
 interpolation. For instance, the norm $\|\hat{\mathcal{D}}\bm{\sigma}_{\rm
   B}\|^2$ can be rewritten using these coefficients as follows:
 \begin{equation}
   \label{eq:norm-and-dot-coefficients}
   \begin{split}
   \|\hat{\mathcal{D}}\bm{\sigma}_{\rm Born}\|^2 &= \int\limits^{\sqrt{s_{\rm max}}}_{\sqrt{s_{\rm min}}}\left(\hat{\mathcal{D}}_{E}\sigma_{\rm Born}(E^2)\right)^2dE\approx
   \sum\limits_{i,j,k}J_k\hat{\mathcal{D}}_{ki}\hat{\mathcal{D}}_{kj}\sigma_{\rm Born}(s_i)\sigma_{\rm Born}(s_j)\\
   &=\sum\limits_{i,j}\hat{\mathcal{Q}}_{ij}\sigma_{\rm Born}(s_i)\sigma_{\rm Born}(s_j) = \bm{\sigma}^{\intercal}_{\rm Born}\hat{\mathcal{Q}}\bm{\sigma}_{\rm Born},
   \end{split}
 \end{equation}
 where $\hat{\mathcal{D}}_{E}$ is a continuously defined derivative operator
 ($\hat{\mathcal{D}}_{E}= \frac{d}{dE}$). The matrix $\hat{\mathcal{D}}$
 corresponds to this operator. The summation in
 eq.~\eqref{eq:norm-and-dot-coefficients} is carried out over all c.m.\ energy
 indices, $\sqrt{s_{\rm min}}$ and $\sqrt{s_{\rm max}}$ represent the maximum
 and minimum c.m.\ energies, respectively.

 The solution to eq.~\eqref{eq:tikhonov-regularized-equation} can be written in the following form:
 \begin{equation}
   \label{eq:tikhonov-regularized-solution}
   \begin{split}
     \bm{\sigma}^{\rm Tikh}_{\rm Born}&=\hat{\mathcal{A}}^{+}_{\lambda}\bm{\sigma}_{\rm vis},\\
     \hat{\mathcal{A}}^{+}_{\lambda} &= \left( \hat{\mathcal{A}}^{\intercal}\hat{\Lambda}^{-1}\hat{\mathcal{A}}
       + \lambda\hat{\mathcal{Q}}\right)^{-1}\hat{\mathcal{A}}^{\intercal}\hat{\Lambda}^{-1}.
   \end{split}
 \end{equation}
 Note that equations~\eqref{eq:tikhonov-regularized-equation} and
 \eqref{eq:tikhonov-regularized-solution} can be formally rewritten as follows:
 \begin{equation}
   \begin{split}
     \hat{\mathcal{R}}_{\lambda}\bm{\sigma}^{\rm Tikh}_{\rm Born}&=\bm{\sigma}_{\rm vis},\\
     \hat{\mathcal{R}}_{\lambda}&= \hat{\mathcal{A}} +
     \lambda\hat{\Lambda}\left(\hat{\mathcal{A}}^{-1}\right)^{\intercal}\hat{\mathcal{Q}},\\
     \bm{\sigma}^{\rm Tikh}_{\rm Born}&=\hat{\mathcal{A}}^{+}_{\lambda}\bm{\sigma}_{\rm vis}=\hat{\mathcal{R}}^{-1}_{\lambda}\bm{\sigma}_{\rm vis}.
 \end{split}
\end{equation}
However, the matrix $\hat{\mathcal{R}}_{\lambda}$ contains the inverse of the
matrix $\hat{\mathcal{A}}$. Since the matrix $\hat{\mathcal{A}}$ can be
ill-conditioned, the numerical inversion of this matrix can be inaccurate. On
the contrary, it can be shown that the matrix
$\hat{\mathcal{A}}^{\intercal}\hat{\Lambda}^{-1}\hat{\mathcal{A}} +
\lambda\hat{\mathcal{Q}}$ is well-conditioned, if the value of the
regularization parameter $\lambda$ is large enough. Thus, the inverse of this
matrix can be found numerically with high accuracy. Taking into account the
above properties of matrices $\hat{\mathcal{A}}$ and
$\hat{\mathcal{A}}^{\intercal}\hat{\Lambda}^{-1}\hat{\mathcal{A}} +
\lambda\hat{\mathcal{Q}}$, eq.~\eqref{eq:tikhonov-regularized-solution} should
be used to obtain the regularized numerical solution $\bm{\sigma}^{\rm Tikh}_{\rm
  Born}$.

The covariance matrix $\hat{\mathcal{M}}^{\rm Tikh}_{\lambda}$ of the
regularized numerical solution can be obtained in the similar way as the
covariance matrix~\eqref{eq:naive-method-covmat} for the numerical solution
obtained using the naive method. The covariance matrix $\hat{\mathcal{M}}^{\rm
  Tikh}_{\lambda}$ is as follows:
\begin{equation}
  \hat{\mathcal{M}}^{\rm Tikh}_{\lambda} = \hat{\mathcal{A}}^{+}_{\lambda}\hat{\Lambda}\left(\hat{\mathcal{A}}^{+}_{\lambda}\right)^{\intercal},
\end{equation}
but, as noted above, its diagonal elements do not represent $68.27\%$ confidence
intervals for the Born cross section at different c.m.\ energies since the
regularized numerical solution $\bm{\sigma}^{\rm Tikh}_{\rm Born}$ is biased.
Indeed, the expectation of the regularized solution $\bm{\sigma}^{\rm Tikh}_{\rm
  Born}$ can be written in the following form~\cite{Kuusela2012}:
 \begin{equation}
   \label{eq:tikhonov-bias}
   \begin{split}
     E\left[\bm{\sigma}^{\rm Tikh}_{\rm Born}\right] &= \hat{\mathcal{A}}^{+}_{\lambda}E\left[\bm{\sigma}_{\rm vis}\right]
     = \hat{\mathcal{A}}^{+}_{\lambda}\hat{\mathcal{A}}E\left[\bm{\sigma}_{\rm Born}\right]
     = \left(\hat{\mathcal{A}}^{-1}\hat{\mathcal{R}}\right)^{-1}E\left[\bm{\sigma}_{\rm Born}\right]\\
     &= E\left[\bm{\sigma}_{\rm Born}\right] - \left( \hat{\mathcal{I}}  - \left( \hat{\mathcal{I}} + \lambda\hat{\mathcal{A}}^{-1}\hat{\Lambda}\hat{\mathcal{A}}^{-1}\hat{\mathcal{Q}}\right)^{-1}\right)E\left[\bm{\sigma}_{\rm Born}\right],
   \end{split}
 \end{equation}
 where the last term is the bias of the regularized solution and the matrix
 $\hat{\mathcal{I}}$ is the identity matrix. It is seen from the last equation
 that at $\lambda=0$, the bias term disappears.

 \subsection{Numerical experiments and L-curve criterion}
\begin{figure}[tbp]
   \centering
   \begin{subfigure}[t]{0.47\textwidth}
     \centering
     \includegraphics[width=\textwidth]{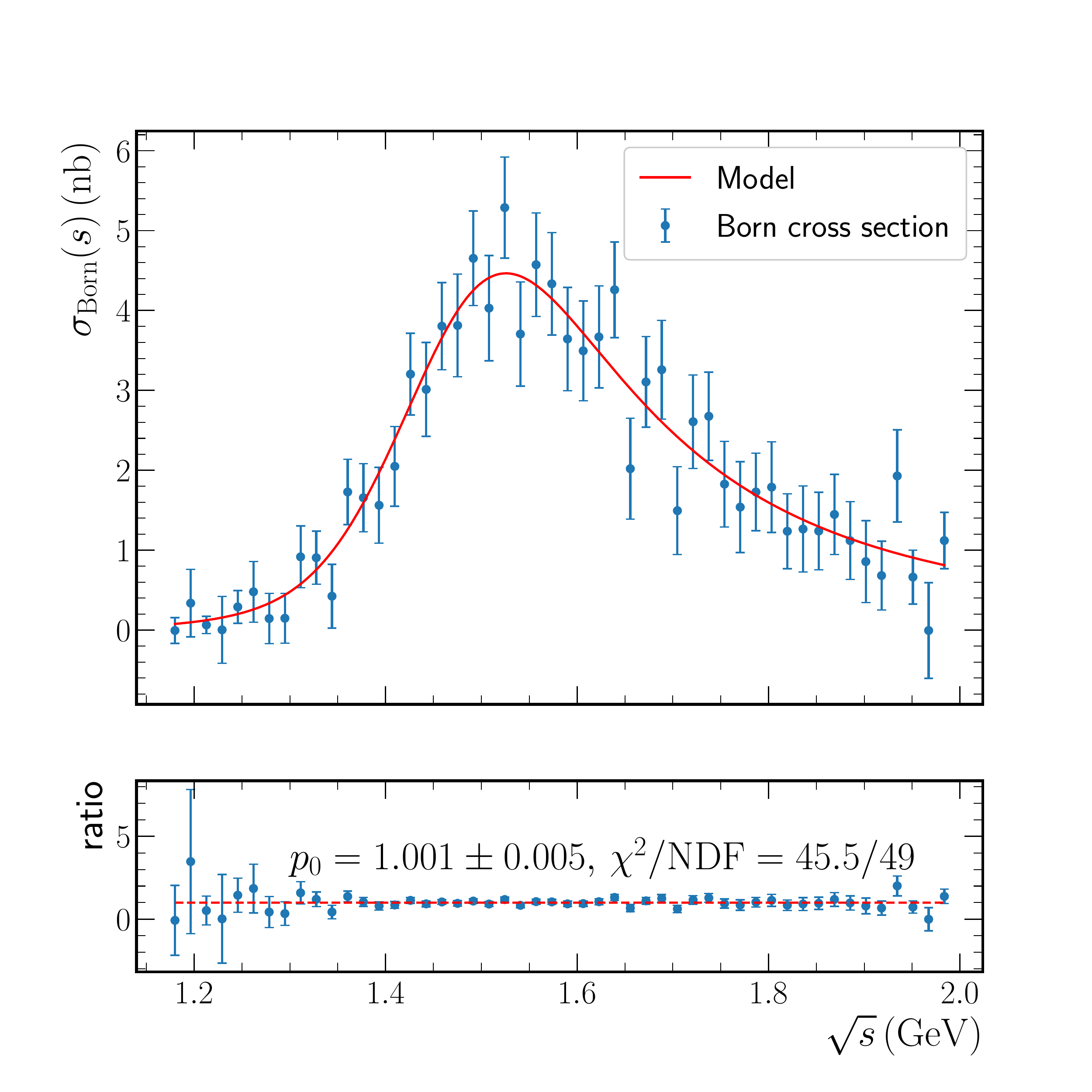}
     \caption{\label{fig:bcs_etapipi_tikh_enspread1p0em2_alpha1p0em3}
       $\lambda=10^{-3}\text{ GeV}^2\text{nb}^{-2}$.}
   \end{subfigure}
   \hfill
   \begin{subfigure}[t]{0.47\textwidth}
     \centering
     \includegraphics[width=\textwidth]{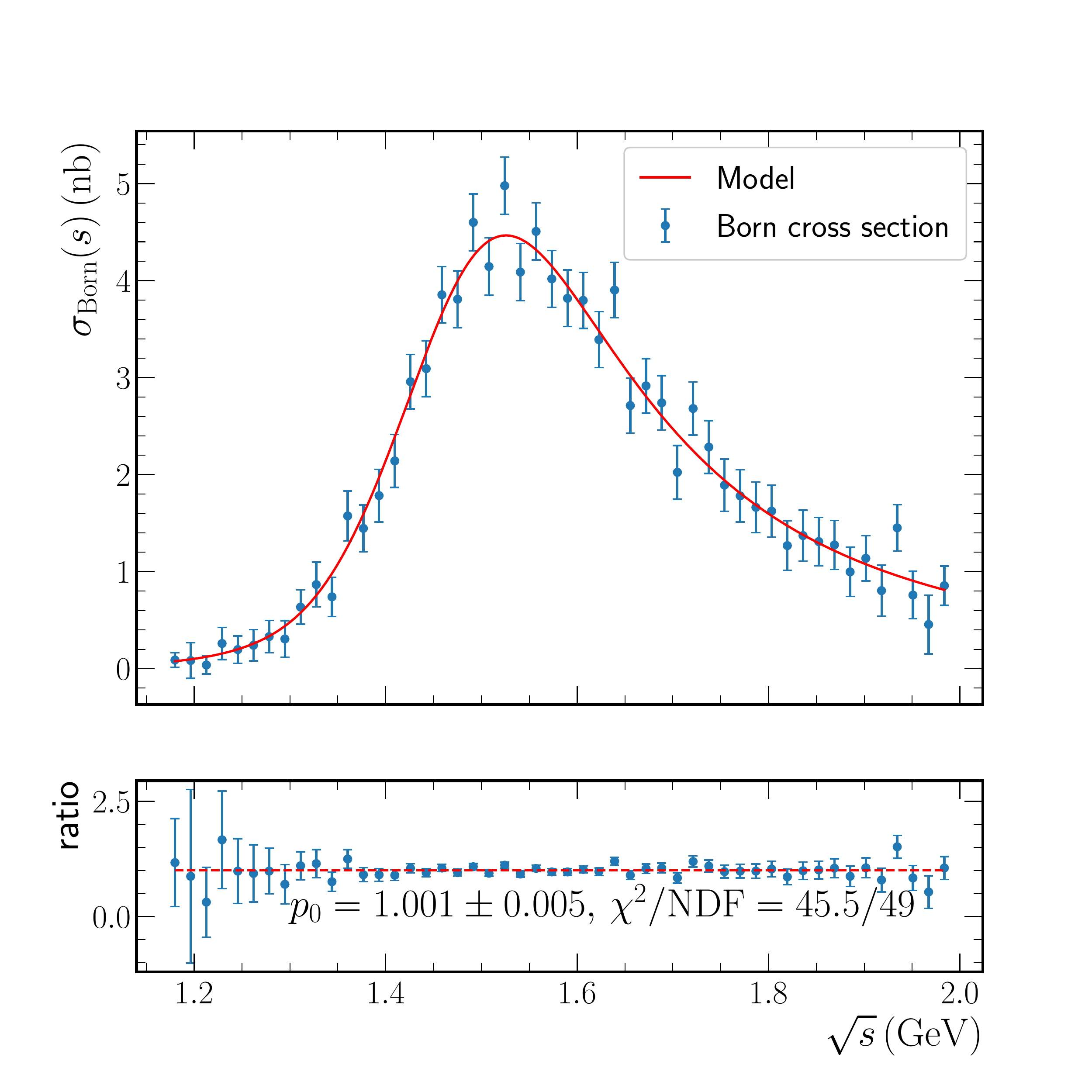}
     \caption{\label{fig:bcs_etapipi_tikh_enspread1p0em2_alpha1p0em2}
       $\lambda=10^{-2}\text{ GeV}^2\text{nb}^{-2}$.}
   \end{subfigure}
   \vfill
   \begin{subfigure}[t]{0.47\textwidth}
     \centering
     \includegraphics[width=\textwidth]{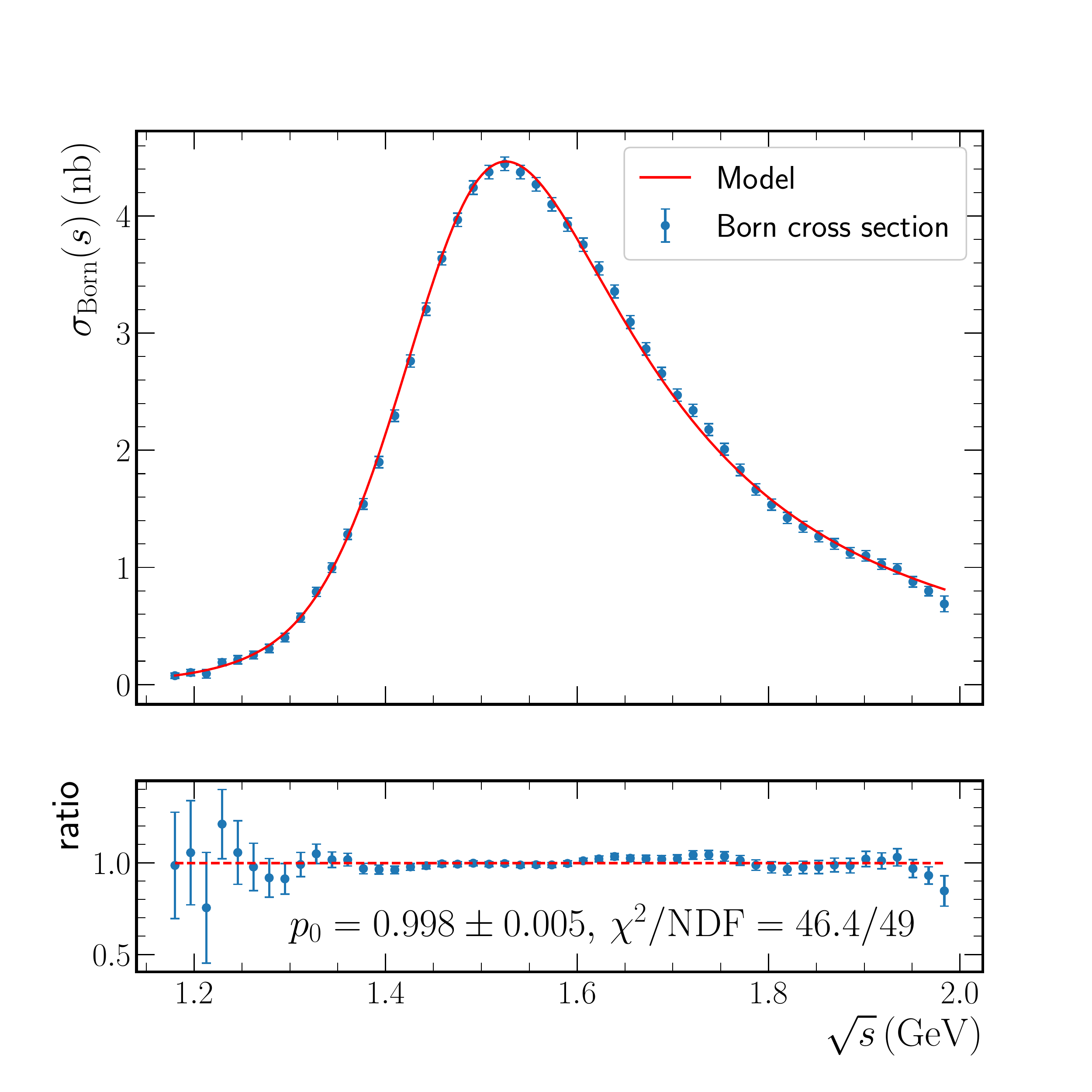}
     \caption{\label{fig:bcs_etapipi_tikh_enspread1p0em2_alpha1p0em0}
       $\lambda=1\text{ GeV}^2\text{nb}^{-2}$.}
   \end{subfigure}
   \hfill
   \begin{subfigure}[t]{0.47\textwidth}
     \centering
     \includegraphics[width=\textwidth]{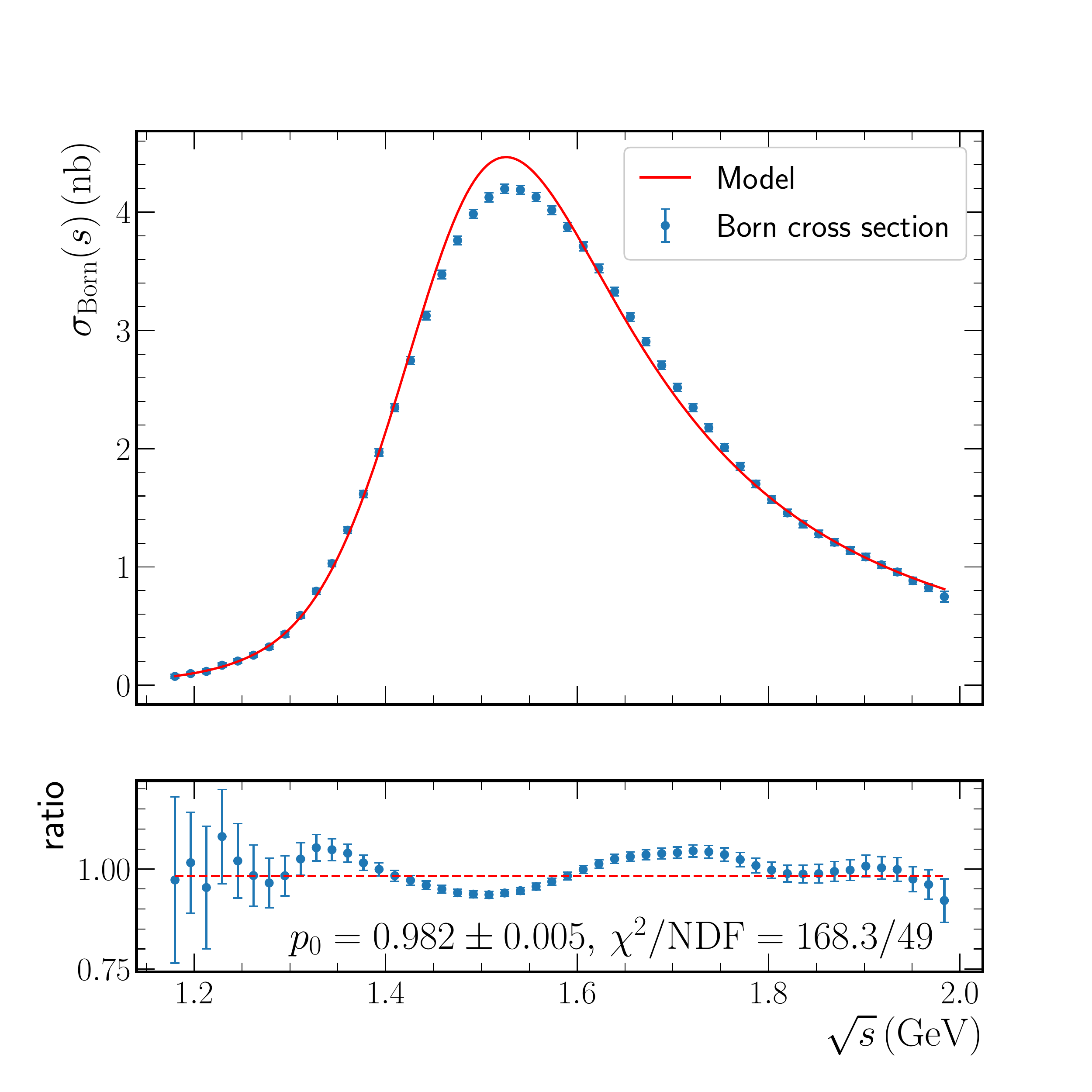}
     \caption{\label{fig:bcs_etapipi_tikh_enspread1p0em2_alpha5p0em0}
       $\lambda=5\text{ GeV}^2\text{nb}^{-2}$.}
   \end{subfigure}
   \caption{\label{fig:bcs_etapipi_tikh_enspread1p0em2_nonopt} Comparison of the
     numerical solution of
     eq.~\eqref{eq:visible-Born-relationship-kuraev-fadin-2-blured} and the
     model Born cross section of the $e^+e^-\rightarrow\eta\pi^+\pi^-$ process.
     The numerical solution is obtained using the Tikhonov regularization
     method. Each sub-figure corresponds to a separate regularization parameter
     $\lambda$. For each sub-figure, the parameter $\sigma_E$ is
     $20\text{ MeV}$.}
 \end{figure}

 Let us consider a numerical experiment with the
 $e^+e^-\rightarrow\eta\pi^+\pi^-$ cross section. We assume that there are $50$
 equally spaced c.m.\ energy points in the range from $1.18$ to $2.00$ GeV. We
 also assume that the visible cross section is generated at these points using
 some diagonal covariance matrix in the same way as it is done in
 section~\ref{sec:naive-method}. Next,
 eq.~\eqref{eq:tikhonov-regularized-equation} is solved using the generated
 visible cross section for different values of the regularization parameter
 $\lambda$. We also assume that in this numerical experiment the parameter
 $\sigma_E$ is $20\text{ MeV}$. Comparisons of the corresponding numerical
 solutions with the model Born cross section are shown in
 figure~\ref{fig:bcs_etapipi_tikh_enspread1p0em2_nonopt}. The model Born cross
 section is shown in this figure as a solid curve, and the numerical solution is
 shown in the form of points with error bars. The lower part of each sub-figure
 shows the ratio of the numerical solution to the model Born cross section in
 the form of points with error bars.
 Figure~\ref{fig:bcs_etapipi_tikh_enspread1p0em2_nonopt} shows that as the
 regularization parameter increases, the dependence of the numerical solution on
 the c.m.\ energy becomes more regular. However, for too large values of the
 regularization parameter, the numerical solution is systematically lower than
 the model Born cross section. Therefore, the problem of the optimal choice of
 the regularization parameter arises in a natural way. There are several
 semi-heuristic methods to solve this problem. In this work, the L-curve
 criterion is used. A more detailed description of the methods for choosing the
 optimal regularization parameter can be found in the
 works~\cite{Hansen1993,Hansen2010DiscreteIP,Kuusela2012}.

\begin{figure}[tbp]
   \centering
   \begin{subfigure}[t]{0.47\textwidth}
     \centering
     \includegraphics[width=\textwidth]{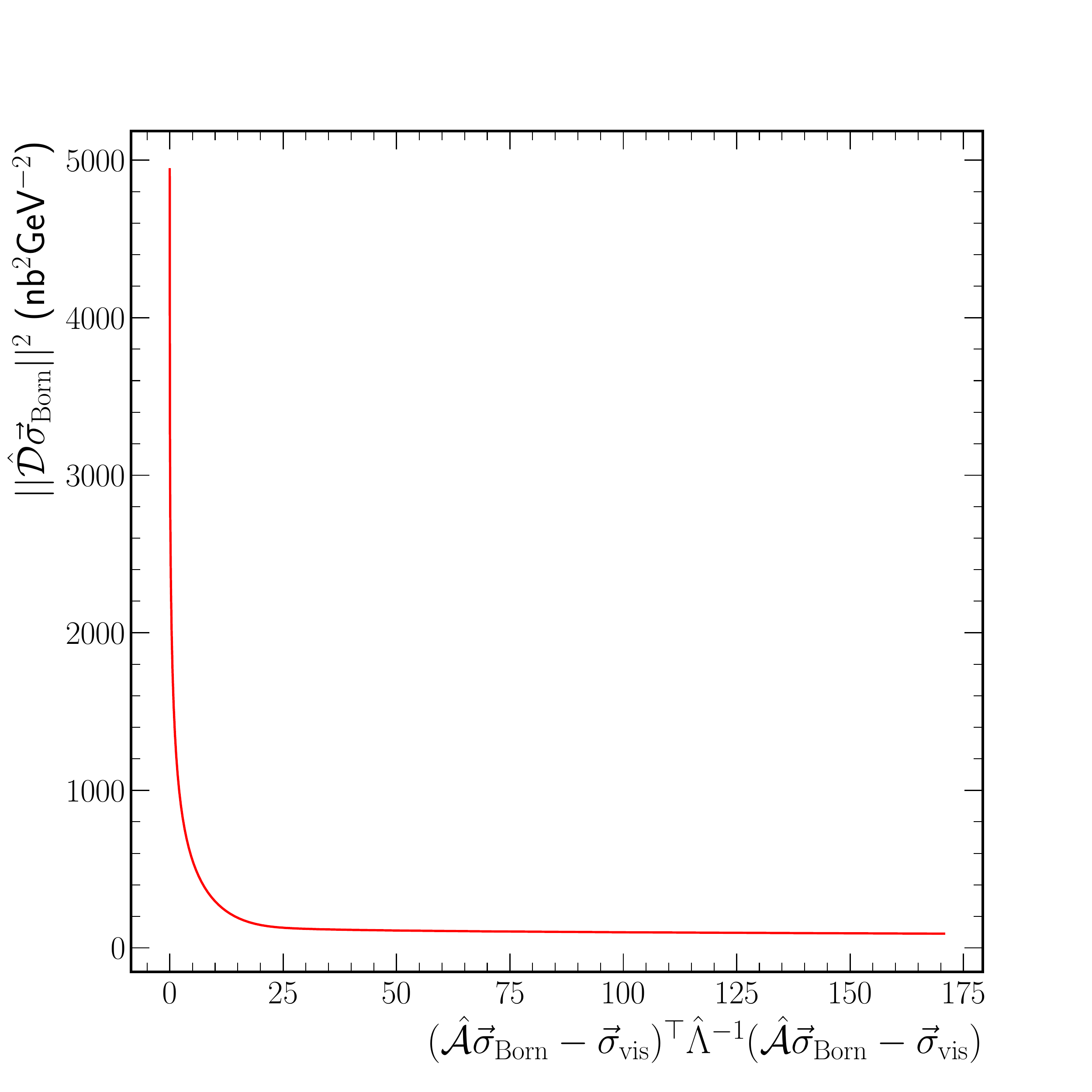}
     \caption{\label{fig:lcurve_etapipi_enspread1p0em2_errkoeff5em2} L-curve.}
   \end{subfigure}
   \hfill
   \begin{subfigure}[t]{0.47\textwidth}
     \centering
     \includegraphics[width=\textwidth]{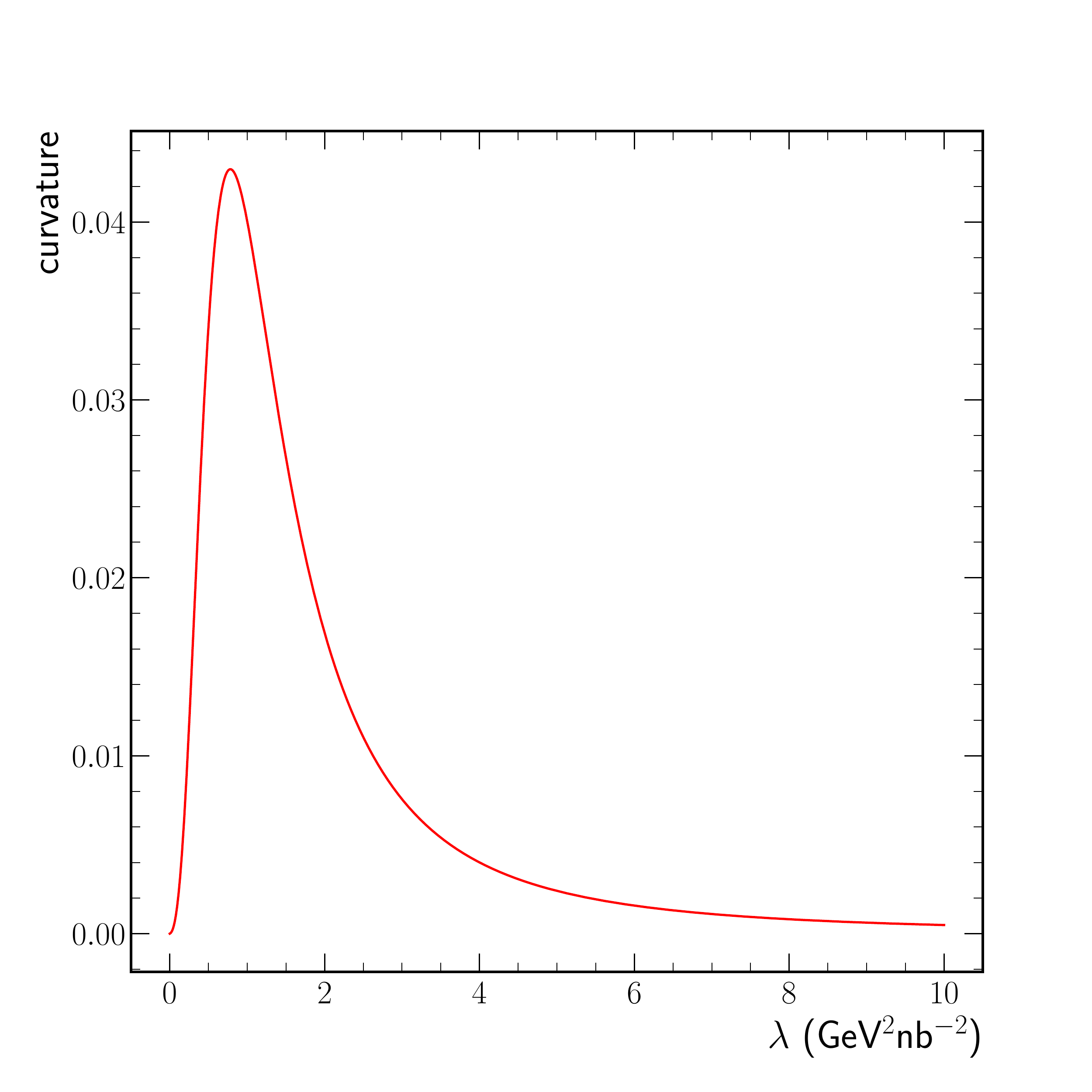}
     \caption{\label{fig:lcurve_curvature_etapipi_enspread1p0em2_errkoeff5em2}
       Curvature of the L-curve. The maximum curvature is achieved with the
       regularization parameter of $0.78\text{ GeV}^2\text{nb}^{-2}$.}
   \end{subfigure}
   \caption{\label{fig:lcurve_criterion_etapipi_enspread1p0em2} An example of an
     L-curve and its curvature. The L-curve and its curvature are obtained by
     solving eq.~\eqref{eq:visible-Born-relationship-kuraev-fadin-2-blured} with
     the parameter $\sigma_E = 20\text{ MeV}$. }
 \end{figure}

 Consider a graph in which the values of the first term from
 functional~\eqref{eq:chi2-tikhonov-functional} are plotted along the abscissa,
 and the values of the second term divided by the regularization parameter are
 plotted along the ordinate. Both the first and second terms depend on the
 regularization parameter, therefore the coordinates of the points on this graph
 depend on this parameter. The continuous dependence of these coordinates on the
 regularization parameter is called an L-curve. An example of an L-curve is
 shown in figure~\ref{fig:lcurve_etapipi_enspread1p0em2_errkoeff5em2}. There are
 two typical parts that can be distinguished in the L-curve. The first part of
 the L-curve is almost vertical and corresponds to small almost constant values
 of the first term of the functional~\eqref{eq:chi2-tikhonov-functional}. The value
 of the regularization parameter in this part of the L-curve is insufficient to
 suppress the effects of ill-posedness of the problem, therefore the value of
 the term $\|\hat{\mathcal{D}}\bm{\sigma}_{\rm Born}\|^2$ in this part of the
 L-curve can be large. On the contrary, the second part of the L-curve is almost
 horizontal and corresponds to smaller almost constant values of the term
 $\|\hat{\mathcal{D}}\bm{\sigma}_{\rm Born}\|^2$ and large values of the first
 term of the functional~\eqref{eq:chi2-tikhonov-functional}. In this part of the
 L-curve, the value of the regularization parameter is too large, which leads to
 a systematically underestimated numerical solution in comparison with the exact
 Born cross section. Therefore, the optimal regularization parameter corresponds
 to the region of the L-curve located between the two considered parts. In this
 region, the curvature of the L-curve reaches a maximum. Thus, the optimal value
 of the regularization parameter can be estimated as the value of this parameter
 at which the curvature of the L-curve reaches its maximum. An example of the
 curvature of the L-curve is shown in
 figure~\ref{fig:lcurve_curvature_etapipi_enspread1p0em2_errkoeff5em2}.

 It should be noted that very often the L-curve is brought to a log-log scale in
 order to emphasize its typical features. The L-curve curvature is also
 calculated using the L-curve plot at this scale. It was checked that in the
 case of the considered problem, the L-curve criteria in linear and log-log
 scales give approximately the same results. However, in the case of the log-log
 scale, an additional local maximum of the curvature appears near small values
 of the regularization parameter, which can be falsely interpreted as the
 maximum curvature at which the optimal value of the regularization parameter is
 achieved. For this reason, in this work, the L-curve is plotted on a linear
 scale and its curvature is calculated on the same scale.

 The maximum of the L-curve curvature shown in
 figure~\ref{fig:lcurve_criterion_etapipi_enspread1p0em2} is achieved at a
 regularization parameter of approximately $0.78\text{ GeV}^2\text{nb}^{-2}$. A
 comparison of the numerical solution corresponding to this parameter with the
 model Born cross section is shown in
 figure~\ref{fig:bcs_ratio_etapipi_errkoeff5em2_enspread1p0em2_method_tikhonov_optalpha0p721}.
 This figure shows that the regularized numerical solution is in good agreement
 with the model Born cross section, although the error bars of this solution
 do not represent the correct $68.27\%$ confidence intervals.

 \section{\label{sec::ISRSolver}ISRSolver package}
 Examples of numerical solutions to
 equations~\eqref{eq:visible-Born-relationship-kuraev-fadin-2} and
 \eqref{eq:visible-Born-relationship-kuraev-fadin-2-blured} considered in this
 paper were obtained using the ISRSolver package. This package is written using
 the C++ programming language and also has a Python API for calling some
 functions. The package includes tools for obtaining a numerical solution using
 both the naive method and the Tikhonov regularization method. Tools for
 verifying these numerical solutions are also included. It should be noted that
 the ISRSolver can be used as a library, for example, to find the convolution of
 the kernel with an arbitrary function or to implement conventional methods for
 obtaining a Born cross section. The source code of the package can be found in
 the repository \url{https://github.com/sergeigribanov/ISRSolver}.

 \section{Summary}
 
 In this work, we considered the problem of finding a numerical solution to
 eq.~\eqref{eq:visible-Born-relationship-kuraev-fadin-2} or
 eq.~\eqref{eq:visible-Born-relationship-kuraev-fadin-2-blured} using visible
 cross-section data. The problems given by these equations are ill-posed.
 However, due to the fact that the condition number of the matrix of the
 integral operator from eq.~\eqref{eq:visible-Born-relationship-kuraev-fadin-2}
 is comparable to unity, the problem specified by this equation can be solved
 with a good accuracy using the naive method, i.e.\ without using
 regularization. The condition number in this case is of the order of unity due
 to the fact that the kernel of the integral operator is a function that rapidly
 decreases with increasing $x$. The work also shows that the naive method can be
 used to find a numerical solution to
 eq.~\eqref{eq:visible-Born-relationship-kuraev-fadin-2-blured} at values of the
 c.m.\ energy spread that are small compared to the distances between
 c.m.\ energy points. Otherwise, the numerical solution found using the naive
 method has a large scatter due to the ill-posedness of the considered problem.
 The naive method consists in the direct solving of a system of linear equations
 that approximately describes the original integral equation. In the case of
 this work the integral operator matrix is a full rank square matrix. Therefore,
 according to the work~\cite{Kuusela2012}, this covariance matrix describes the
 variability of a Born cross section.

 If the c.m.\ energy spread is greater or comparable to the distances between
 c.m.\ energy points, the Tikhonov regularization method can be used to
 solve eq.~\eqref{eq:visible-Born-relationship-kuraev-fadin-2-blured}. However,
 the applicability of this method is limited by the fact that the diagonal
 elements of the covariance matrix of the regularized numerical solution do not
 represent $68.27\%$ confidence intervals~\cite{Kuusela2012} for a Born cross
 section at different c.m.\ energies, since this numerical solution is biased.

 The advantages of the naive method are its model independence and the
 possibility of obtaining the covariance matrix of a Born cross section in a
 simple way. The main proposal of this work is that the naive method can be used
 to find a Born cross section using visible cross section data in
 cases where it is possible, i.e.\  when there is a sufficient density of
 c.m.\ energy points to interpolate a cross section, and the
 c.m.\ energy spread is small or not taken into account. Otherwise,
 some of the conventional methods should be used, for example, a
 model-dependent fit to a visible cross section using
 eq.~\eqref{eq:visible-Born-relationship-kuraev-fadin-2} or
 eq.~\eqref{eq:visible-Born-relationship-kuraev-fadin-2-blured}.

 \section{Acknowledgments}
 The authors are grateful to V.P.\ Druzhinin, A.A.\ Korol and the members of the CMD-3
 collaboration for helpful discussions and advice. The work has been
 partially supported by the Russian Foundation for Basic Research grant No. 20-02-00496 A.

 \bibliography{article-ISR-inverse-problem} 
\end{document}